\documentclass[showpacs,twocolumn,pra,superscriptaddress,notitlepage]{revtex4-1}
\usepackage{qcircuit}
\usepackage[dvips]{graphicx}
\usepackage{amsmath,amssymb,amsthm,mathrsfs,amsfonts,dsfont}
\usepackage{subfigure, epsfig}
\usepackage{braket}
\usepackage{bm}
\usepackage{fancyhdr}
\usepackage{enumerate}
\usepackage{color}
\usepackage{multirow}
\usepackage{hyperref}
\hypersetup{colorlinks=true, linkcolor=blue, citecolor=blue, urlcolor=black }

\newtheorem{proposition}{Proposition}

\newcommand{\tr}{\mathrm{Tr}}
\newcommand{\comments}[1]{}

\newcommand{\Hamilt}{{H}}

\begin{document}
\title{Quantum simulation with hybrid tensor networks}

\begin{abstract}
Tensor network theory and quantum simulation are respectively the key classical and quantum computing methods in understanding quantum many-body physics. 
Here, we introduce the framework of hybrid tensor networks with building blocks consisting of measurable quantum states and classically contractable tensors, inheriting both their distinct features in efficient representation of many-body wave functions.  With the example of hybrid tree tensor networks, we demonstrate efficient quantum simulation using a quantum computer whose size is significantly smaller than the one of the target system. We numerically benchmark our method for finding the ground state of 1D and 2D spin systems of up to $8\times 8$ and $9\times 8$ qubits with operations only acting on $8+1$ and $9+1$ qubits,~respectively. 
Our approach sheds light on simulation of large practical problems with intermediate-scale quantum computers, with potential applications in chemistry, quantum many-body physics, quantum field theory, and quantum gravity thought experiments.
\end{abstract}
\date{\today}

\author{Xiao Yuan}
\email{xiaoyuan@pku.edu.cn}
\affiliation{Center on Frontiers of Computing Studies, Department of Computer Science, Peking University, Beijing 100871, China}
\affiliation{Stanford Institute for Theoretical Physics, Stanford University, Stanford California 94305, USA}

\author{Jinzhao Sun}
\email{jinzhao.sun@physics.ox.ac.uk}
\affiliation{Clarendon Laboratory, University of Oxford, Parks Road, Oxford OX1 3PU, United Kingdom}

\author{Junyu Liu}
\email{jliu2@caltech.edu}
\affiliation{Walter Burke Institute for Theoretical Physics, California Institute of Technology, Pasadena, California 91125, USA}
\affiliation{Institute for Quantum Information and Matter, California Institute of Technology, Pasadena, CA 91125, USA}

\author{Qi Zhao}
\email{zhaoq@umd.edu}
\affiliation{Joint Center for Quantum Information and Computer Science, University of Maryland, College Park, Maryland 20742, USA}

\author{You Zhou}
\email{you\_zhou@g.harvard.edu}
\affiliation{Department of Physics, Harvard University, Cambridge, Massachusetts 02138, USA}

\maketitle

A major challenge in studying quantum many-body physics stems from the hardness of efficient representation of quantum wave functions.
The tensor network~(TN) theory, originated from the density matrix renormalization group (DMRG) for 1D Hamiltonians~\cite{PhysRevLett.69.2863,PhysRevB.55.2164}, provides a potential solution by describing the state with a network consisting of low-rank tensors~\cite{orus2019tensor}. 
Despite its notable success in various problems, 
{the TN theory is inadequate to represent arbitrary systems, such as thoses behaving volume-law of entanglement.}
This motivates an alternative approach of quantum simulation, which uses a controlled quantum hardware to represent the target quantum system naturally~\cite{Feynman1982}. Quantum simulation can be used for studying complex many-body systems, such as 
quantum chemistry and the Hubbard model~\cite{RevModPhys.92.015003,cao2019quantum}. While conventional quantum simulation algorithms require universal quantum computing, which is challenging to current technology~\cite{PhysRevA.95.032338}, whether near-term quantum devices~\footnote{{Near-term quantum devices refers to quantum hardware processing tens to hundreds of qubits with relatively noisy operations. These types of devices are in hand now and will be greatly improved in the near future, although they are yet insufficient to realize universal quantum computing. A potential use of near-term quantum devices is to demonstrate quantum advantage and develop specific applications, such as quantum chemistry and materials.}} are capable of solving realistic problems remains open~\cite{preskill2018quantum,altman2019quantum,wang2020noise,cerezo2020impact,endoreview,cerezo2020variational,bharti2021noisy}. Major technological challenges include whether we can control a sufficient number of qubits and whether the gate fidelity is sufficiently low to guarantee the calculation accuracy.

Here,  we propose a hybrid TN approach to address these challenges. Leveraging the ability of TNs and quantum computers in efficient classical and quantum representation of quantum states, we introduce a framework of hybrid TN, which enables simulation of large systems using a small quantum processor with a shallow circuit. 
Previous studies along this line include chemistry computation beyond the active-space approximation~\cite{PhysRevX.10.011004}, concatenation of quantum states to a matrix product state (MPS)~\cite{barratt2020parallel}, etc. Our result unifies these existing task-tailored schemes; 
Yet, more importantly, it provides the basis for general hybrid classical-quantum representation of many-body wave functions that is applicable to broad problems. 
We show this by considering an example of hybrid tree TNs and demonstrating its application in studying static and dynamic problems of quantum systems~\cite{yuan2019theory,hackl2020geometry}. We numerically test our method in finding ground states of 1D spin clusters and 2D spin lattices with up to $8\times8$ and $9\times 8$ qubits.

\begin{figure}[t]\centering
  \includegraphics[width=.95\linewidth]{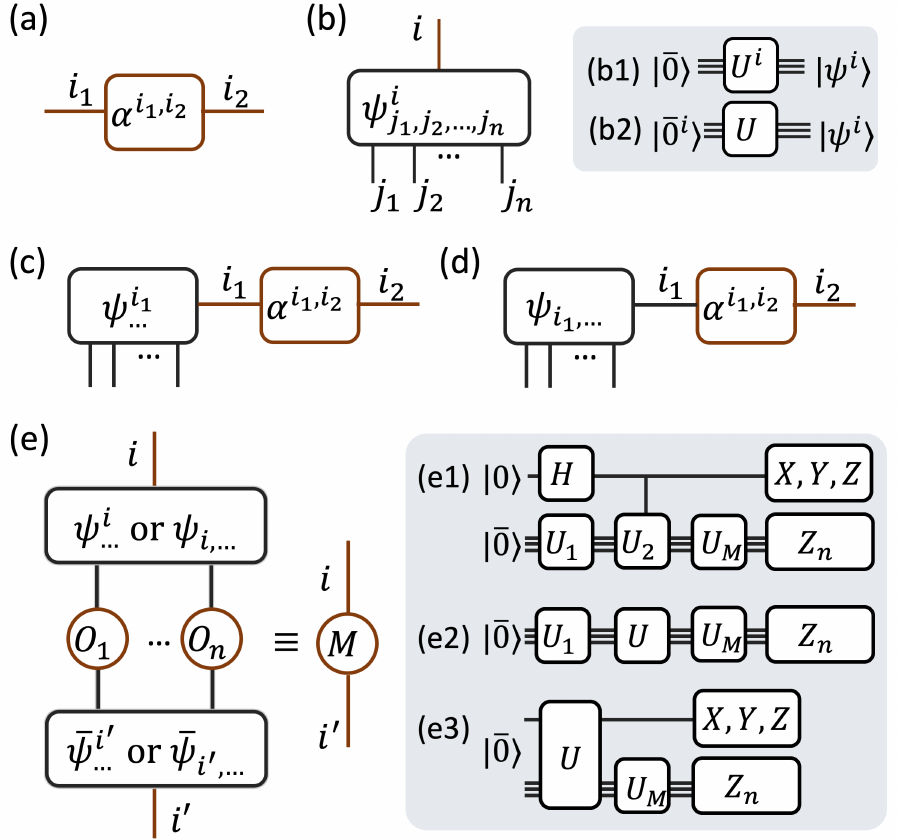}
  \caption{Hybrid tensors and tensor contraction. 
     (a) A low-rank classical tensor.
   (b) An $n+1$ rank tensor with $n$ indices representing an $n$-partite quantum system and $1$ classical index. Each state could be prepared with (b1) different unitary as $\ket{\psi^i}=U^i\ket{\bar 0}$ or (b2) different initial states as $\ket{\psi^i}=U\ket{\bar 0^i}$. 
   (c, d)  Tensor contraction between a quantum and a classical tensor. The contracted index could be (c) classical or (d) quantum, which shares the same mathematical definition, but is contracted in different ways.
  (e) Expectation values of local observables for a rank-$(n+1)$ tensor as a hermitian observable $M^{i',i}=\bra{\psi^{i'}}O_1\otimes O_2\otimes \cdots\otimes O_n\ket{\psi^i}$. 
  (e1) Suppose the index $i$ is classical and $\ket{\psi^i}=U^i\ket{\bar 0}$, we get each $M^{i',i}$ by measuring the ancillary qubit in the three Pauli bases and the other $n$ qubits in the $Z$ basis, with $U_1=U^i$, $U_2=U^{i'}(U^i)^\dag$, and $U_M$ being the unitary that rotates to the observable basis.
  (e2) Suppose the index $i$ is classical and $\ket{\psi^i}=U\ket{\bar 0^i}$, we use $U_1$ to prepare four input states $\ket{\bar0^i}, \ket{\bar0^{i'}}, (\ket{\bar 0^i}+ \ket{\bar 0^{i'}})/\sqrt{2}, (\ket{\bar 0^i}+i \ket{\bar 0^{i'}})/\sqrt{2}$ and each $M^{i',i}$ corresponds to a linear combination of the measurement results. 
  (e3) Suppose the index $i$ is quantum, after applying the unitary $U$ for preparing the state $\ket{\psi}=U\ket{\bar 0}$, we measure $n$ qubits in the $Z$ basis and the other qubit in the Pauli $X$, $Y$, and $Z$ bases. 
  }\label{fig:FigExpectation}
\end{figure}

\emph{\textbf{Framework}.---}We first introduce the framework of hybrid tensor networks. We focus on qubits and the results can be straightforwardly generalized to higher dimensions. 
A rank-$n$ tensor, when regarded as a multidimension array, can be represented as $T_{j_1,j_2,\dots,j_n}$ with $n$ indices. 
The amplitude of an $n$-partite quantum state in the computational basis corresponds to a rank-$n$ tensor 	$\ket{\psi}=\sum_{j_1,j_2,\dots,j_n}\psi_{j_1,j_2,\dots,j_n}\ket{j_1}\ket{j_2}\dots\ket{j_n}$.
A classical TN consists of low-rank tensors,  see Fig.~\ref{fig:FigExpectation}(a), which can efficiently describe physical states that lie in a small subset of the whole Hilbert space. 
For example, a MPS~\cite{schollwock2011density} 
	$\ket{\psi} = \sum_{j_1\cdots j_n} \tr[A^{j_1}\dots A^{j_n}]\ket{j_1\dots j_n}$ consists of rank-3 tensors with a small bond dimension $\kappa$ of each matrix $A^{j_k}$, and compresses the state dimension from $O(2^n)$ to $O(n\kappa^2)$. 
A quantum computer prepares  states $\ket{\psi}$ by applying a unitary circuit to some initial states. 
We can further add a classical index to the $n$-qubit state to form a rank-$(n+1)$ tensor $\{\ket{\psi^i}\}$; see Fig.~\ref{fig:FigExpectation}(b).

Regarding low-rank tensors as classical tensors (superscript index) and quantum states as quantum tensors (subscript index), we define hybrid TNs as networks constructed by connecting both classical and quantum tensors. 
For example, the tensor $A^{i_1,i_2}$ represents a classical tensor with two classical indices and $\psi^{i}_{j_1,j_2,\dots,j_n}$ represents a set of $n$-partite quantum states.
Two tensors, being either classical or quantum, are connected by following the conventional contraction rule, such as $C^{i_1,i_3}=\sum_{i_2}A^{i_1,i_2}B^{i_2,i_3}$. For example, we show the connections of a quantum and a classical tensor in Fig.~\ref{fig:FigExpectation}(c, d) and refer to \cite{NoteX} for general cases.

\begin{figure*}[t]\centering
  \includegraphics[width=.9\linewidth]{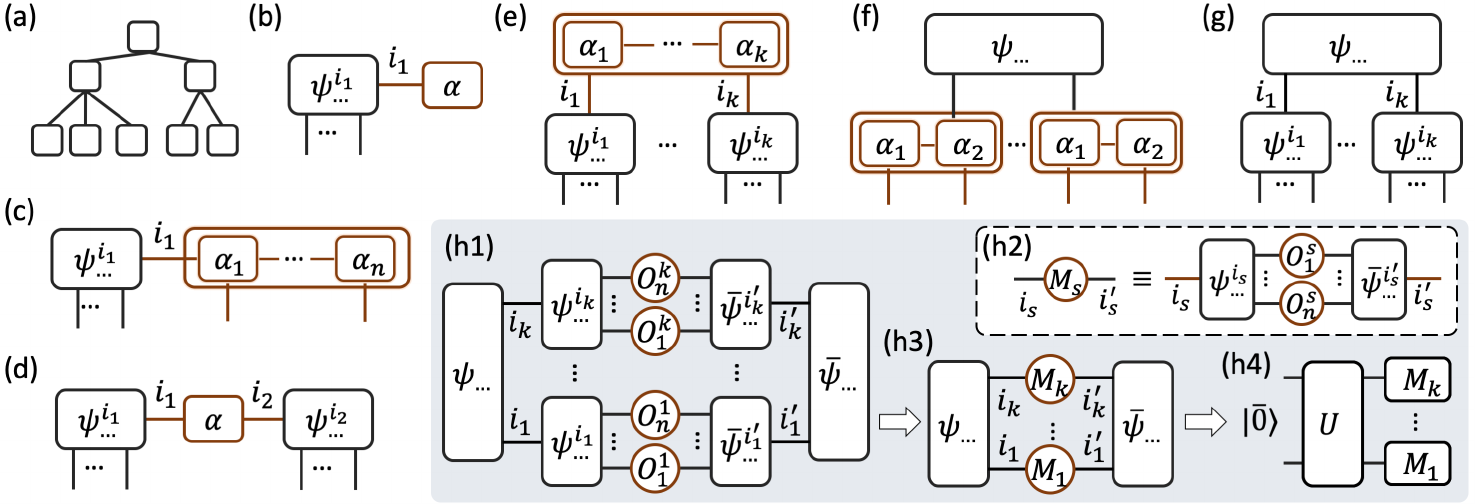}
  \caption{Hybrid tree tensor networks. 
  (a) An example tree structure.
  (b) Connection of a quantum tensor and a classical tensor.
  (c) Connection of a quantum tensor and a classical TN.
  (d) Connection of two quantum tensors via a classical tensor. 
  (e) Generalization of (d) with multiple subsystems.
  (f) Using classical tensors to represent local correlation and a quantum tensor to represent correlations between subsystems.
  (g) A quantum-quantum network. 
  (h1-4) An example for calculating expectation values of (g). 
To calculate (h1) the expectation value of local observables $\otimes_{i=1}^k\otimes_{j=1}^n O_j^i$, we first calculate (h2) the observable $M_s^{i_s',i_s}$ for each tensor on the second layer with quantum circuits shown in Fig.~\ref{fig:FigExpectation}(e1,e2), which converts to the contraction (h3) and the measurement of {the quantum circuit} (h4).
  }\label{fig:ansatz}
\end{figure*}

While the mathematical rules are the same, classical and quantum tensors are contracted in two different ways via tensor contraction and quantum state measurement,~respectively. 
For a rank-$(n+1)$ quantum tensor, we show how to calculate expectation values of local observables in Fig.~\ref{fig:FigExpectation}(e). 
{Calculating expectation values of general hybrid TNs works similarly, although the  complexity highly depends on the network and the contraction order  (see~\cite{NoteX} for details). } 

\vspace{.2cm}

\emph{\textbf{Hybrid tree TN}.---}Contracting a general-structured TN may have exponential complexity, explaining why conventional TN theories consider networks with specific topology, including 1D MPS~\cite{fannes1992finitely,klumper1991equivalence,klumper1993matrix}, 2D projected entangled pair states {(PEPS, approximate contraction)}~\cite{verstraete2004renormalization}, tree TNs (TTN)~\cite{PhysRevA.74.022320}, multiscale entanglement renormalization ansatz (MERA)~\cite{PhysRevLett.99.220405}, etc. 
Here we consider hybrid TNs with a tree structure such as in Fig.~\ref{fig:ansatz}(a), which admits an efficient tensor contraction. 
Each node is either a quantum tensor or any efficiently contractable classical TN.

We consider several tree structures with depth 2.  By connecting a classical tensor to a quantum tensor, we extend the state subspace as in Fig.~\ref{fig:ansatz}(b) or represent virtual qubits as in Fig.~\ref{fig:ansatz}(c). 
Specifically, denoting the classical tensor as $\alpha^i$, the network in Fig.~\ref{fig:ansatz}(b) describes a subspace
$\{\ket{\psi} = \sum_i \alpha^i \ket{\psi^i}\}$,
which, when applied in quantum simulation, is a generalization of the subspace expansion method that has been widely used for finding excited energy spectra~\cite{PhysRevA.95.042308}, error mitigation~\cite{PhysRevX.8.011021}, and error correction~\cite{mcclean2020decoding}.
For the network in Fig.~\ref{fig:ansatz}(c), 
it describes the scenario where we use a quantum state and a classical tensor to respectively represent the active and virtual space or multidegrees of freedom, as in quantum chemistry and condensed matter~\cite{PhysRevX.10.011004,wouters2016practical,lanata2015phase,rohringer2018diagrammatic}. We can further connect two quantum tensors via a classical tensor as in Fig.~\ref{fig:ansatz}(d), representing weakly interacted two subsystems as considered in Ref.~\cite{barratt2020parallel}.

Its generalization to multiple subsystems is given in   Fig.~\ref{fig:ansatz}(e), where entanglement of local subsystems is described by quantum states while the correlation between local subsystems is described classically. 
Such a hybrid TN can be useful for describing weakly coupled subsystems, such as clustered systems.
We can also use classical tensors to represent local correlations while a quantum tensor to represent the nonlocal correlation, as shown in Fig.~\ref{fig:ansatz}(f), which may be useful for studying topological order with long-range entanglement~\cite{PhysRevB.82.155138,PhysRevB.83.035107}. The construction of tree networks can be understood as an effective renormalization procedure, and other classical TNs such as the multis caleentanglement renormalization  ansatz can be similarly used \cite{NoteX}.
In addition to representing either local correlations or non-local correlations with classical tensors, we can represent both of them with quantum states, as shown in Fig.~\ref{fig:ansatz}(g), and expectation values of local observables can be efficiently obtained in Fig.~\ref{fig:ansatz}(h). 

{Our results can be naturally generalized to an arbitrary tree structure. For a tree with maximal depth $D$, maximal degree $g$, and bond dimension $\kappa$, hybrid TTNs represent a system of $N=\mathcal O(g^{D-1})$ qubits. The number of circuits and the classical cost (using MPS) for measuring local observables scale as $\mathcal O(N\kappa^2)$ and $\mathcal O(Ng\kappa^4)$, respectively. We also show the cost for trees with loops and its capability in representing entanglement beyond the area law in \cite{NoteX}.}
Since the hybrid TTN represents a large set of quantum states and admits efficient calculation of local observables, it can be used for variational quantum simulation for solving static and dynamic problems of large quantum systems.

\begin{figure*}[t]\centering \includegraphics[width=1\linewidth] {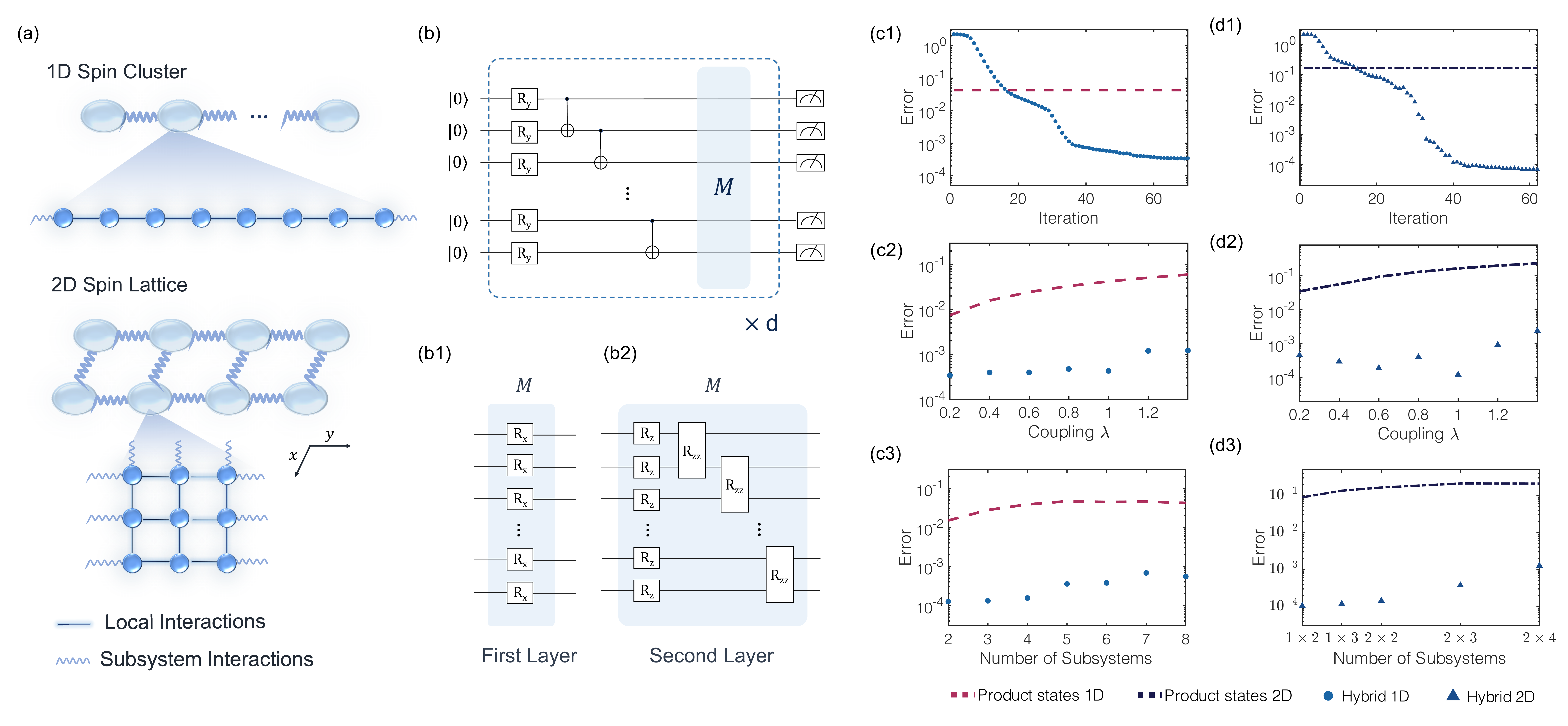}
\caption{
Numerical simulation for 1D and 2D quantum systems with hybrid TTN.
{
(a) 1D spin cluster and 2D spin lattice with interactions (thin lines) on the boundary. The interactions of subsystems are represented by thick lines. We group 8 adjacent qubits and $3 \times 3$  qubits on a square sublattice as subsystems for the 1D and 2D systems, respectively. 
(b) The ansatz circuit for the quantum tensors in Fig.~\ref{fig:ansatz}(g). 
The circuits of both layers share similar structures with $d$ repetitions of circuits in the dashed box. 
Here, $R_{\alpha}$ ($\alpha\in \{\hat X,\hat Y,\hat Z\}$)
represents single-qubit rotation around $\alpha$-axis and the two-qubit gate is $R_{ZZ}(\theta_i)=e^{-i\theta_i \hat Z\otimes \hat  Z}$. 
The rotation angle (parameter) for each gate is initialized from a small random value and updated in each variational cycle.
The circuit depths for $V$ (first layer tensor) and $U$ (second layer tensor) are  $d (V)=6$ and $d (U)=8$. The additional unitary $M$ is inserted at the $1$st and $[d/2+1]$th block of the first layer (b1) and the second layer (b2).
(c)-(d) Simulation results of the ground state energy. For the 1D and 2D cases, we compare $E$ to the reference results $E_0=E_{\rm MPS}$ and $E_0=E_{\rm PEPS}$ obtained from a standard DMRG with bond dimension $\kappa= 32$ and 
from PEPS imaginary time evolution with bond dimension $\kappa = 5$ and maximum allowed bond dimension of $\tilde \kappa = 64$ during the contraction.
We use the relative error $1-E/E_0$ to characterize the  accuracy.
The red dashed line (1D) and blue dash-dotted line (2D) correspond to the energy using tensor products of the ground state of local subsystems. The cyan dot (1D) and blue triangle (2D) are results obtained with hybrid TNs. 
(c1, d1) Convergence towards the ground state for the 1D $8\times 8$ and 2D $9\times 4$ systems with $\lambda=1$, respectively.
(c2, d2) Error versus different subsystem coupling strength $\lambda$ for the 1D $8\times 8$ and 2D $9\times 4$ systems, respectively. 
(c3, d3) Errors with different numbers  of local subsystems with $\lambda=1$, respectively. 
}}
  \label{fig:numer}
\end{figure*}

\emph{\textbf{Numerical simulation}.---}We test the effectiveness of hybrid TNs in finding ground states of 1D and 2D spin lattice systems with nearest-neighbor interactions and external fields in Fig.~\ref{fig:numer}. 
For 1D spin clusters, we regard each adjacent $n=8$ qubits as a subsystem and consider $k=2,3,\dots,8$ subsystems with $n\times k$ qubits. 
A general form of the Hamiltonian is $H = \sum\limits_{j = 1}^{k} {{H_j}}  + \lambda {H_{{\rm{int}}}}$, where $H_j=   \sum\limits_{i = 1}^{7} { {{f}{\hat{Z}_{8j+i}}{\hat{Z}_{8j+i + 1}}} }  +\sum\limits_{i = 1}^{8} {\left( {g{\hat{X}_{8j+i}}+ h{\hat{Z}_{8j+i}}} \right)}$ and $H_{{\rm{int}}}=\sum_{j = 1}^{k-1} {  {f_j}{\hat{Z}_{8j}}{\hat{Z}_{8j + 1}}}$  represent the Hamiltonian of the $j$th subsystem and their interactions, respectively, with interaction strength $\lambda$.
Here $\hat X_i$ and $\hat Z_i$ are Pauli operators acting on the $i$th qubit.
{
For the 2D $n\times k$ spin lattice, we group each $n=3 \times 3$ qubits on a small square lattice as a subsystem and  consider $k=N_x \times N_y$ subsystems with $N_x$~($N_y$) subsystems along $x$ ($y$) direction. 
The 2D Hamiltonian is $H = \sum_{\braket{i,j}} f_{ij} \hat{Z}_i \hat{Z}_j + \sum_i \left( g{\hat{X}_{i}}+ h{\hat{Z}_{i}} \right)$, where $\braket{i,j}$ represents all the nearest-neighbor pairs on a square lattice.
We consider that the interactions in each subsystem are identical $f=1$, while interactions on the boundary of nearest-neighbor subsystem $\{f_j\}$ or $\{f_{i,j}\}$ are generated randomly from $[0,1]$, as shown in Fig.~\ref{fig:numer}(a). 
The parameters of the external fields are set as $h =1/\pi= 0.32$ and $g = 0.5$.
}

Considering the hybrid TTN of Fig.~\ref{fig:ansatz}(g), 
the first layer state and the $j$th subsystem of the second layer are generated as $\ket{\psi}=V(\vec\theta_0)\ket{\bar 0^0}=\sum\alpha _{{i_1}, \ldots ,{i_{k}}}\ket{{{i_1}, \ldots ,{i_{k}}}}$ and  $\ket{\psi^{i_j}_j(\vec\theta_j)}=U(\vec\theta_j)\ket{\bar 0^{i_j}}$, respectively, with $V$ and $U$ shown in Fig.~\ref{fig:numer}(b) and initial states $\ket{\bar 0^{i_j}}=\ket{i_j}^{\otimes n}$, $i_j\in\{0,1\}$. The hybrid TTN represents a quantum state $    \ket{\tilde \psi(\vec \theta)} = \sum\limits_{{i_1} \ldots {i_{k}}} {{\alpha _{{i_1}, \ldots ,{i_{k}}}}}(\vec\theta_0) \ket{{\psi _1^{{i_1}}}(\vec\theta_1) }  \otimes  \cdots  \otimes \ket{ {\psi_{k}^{{i_{k}}}}(\vec\theta_k)}$
with $\vec\theta=(\vec\theta_0,\vec\theta_1,\dots,\vec\theta_k)$ representing all the parameters. 
The state is automatically normalized since $\braket{ {\psi_{j}^{{i_{j}'}}}| {\psi_{j}^{{i_{j}}}}}=\delta_{i_{j}',i_{j}'}$. 
For parameters $\vec\theta$, we obtain the energy expectation value  $E(\vec\theta)=\braket{\tilde \psi(\vec \theta)|H|\tilde \psi(\vec \theta)}$ by following the contraction rule of Fig.~\ref{fig:ansatz}(h). 
We use  variational imaginary time evolution to minimize the energy $E(\vec\theta)$, which requires an ancillary qubit (see~\cite{mcardle2019variational,NoteX}). 
Thus the quantum systems needed for simulating the $8\times k$-qubit 1D and $9\times k$-qubit 2D systems need $8+1$ and $9+1$ qubits, respectively. 

{
We benchmark the calculation by comparing with open-boundary MPS for 1D systems and imaginary time evolution PEPS for 2D systems. 
We consider the relative error $1-E/E_0$ with the ground state  energy $E$ from hybrid TTN calculation, and $E_0$ from MPS or PEPS. 
In Fig.~\ref{fig:numer}(c1, d1), we study the convergence of ground state energy of 1D (c1) and 2D (d1) systems with coupling strength $\lambda=1$ on $8\times 8$ and $9\times 4$ qubits respectively, and show a relative error below $10^{-3}$.
Next, we study how the coupling strength or the number of subsystems affect the efficacy of hybrid TTN. We  present the calculation error with respect to different $\lambda$ for the $8\times8$-qubit 1D and $9\times 4$-qubit 2D systems in Fig.~\ref{fig:numer}(c2) and (d2), respectively. We find that although the error fluctuates with different coupling strength, which might owe to instability from the optimization, the error remains consistent around $10^{-3}$.
In Fig.~\ref{fig:numer}(c3, d3), we show the calculation error for the 1D with $k$ subsystems (c3) and 2D with $N_x \times N_y$ subsystems (d3) for $\lambda=1$, and we can achieve a desired simulation accuracy. 
These results with different coupling strength and number of subsystems verify the effectiveness and robustness of hybrid TTN method. We refer to Ref.~\cite{NoteX} for addtional simulation results and implemention details.
}

\emph{\textbf{Applications}.---}
{While we are not expecting the hybrid TN applies universally to arbitrary quantum systems in a similar way to universal quantum computers, we do anticipate hybrid TNs find its applications in a wide class of problems such as 
chemistry, many-body physics, quantum field theory, and quantum gravity~\cite{NoteX}. 
Assisted by classical computers, hybrid TN could more efficiently represent multipartite quantum states and bolster up the power of near-term quantum computers to significantly alleviate the limitations on the number of controllable qubits and circuit depth. }

{Ideas corresponding to simple hybrid TTNs of Fig.~\ref{fig:ansatz}(b, c) have been studied for representing excited energy eigenstates~\cite{PhysRevA.95.042308} and active + virtual orbitals~\cite{PhysRevX.10.011004} in electronic structure calculation. While the scheme in Ref.~\cite{PhysRevX.10.011004} assumed the configuration interaction ansatz for the virtual orbitals, a general classical TN may be used instead to improve the approximation (see Ref.~\cite{szalay2015tensor}). Another application of the hybrid TN is to go beyond the Born–Oppenheimer (BO) approximation, which may have applications in understanding radiationless decay between electronic states~\cite{worth2004beyond}, relativistic effects~\cite{reiher2014relativistic}, or conical intersections~\cite{domcke2004conical, domcke2012spectroscopy, ryabinkin2017conical}.
Hybrid TNs could also be used for investigating
the cluster systems~\cite{garlatti2017portraying,timco2009engineering}, toy models for high energy physics~\cite{Gu:2016oyy,Hayden:2007cs,Maldacena:2017axo,Maldacena:2018lmt,Yoshida:2017non,Maldacena:2016hyu,Maldacena:1997re}, correlated materials~\cite{lanata2015phase,ma2020quantum,wouters2016practical,rohringer2018diagrammatic}, as well as for exploring emergent quantum phenomena~\cite{zhou2017quantum,liu2019confined,sato2017topological,chubukov2008magnetism}, including searching for Majorana zero-modes and topological phase transitions~\cite{lutchyn2018majorana,you2018subsystem,alicea2012new,sau2010non,beenakker2013search}. We refer to Ref.~\cite{NoteX} for detailed discussions. }

\emph{\textbf{Discussion}.---}We proposed a framework of hybrid tensor networks and studied its application in variational quantum simulation.
Targeting at practical problems that have both classical and genuine quantum effects, hybrid TNs integrate the power of classical TN theories and quantum computing, and hence enable quantum simulation of large-scale problems with small quantum processors and shallower circuits. 
Our work is different from proposals of using a quantum computer to contract a classical TN~\cite{kim2017robust, kim2017noise,kim2017holographic,PhysRevResearch.1.023025}, whose generalization to hybrid TN could be a further work.  
Besides TNs, there also exist other powerful classical methods, such as quantum Monte Carlo~\cite{RevModPhys.73.33,RevModPhys.87.1067} and machine learning with neural networks~\cite{carleo2017solving,li2017deep}. A future direction is to investigate the combination of these methods with quantum computing. 
Another independent approach of simulating large quantum systems with small quantum computers is to decompose multiqubit gates into a mixture of single-qubit gates~\cite{peng2019simulating,bravyi2016trading,sun2021perturbative,mitarai2019constructing,2020arXiv200611174M}, whose combination with our method may lead to an interesting future direction.
After showing advantages over classical supercomputers in certain tasks~\cite{arute2019quantum, huang2020classical}, the next milestone is to solve practically meaningful and classically intractable tasks. Our work sheds light on the avenue for achieving this goal with near-term hardware.

\begin{acknowledgements}
\emph{Acknowledgements.}---We thank Suguru Endo, Patrick Hayden, Arthur Jaffe, Sam McArdle, John Preskill, Vlatko Vedral, and Ying Li for insightful, related discussions and comments. J.S. thanks Chenbing Wang for useful discussions on the numerics.
X.Y acknowledges support from the Simons Foundation.
J.L. is supported in part by the Institute for Quantum Information and Matter (IQIM), an NSF Physics Frontiers Center (NSF Grant PHY-1125565) with support from the Gordon and Betty Moore Foundation (GBMF-2644), and by the Walter Burke Institute for Theoretical Physics. Q.Z. acknowledges the support by the Department of Defense through the Hartree Postdoctoral Fellowship at QuICS. Y.Z. was supported in part by the Templeton Religion Trust under grant TRT 0159 and by the ARO under contract W911NF1910302.

\noindent\emph{Note added.}---Recently, a relevant work was posted by \textcite{fujii2020deep}. They suggest a divide-and-conquer method for solving a larger problem with smaller size quantum computers in a similar vein to Fig.~\ref{fig:ansatz}(g) (See \cite{NoteX} for details). While their work used a different language and focused on different examples, our results are consistent and can be compared.

\end{acknowledgements}

\bibliographystyle{apsrev4-1}
\bibliography{tensor.bib}

\begin{thebibliography}{89}%
\makeatletter
\providecommand \@ifxundefined [1]{%
 \@ifx{#1\undefined}
}%
\providecommand \@ifnum [1]{%
 \ifnum #1\expandafter \@firstoftwo
 \else \expandafter \@secondoftwo
 \fi
}%
\providecommand \@ifx [1]{%
 \ifx #1\expandafter \@firstoftwo
 \else \expandafter \@secondoftwo
 \fi
}%
\providecommand \natexlab [1]{#1}%
\providecommand \enquote  [1]{``#1''}%
\providecommand \bibnamefont  [1]{#1}%
\providecommand \bibfnamefont [1]{#1}%
\providecommand \citenamefont [1]{#1}%
\providecommand \href@noop [0]{\@secondoftwo}%
\providecommand \href [0]{\begingroup \@sanitize@url \@href}%
\providecommand \@href[1]{\@@startlink{#1}\@@href}%
\providecommand \@@href[1]{\endgroup#1\@@endlink}%
\providecommand \@sanitize@url [0]{\catcode `\\12\catcode `\$12\catcode
  `\&12\catcode `\#12\catcode `\^12\catcode `\_12\catcode `\%12\relax}%
\providecommand \@@startlink[1]{}%
\providecommand \@@endlink[0]{}%
\providecommand \url  [0]{\begingroup\@sanitize@url \@url }%
\providecommand \@url [1]{\endgroup\@href {#1}{\urlprefix }}%
\providecommand \urlprefix  [0]{URL }%
\providecommand \Eprint [0]{\href }%
\providecommand \doibase [0]{http://dx.doi.org/}%
\providecommand \selectlanguage [0]{\@gobble}%
\providecommand \bibinfo  [0]{\@secondoftwo}%
\providecommand \bibfield  [0]{\@secondoftwo}%
\providecommand \translation [1]{[#1]}%
\providecommand \BibitemOpen [0]{}%
\providecommand \bibitemStop [0]{}%
\providecommand \bibitemNoStop [0]{.\EOS\space}%
\providecommand \EOS [0]{\spacefactor3000\relax}%
\providecommand \BibitemShut  [1]{\csname bibitem#1\endcsname}%
\let\auto@bib@innerbib\@empty
\bibitem [{\citenamefont {White}(1992)}]{PhysRevLett.69.2863}%
  \BibitemOpen
  \bibfield  {author} {\bibinfo {author} {\bibfnamefont {S.~R.}\ \bibnamefont
  {White}},\ }\href {\doibase 10.1103/PhysRevLett.69.2863} {\bibfield
  {journal} {\bibinfo  {journal} {Phys. Rev. Lett.}\ }\textbf {\bibinfo
  {volume} {69}},\ \bibinfo {pages} {2863} (\bibinfo {year}
  {1992})}\BibitemShut {NoStop}%
\bibitem [{\citenamefont {Rommer}\ and\ \citenamefont
  {\"Ostlund}(1997)}]{PhysRevB.55.2164}%
  \BibitemOpen
  \bibfield  {author} {\bibinfo {author} {\bibfnamefont {S.}~\bibnamefont
  {Rommer}}\ and\ \bibinfo {author} {\bibfnamefont {S.}~\bibnamefont
  {\"Ostlund}},\ }\href {\doibase 10.1103/PhysRevB.55.2164} {\bibfield
  {journal} {\bibinfo  {journal} {Phys. Rev. B}\ }\textbf {\bibinfo {volume}
  {55}},\ \bibinfo {pages} {2164} (\bibinfo {year} {1997})}\BibitemShut
  {NoStop}%
\bibitem [{\citenamefont {Or{\'u}s}(2019)}]{orus2019tensor}%
  \BibitemOpen
  \bibfield  {author} {\bibinfo {author} {\bibfnamefont {R.}~\bibnamefont
  {Or{\'u}s}},\ }\href@noop {} {\bibfield  {journal} {\bibinfo  {journal}
  {Nature Reviews Physics}\ }\textbf {\bibinfo {volume} {1}},\ \bibinfo {pages}
  {538} (\bibinfo {year} {2019})}\BibitemShut {NoStop}%
\bibitem [{\citenamefont {Feynman}(1982)}]{Feynman1982}%
  \BibitemOpen
  \bibfield  {author} {\bibinfo {author} {\bibfnamefont {R.~P.}\ \bibnamefont
  {Feynman}},\ }\href {\doibase 10.1007/BF02650179} {\bibfield  {journal}
  {\bibinfo  {journal} {International Journal of Theoretical Physics}\ }\textbf
  {\bibinfo {volume} {21}},\ \bibinfo {pages} {467} (\bibinfo {year}
  {1982})}\BibitemShut {NoStop}%
\bibitem [{\citenamefont {McArdle}\ \emph {et~al.}(2020)\citenamefont
  {McArdle}, \citenamefont {Endo}, \citenamefont {Aspuru-Guzik}, \citenamefont
  {Benjamin},\ and\ \citenamefont {Yuan}}]{RevModPhys.92.015003}%
  \BibitemOpen
  \bibfield  {author} {\bibinfo {author} {\bibfnamefont {S.}~\bibnamefont
  {McArdle}}, \bibinfo {author} {\bibfnamefont {S.}~\bibnamefont {Endo}},
  \bibinfo {author} {\bibfnamefont {A.}~\bibnamefont {Aspuru-Guzik}}, \bibinfo
  {author} {\bibfnamefont {S.~C.}\ \bibnamefont {Benjamin}}, \ and\ \bibinfo
  {author} {\bibfnamefont {X.}~\bibnamefont {Yuan}},\ }\href {\doibase
  10.1103/RevModPhys.92.015003} {\bibfield  {journal} {\bibinfo  {journal}
  {Rev. Mod. Phys.}\ }\textbf {\bibinfo {volume} {92}},\ \bibinfo {pages}
  {015003} (\bibinfo {year} {2020})}\BibitemShut {NoStop}%
\bibitem [{\citenamefont {Cao}\ \emph {et~al.}(2019)\citenamefont {Cao},
  \citenamefont {Romero}, \citenamefont {Olson}, \citenamefont {Degroote},
  \citenamefont {Johnson}, \citenamefont {Kieferov{\'a}}, \citenamefont
  {Kivlichan}, \citenamefont {Menke}, \citenamefont {Peropadre}, \citenamefont
  {Sawaya} \emph {et~al.}}]{cao2019quantum}%
  \BibitemOpen
  \bibfield  {author} {\bibinfo {author} {\bibfnamefont {Y.}~\bibnamefont
  {Cao}}, \bibinfo {author} {\bibfnamefont {J.}~\bibnamefont {Romero}},
  \bibinfo {author} {\bibfnamefont {J.~P.}\ \bibnamefont {Olson}}, \bibinfo
  {author} {\bibfnamefont {M.}~\bibnamefont {Degroote}}, \bibinfo {author}
  {\bibfnamefont {P.~D.}\ \bibnamefont {Johnson}}, \bibinfo {author}
  {\bibfnamefont {M.}~\bibnamefont {Kieferov{\'a}}}, \bibinfo {author}
  {\bibfnamefont {I.~D.}\ \bibnamefont {Kivlichan}}, \bibinfo {author}
  {\bibfnamefont {T.}~\bibnamefont {Menke}}, \bibinfo {author} {\bibfnamefont
  {B.}~\bibnamefont {Peropadre}}, \bibinfo {author} {\bibfnamefont {N.~P.}\
  \bibnamefont {Sawaya}},  \emph {et~al.},\ }\href@noop {} {\bibfield
  {journal} {\bibinfo  {journal} {Chemical reviews}\ }\textbf {\bibinfo
  {volume} {119}},\ \bibinfo {pages} {10856} (\bibinfo {year}
  {2019})}\BibitemShut {NoStop}%
\bibitem [{\citenamefont {O'Gorman}\ and\ \citenamefont
  {Campbell}(2017)}]{PhysRevA.95.032338}%
  \BibitemOpen
  \bibfield  {author} {\bibinfo {author} {\bibfnamefont {J.}~\bibnamefont
  {O'Gorman}}\ and\ \bibinfo {author} {\bibfnamefont {E.~T.}\ \bibnamefont
  {Campbell}},\ }\href {\doibase 10.1103/PhysRevA.95.032338} {\bibfield
  {journal} {\bibinfo  {journal} {Phys. Rev. A}\ }\textbf {\bibinfo {volume}
  {95}},\ \bibinfo {pages} {032338} (\bibinfo {year} {2017})}\BibitemShut
  {NoStop}%
\bibitem [{Note1()}]{Note1}%
  \BibitemOpen
  \bibinfo {note} {{Near-term quantum devices refers to quantum hardware
  processing tens to hundreds of qubits with relatively noisy operations. These
  types of devices are in hand now and will be greatly improved in the near
  future, although they are yet insufficient to realize universal quantum
  computing. A potential use of near-term quantum devices is to demonstrate
  quantum advantage and develop specific applications, such as quantum
  chemistry and materials.}}\BibitemShut {Stop}%
\bibitem [{\citenamefont {Preskill}(2018)}]{preskill2018quantum}%
  \BibitemOpen
  \bibfield  {author} {\bibinfo {author} {\bibfnamefont {J.}~\bibnamefont
  {Preskill}},\ }\href@noop {} {\bibfield  {journal} {\bibinfo  {journal}
  {Quantum}\ }\textbf {\bibinfo {volume} {2}},\ \bibinfo {pages} {79} (\bibinfo
  {year} {2018})}\BibitemShut {NoStop}%
\bibitem [{\citenamefont {Altman}\ \emph {et~al.}(2019)\citenamefont {Altman},
  \citenamefont {Brown}, \citenamefont {Carleo}, \citenamefont {Carr},
  \citenamefont {Demler}, \citenamefont {Chin}, \citenamefont {DeMarco},
  \citenamefont {Economou}, \citenamefont {Eriksson}, \citenamefont {Fu} \emph
  {et~al.}}]{altman2019quantum}%
  \BibitemOpen
  \bibfield  {author} {\bibinfo {author} {\bibfnamefont {E.}~\bibnamefont
  {Altman}}, \bibinfo {author} {\bibfnamefont {K.~R.}\ \bibnamefont {Brown}},
  \bibinfo {author} {\bibfnamefont {G.}~\bibnamefont {Carleo}}, \bibinfo
  {author} {\bibfnamefont {L.~D.}\ \bibnamefont {Carr}}, \bibinfo {author}
  {\bibfnamefont {E.}~\bibnamefont {Demler}}, \bibinfo {author} {\bibfnamefont
  {C.}~\bibnamefont {Chin}}, \bibinfo {author} {\bibfnamefont {B.}~\bibnamefont
  {DeMarco}}, \bibinfo {author} {\bibfnamefont {S.~E.}\ \bibnamefont
  {Economou}}, \bibinfo {author} {\bibfnamefont {M.~A.}\ \bibnamefont
  {Eriksson}}, \bibinfo {author} {\bibfnamefont {K.-M.~C.}\ \bibnamefont {Fu}},
   \emph {et~al.},\ }\href@noop {} {\bibfield  {journal} {\bibinfo  {journal}
  {arXiv preprint arXiv:1912.06938}\ } (\bibinfo {year} {2019})}\BibitemShut
  {NoStop}%
\bibitem [{\citenamefont {Wang}\ \emph {et~al.}(2020)\citenamefont {Wang},
  \citenamefont {Fontana}, \citenamefont {Cerezo}, \citenamefont {Sharma},
  \citenamefont {Sone}, \citenamefont {Cincio},\ and\ \citenamefont
  {Coles}}]{wang2020noise}%
  \BibitemOpen
  \bibfield  {author} {\bibinfo {author} {\bibfnamefont {S.}~\bibnamefont
  {Wang}}, \bibinfo {author} {\bibfnamefont {E.}~\bibnamefont {Fontana}},
  \bibinfo {author} {\bibfnamefont {M.}~\bibnamefont {Cerezo}}, \bibinfo
  {author} {\bibfnamefont {K.}~\bibnamefont {Sharma}}, \bibinfo {author}
  {\bibfnamefont {A.}~\bibnamefont {Sone}}, \bibinfo {author} {\bibfnamefont
  {L.}~\bibnamefont {Cincio}}, \ and\ \bibinfo {author} {\bibfnamefont {P.~J.}\
  \bibnamefont {Coles}},\ }\href@noop {} {\bibfield  {journal} {\bibinfo
  {journal} {arXiv preprint arXiv:2007.14384}\ } (\bibinfo {year}
  {2020})}\BibitemShut {NoStop}%
\bibitem [{\citenamefont {Cerezo}\ and\ \citenamefont
  {Coles}(2020)}]{cerezo2020impact}%
  \BibitemOpen
  \bibfield  {author} {\bibinfo {author} {\bibfnamefont {M.}~\bibnamefont
  {Cerezo}}\ and\ \bibinfo {author} {\bibfnamefont {P.~J.}\ \bibnamefont
  {Coles}},\ }\href@noop {} {\bibfield  {journal} {\bibinfo  {journal} {arXiv
  preprint arXiv:2008.07454}\ } (\bibinfo {year} {2020})}\BibitemShut {NoStop}%
\bibitem [{\citenamefont {Endo}\ \emph {et~al.}(2021)\citenamefont {Endo},
  \citenamefont {Cai}, \citenamefont {Benjamin},\ and\ \citenamefont
  {Yuan}}]{endoreview}%
  \BibitemOpen
  \bibfield  {author} {\bibinfo {author} {\bibfnamefont {S.}~\bibnamefont
  {Endo}}, \bibinfo {author} {\bibfnamefont {Z.}~\bibnamefont {Cai}}, \bibinfo
  {author} {\bibfnamefont {S.~C.}\ \bibnamefont {Benjamin}}, \ and\ \bibinfo
  {author} {\bibfnamefont {X.}~\bibnamefont {Yuan}},\ }\href {\doibase
  10.7566/JPSJ.90.032001} {\bibfield  {journal} {\bibinfo  {journal} {Journal
  of the Physical Society of Japan}\ }\textbf {\bibinfo {volume} {90}},\
  \bibinfo {pages} {032001} (\bibinfo {year} {2021})},\ \Eprint
  {http://arxiv.org/abs/https://doi.org/10.7566/JPSJ.90.032001}
  {https://doi.org/10.7566/JPSJ.90.032001} \BibitemShut {NoStop}%
\bibitem [{\citenamefont {Cerezo}\ \emph {et~al.}(2020)\citenamefont {Cerezo},
  \citenamefont {Arrasmith}, \citenamefont {Babbush}, \citenamefont {Benjamin},
  \citenamefont {Endo}, \citenamefont {Fujii}, \citenamefont {McClean},
  \citenamefont {Mitarai}, \citenamefont {Yuan}, \citenamefont {Cincio},\ and\
  \citenamefont {Coles}}]{cerezo2020variational}%
  \BibitemOpen
  \bibfield  {author} {\bibinfo {author} {\bibfnamefont {M.}~\bibnamefont
  {Cerezo}}, \bibinfo {author} {\bibfnamefont {A.}~\bibnamefont {Arrasmith}},
  \bibinfo {author} {\bibfnamefont {R.}~\bibnamefont {Babbush}}, \bibinfo
  {author} {\bibfnamefont {S.~C.}\ \bibnamefont {Benjamin}}, \bibinfo {author}
  {\bibfnamefont {S.}~\bibnamefont {Endo}}, \bibinfo {author} {\bibfnamefont
  {K.}~\bibnamefont {Fujii}}, \bibinfo {author} {\bibfnamefont {J.~R.}\
  \bibnamefont {McClean}}, \bibinfo {author} {\bibfnamefont {K.}~\bibnamefont
  {Mitarai}}, \bibinfo {author} {\bibfnamefont {X.}~\bibnamefont {Yuan}},
  \bibinfo {author} {\bibfnamefont {L.}~\bibnamefont {Cincio}}, \ and\ \bibinfo
  {author} {\bibfnamefont {P.~J.}\ \bibnamefont {Coles}},\ }\href@noop {} {\
  (\bibinfo {year} {2020})},\ \Eprint {http://arxiv.org/abs/2012.09265}
  {arXiv:2012.09265 [quant-ph]} \BibitemShut {NoStop}%
\bibitem [{\citenamefont {Bharti}\ \emph {et~al.}(2021)\citenamefont {Bharti},
  \citenamefont {Cervera-Lierta}, \citenamefont {Kyaw}, \citenamefont {Haug},
  \citenamefont {Alperin-Lea}, \citenamefont {Anand}, \citenamefont {Degroote},
  \citenamefont {Heimonen}, \citenamefont {Kottmann}, \citenamefont {Menke},
  \citenamefont {Mok}, \citenamefont {Sim}, \citenamefont {Kwek},\ and\
  \citenamefont {Aspuru-Guzik}}]{bharti2021noisy}%
  \BibitemOpen
  \bibfield  {author} {\bibinfo {author} {\bibfnamefont {K.}~\bibnamefont
  {Bharti}}, \bibinfo {author} {\bibfnamefont {A.}~\bibnamefont
  {Cervera-Lierta}}, \bibinfo {author} {\bibfnamefont {T.~H.}\ \bibnamefont
  {Kyaw}}, \bibinfo {author} {\bibfnamefont {T.}~\bibnamefont {Haug}}, \bibinfo
  {author} {\bibfnamefont {S.}~\bibnamefont {Alperin-Lea}}, \bibinfo {author}
  {\bibfnamefont {A.}~\bibnamefont {Anand}}, \bibinfo {author} {\bibfnamefont
  {M.}~\bibnamefont {Degroote}}, \bibinfo {author} {\bibfnamefont
  {H.}~\bibnamefont {Heimonen}}, \bibinfo {author} {\bibfnamefont {J.~S.}\
  \bibnamefont {Kottmann}}, \bibinfo {author} {\bibfnamefont {T.}~\bibnamefont
  {Menke}}, \bibinfo {author} {\bibfnamefont {W.-K.}\ \bibnamefont {Mok}},
  \bibinfo {author} {\bibfnamefont {S.}~\bibnamefont {Sim}}, \bibinfo {author}
  {\bibfnamefont {L.-C.}\ \bibnamefont {Kwek}}, \ and\ \bibinfo {author}
  {\bibfnamefont {A.}~\bibnamefont {Aspuru-Guzik}},\ }\href@noop {} {\
  (\bibinfo {year} {2021})},\ \Eprint {http://arxiv.org/abs/2101.08448}
  {arXiv:2101.08448 [quant-ph]} \BibitemShut {NoStop}%
\bibitem [{\citenamefont {Takeshita}\ \emph {et~al.}(2020)\citenamefont
  {Takeshita}, \citenamefont {Rubin}, \citenamefont {Jiang}, \citenamefont
  {Lee}, \citenamefont {Babbush},\ and\ \citenamefont
  {McClean}}]{PhysRevX.10.011004}%
  \BibitemOpen
  \bibfield  {author} {\bibinfo {author} {\bibfnamefont {T.}~\bibnamefont
  {Takeshita}}, \bibinfo {author} {\bibfnamefont {N.~C.}\ \bibnamefont
  {Rubin}}, \bibinfo {author} {\bibfnamefont {Z.}~\bibnamefont {Jiang}},
  \bibinfo {author} {\bibfnamefont {E.}~\bibnamefont {Lee}}, \bibinfo {author}
  {\bibfnamefont {R.}~\bibnamefont {Babbush}}, \ and\ \bibinfo {author}
  {\bibfnamefont {J.~R.}\ \bibnamefont {McClean}},\ }\href {\doibase
  10.1103/PhysRevX.10.011004} {\bibfield  {journal} {\bibinfo  {journal} {Phys.
  Rev. X}\ }\textbf {\bibinfo {volume} {10}},\ \bibinfo {pages} {011004}
  (\bibinfo {year} {2020})}\BibitemShut {NoStop}%
\bibitem [{\citenamefont {Barratt}\ \emph {et~al.}(2020)\citenamefont
  {Barratt}, \citenamefont {Dborin}, \citenamefont {Bal}, \citenamefont
  {Stojevic}, \citenamefont {Pollmann},\ and\ \citenamefont
  {Green}}]{barratt2020parallel}%
  \BibitemOpen
  \bibfield  {author} {\bibinfo {author} {\bibfnamefont {F.}~\bibnamefont
  {Barratt}}, \bibinfo {author} {\bibfnamefont {J.}~\bibnamefont {Dborin}},
  \bibinfo {author} {\bibfnamefont {M.}~\bibnamefont {Bal}}, \bibinfo {author}
  {\bibfnamefont {V.}~\bibnamefont {Stojevic}}, \bibinfo {author}
  {\bibfnamefont {F.}~\bibnamefont {Pollmann}}, \ and\ \bibinfo {author}
  {\bibfnamefont {A.~G.}\ \bibnamefont {Green}},\ }\href@noop {} {\bibfield
  {journal} {\bibinfo  {journal} {arXiv preprint arXiv:2003.12087}\ } (\bibinfo
  {year} {2020})}\BibitemShut {NoStop}%
\bibitem [{\citenamefont {Yuan}\ \emph {et~al.}(2019)\citenamefont {Yuan},
  \citenamefont {Endo}, \citenamefont {Zhao}, \citenamefont {Li},\ and\
  \citenamefont {Benjamin}}]{yuan2019theory}%
  \BibitemOpen
  \bibfield  {author} {\bibinfo {author} {\bibfnamefont {X.}~\bibnamefont
  {Yuan}}, \bibinfo {author} {\bibfnamefont {S.}~\bibnamefont {Endo}}, \bibinfo
  {author} {\bibfnamefont {Q.}~\bibnamefont {Zhao}}, \bibinfo {author}
  {\bibfnamefont {Y.}~\bibnamefont {Li}}, \ and\ \bibinfo {author}
  {\bibfnamefont {S.~C.}\ \bibnamefont {Benjamin}},\ }\href@noop {} {\bibfield
  {journal} {\bibinfo  {journal} {Quantum}\ }\textbf {\bibinfo {volume} {3}},\
  \bibinfo {pages} {191} (\bibinfo {year} {2019})}\BibitemShut {NoStop}%
\bibitem [{\citenamefont {Hackl}\ \emph {et~al.}(2020)\citenamefont {Hackl},
  \citenamefont {Guaita}, \citenamefont {Shi}, \citenamefont {Haegeman},
  \citenamefont {Demler},\ and\ \citenamefont {Cirac}}]{hackl2020geometry}%
  \BibitemOpen
  \bibfield  {author} {\bibinfo {author} {\bibfnamefont {L.}~\bibnamefont
  {Hackl}}, \bibinfo {author} {\bibfnamefont {T.}~\bibnamefont {Guaita}},
  \bibinfo {author} {\bibfnamefont {T.}~\bibnamefont {Shi}}, \bibinfo {author}
  {\bibfnamefont {J.}~\bibnamefont {Haegeman}}, \bibinfo {author}
  {\bibfnamefont {E.}~\bibnamefont {Demler}}, \ and\ \bibinfo {author}
  {\bibfnamefont {I.}~\bibnamefont {Cirac}},\ }\href@noop {} {\bibfield
  {journal} {\bibinfo  {journal} {arXiv preprint arXiv:2004.01015}\ } (\bibinfo
  {year} {2020})}\BibitemShut {NoStop}%
\bibitem [{\citenamefont {Schollw{\"o}ck}(2011)}]{schollwock2011density}%
  \BibitemOpen
  \bibfield  {author} {\bibinfo {author} {\bibfnamefont {U.}~\bibnamefont
  {Schollw{\"o}ck}},\ }\href@noop {} {\bibfield  {journal} {\bibinfo  {journal}
  {Annals of Physics}\ }\textbf {\bibinfo {volume} {326}},\ \bibinfo {pages}
  {96} (\bibinfo {year} {2011})}\BibitemShut {NoStop}%
\bibitem [{Not()}]{NoteX}%
  \BibitemOpen
  \href@noop {} {}\bibinfo {note} {See Supplemental Materials for the framework
  of hybrid tensor networks, implementations, examples of tree structures and
  potential applications in details.}\BibitemShut {Stop}%
\bibitem [{\citenamefont {Fannes}\ \emph {et~al.}(1992)\citenamefont {Fannes},
  \citenamefont {Nachtergaele},\ and\ \citenamefont
  {Werner}}]{fannes1992finitely}%
  \BibitemOpen
  \bibfield  {author} {\bibinfo {author} {\bibfnamefont {M.}~\bibnamefont
  {Fannes}}, \bibinfo {author} {\bibfnamefont {B.}~\bibnamefont
  {Nachtergaele}}, \ and\ \bibinfo {author} {\bibfnamefont {R.~F.}\
  \bibnamefont {Werner}},\ }\href@noop {} {\bibfield  {journal} {\bibinfo
  {journal} {Communications in mathematical physics}\ }\textbf {\bibinfo
  {volume} {144}},\ \bibinfo {pages} {443} (\bibinfo {year}
  {1992})}\BibitemShut {NoStop}%
\bibitem [{\citenamefont {Klumper}\ \emph {et~al.}(1991)\citenamefont
  {Klumper}, \citenamefont {Schadschneider},\ and\ \citenamefont
  {Zittartz}}]{klumper1991equivalence}%
  \BibitemOpen
  \bibfield  {author} {\bibinfo {author} {\bibfnamefont {A.}~\bibnamefont
  {Klumper}}, \bibinfo {author} {\bibfnamefont {A.}~\bibnamefont
  {Schadschneider}}, \ and\ \bibinfo {author} {\bibfnamefont {J.}~\bibnamefont
  {Zittartz}},\ }\href@noop {} {\bibfield  {journal} {\bibinfo  {journal}
  {Journal of Physics A: Mathematical and General}\ }\textbf {\bibinfo {volume}
  {24}},\ \bibinfo {pages} {L955} (\bibinfo {year} {1991})}\BibitemShut
  {NoStop}%
\bibitem [{\citenamefont {Kl{\"u}mper}\ \emph {et~al.}(1993)\citenamefont
  {Kl{\"u}mper}, \citenamefont {Schadschneider},\ and\ \citenamefont
  {Zittartz}}]{klumper1993matrix}%
  \BibitemOpen
  \bibfield  {author} {\bibinfo {author} {\bibfnamefont {A.}~\bibnamefont
  {Kl{\"u}mper}}, \bibinfo {author} {\bibfnamefont {A.}~\bibnamefont
  {Schadschneider}}, \ and\ \bibinfo {author} {\bibfnamefont {J.}~\bibnamefont
  {Zittartz}},\ }\href@noop {} {\bibfield  {journal} {\bibinfo  {journal} {EPL
  (Europhysics Letters)}\ }\textbf {\bibinfo {volume} {24}},\ \bibinfo {pages}
  {293} (\bibinfo {year} {1993})}\BibitemShut {NoStop}%
\bibitem [{\citenamefont {Verstraete}\ and\ \citenamefont
  {Cirac}(2004)}]{verstraete2004renormalization}%
  \BibitemOpen
  \bibfield  {author} {\bibinfo {author} {\bibfnamefont {F.}~\bibnamefont
  {Verstraete}}\ and\ \bibinfo {author} {\bibfnamefont {J.~I.}\ \bibnamefont
  {Cirac}},\ }\href@noop {} {\bibfield  {journal} {\bibinfo  {journal} {arXiv
  preprint cond-mat/0407066}\ } (\bibinfo {year} {2004})}\BibitemShut {NoStop}%
\bibitem [{\citenamefont {Shi}\ \emph {et~al.}(2006)\citenamefont {Shi},
  \citenamefont {Duan},\ and\ \citenamefont {Vidal}}]{PhysRevA.74.022320}%
  \BibitemOpen
  \bibfield  {author} {\bibinfo {author} {\bibfnamefont {Y.-Y.}\ \bibnamefont
  {Shi}}, \bibinfo {author} {\bibfnamefont {L.-M.}\ \bibnamefont {Duan}}, \
  and\ \bibinfo {author} {\bibfnamefont {G.}~\bibnamefont {Vidal}},\ }\href
  {\doibase 10.1103/PhysRevA.74.022320} {\bibfield  {journal} {\bibinfo
  {journal} {Phys. Rev. A}\ }\textbf {\bibinfo {volume} {74}},\ \bibinfo
  {pages} {022320} (\bibinfo {year} {2006})}\BibitemShut {NoStop}%
\bibitem [{\citenamefont {Vidal}(2007)}]{PhysRevLett.99.220405}%
  \BibitemOpen
  \bibfield  {author} {\bibinfo {author} {\bibfnamefont {G.}~\bibnamefont
  {Vidal}},\ }\href {\doibase 10.1103/PhysRevLett.99.220405} {\bibfield
  {journal} {\bibinfo  {journal} {Phys. Rev. Lett.}\ }\textbf {\bibinfo
  {volume} {99}},\ \bibinfo {pages} {220405} (\bibinfo {year}
  {2007})}\BibitemShut {NoStop}%
\bibitem [{\citenamefont {McClean}\ \emph {et~al.}(2017)\citenamefont
  {McClean}, \citenamefont {Kimchi-Schwartz}, \citenamefont {Carter},\ and\
  \citenamefont {de~Jong}}]{PhysRevA.95.042308}%
  \BibitemOpen
  \bibfield  {author} {\bibinfo {author} {\bibfnamefont {J.~R.}\ \bibnamefont
  {McClean}}, \bibinfo {author} {\bibfnamefont {M.~E.}\ \bibnamefont
  {Kimchi-Schwartz}}, \bibinfo {author} {\bibfnamefont {J.}~\bibnamefont
  {Carter}}, \ and\ \bibinfo {author} {\bibfnamefont {W.~A.}\ \bibnamefont
  {de~Jong}},\ }\href {\doibase 10.1103/PhysRevA.95.042308} {\bibfield
  {journal} {\bibinfo  {journal} {Phys. Rev. A}\ }\textbf {\bibinfo {volume}
  {95}},\ \bibinfo {pages} {042308} (\bibinfo {year} {2017})}\BibitemShut
  {NoStop}%
\bibitem [{\citenamefont {Colless}\ \emph {et~al.}(2018)\citenamefont
  {Colless}, \citenamefont {Ramasesh}, \citenamefont {Dahlen}, \citenamefont
  {Blok}, \citenamefont {Kimchi-Schwartz}, \citenamefont {McClean},
  \citenamefont {Carter}, \citenamefont {de~Jong},\ and\ \citenamefont
  {Siddiqi}}]{PhysRevX.8.011021}%
  \BibitemOpen
  \bibfield  {author} {\bibinfo {author} {\bibfnamefont {J.~I.}\ \bibnamefont
  {Colless}}, \bibinfo {author} {\bibfnamefont {V.~V.}\ \bibnamefont
  {Ramasesh}}, \bibinfo {author} {\bibfnamefont {D.}~\bibnamefont {Dahlen}},
  \bibinfo {author} {\bibfnamefont {M.~S.}\ \bibnamefont {Blok}}, \bibinfo
  {author} {\bibfnamefont {M.~E.}\ \bibnamefont {Kimchi-Schwartz}}, \bibinfo
  {author} {\bibfnamefont {J.~R.}\ \bibnamefont {McClean}}, \bibinfo {author}
  {\bibfnamefont {J.}~\bibnamefont {Carter}}, \bibinfo {author} {\bibfnamefont
  {W.~A.}\ \bibnamefont {de~Jong}}, \ and\ \bibinfo {author} {\bibfnamefont
  {I.}~\bibnamefont {Siddiqi}},\ }\href {\doibase 10.1103/PhysRevX.8.011021}
  {\bibfield  {journal} {\bibinfo  {journal} {Phys. Rev. X}\ }\textbf {\bibinfo
  {volume} {8}},\ \bibinfo {pages} {011021} (\bibinfo {year}
  {2018})}\BibitemShut {NoStop}%
\bibitem [{\citenamefont {McClean}\ \emph {et~al.}(2020)\citenamefont
  {McClean}, \citenamefont {Jiang}, \citenamefont {Rubin}, \citenamefont
  {Babbush},\ and\ \citenamefont {Neven}}]{mcclean2020decoding}%
  \BibitemOpen
  \bibfield  {author} {\bibinfo {author} {\bibfnamefont {J.~R.}\ \bibnamefont
  {McClean}}, \bibinfo {author} {\bibfnamefont {Z.}~\bibnamefont {Jiang}},
  \bibinfo {author} {\bibfnamefont {N.~C.}\ \bibnamefont {Rubin}}, \bibinfo
  {author} {\bibfnamefont {R.}~\bibnamefont {Babbush}}, \ and\ \bibinfo
  {author} {\bibfnamefont {H.}~\bibnamefont {Neven}},\ }\href@noop {}
  {\bibfield  {journal} {\bibinfo  {journal} {Nature Communications}\ }\textbf
  {\bibinfo {volume} {11}},\ \bibinfo {pages} {1} (\bibinfo {year}
  {2020})}\BibitemShut {NoStop}%
\bibitem [{\citenamefont {Wouters}\ \emph {et~al.}(2016)\citenamefont
  {Wouters}, \citenamefont {Jim{\'e}nez-Hoyos}, \citenamefont {Sun},\ and\
  \citenamefont {Chan}}]{wouters2016practical}%
  \BibitemOpen
  \bibfield  {author} {\bibinfo {author} {\bibfnamefont {S.}~\bibnamefont
  {Wouters}}, \bibinfo {author} {\bibfnamefont {C.~A.}\ \bibnamefont
  {Jim{\'e}nez-Hoyos}}, \bibinfo {author} {\bibfnamefont {Q.}~\bibnamefont
  {Sun}}, \ and\ \bibinfo {author} {\bibfnamefont {G.~K.-L.}\ \bibnamefont
  {Chan}},\ }\href@noop {} {\bibfield  {journal} {\bibinfo  {journal} {Journal
  of chemical theory and computation}\ }\textbf {\bibinfo {volume} {12}},\
  \bibinfo {pages} {2706} (\bibinfo {year} {2016})}\BibitemShut {NoStop}%
\bibitem [{\citenamefont {Lanata}\ \emph {et~al.}(2015)\citenamefont {Lanata},
  \citenamefont {Yao}, \citenamefont {Wang}, \citenamefont {Ho},\ and\
  \citenamefont {Kotliar}}]{lanata2015phase}%
  \BibitemOpen
  \bibfield  {author} {\bibinfo {author} {\bibfnamefont {N.}~\bibnamefont
  {Lanata}}, \bibinfo {author} {\bibfnamefont {Y.}~\bibnamefont {Yao}},
  \bibinfo {author} {\bibfnamefont {C.-Z.}\ \bibnamefont {Wang}}, \bibinfo
  {author} {\bibfnamefont {K.-M.}\ \bibnamefont {Ho}}, \ and\ \bibinfo {author}
  {\bibfnamefont {G.}~\bibnamefont {Kotliar}},\ }\href@noop {} {\bibfield
  {journal} {\bibinfo  {journal} {Physical Review X}\ }\textbf {\bibinfo
  {volume} {5}},\ \bibinfo {pages} {011008} (\bibinfo {year}
  {2015})}\BibitemShut {NoStop}%
\bibitem [{\citenamefont {Rohringer}\ \emph {et~al.}(2018)\citenamefont
  {Rohringer}, \citenamefont {Hafermann}, \citenamefont {Toschi}, \citenamefont
  {Katanin}, \citenamefont {Antipov}, \citenamefont {Katsnelson}, \citenamefont
  {Lichtenstein}, \citenamefont {Rubtsov},\ and\ \citenamefont
  {Held}}]{rohringer2018diagrammatic}%
  \BibitemOpen
  \bibfield  {author} {\bibinfo {author} {\bibfnamefont {G.}~\bibnamefont
  {Rohringer}}, \bibinfo {author} {\bibfnamefont {H.}~\bibnamefont
  {Hafermann}}, \bibinfo {author} {\bibfnamefont {A.}~\bibnamefont {Toschi}},
  \bibinfo {author} {\bibfnamefont {A.}~\bibnamefont {Katanin}}, \bibinfo
  {author} {\bibfnamefont {A.}~\bibnamefont {Antipov}}, \bibinfo {author}
  {\bibfnamefont {M.}~\bibnamefont {Katsnelson}}, \bibinfo {author}
  {\bibfnamefont {A.}~\bibnamefont {Lichtenstein}}, \bibinfo {author}
  {\bibfnamefont {A.}~\bibnamefont {Rubtsov}}, \ and\ \bibinfo {author}
  {\bibfnamefont {K.}~\bibnamefont {Held}},\ }\href@noop {} {\bibfield
  {journal} {\bibinfo  {journal} {Reviews of Modern Physics}\ }\textbf
  {\bibinfo {volume} {90}},\ \bibinfo {pages} {025003} (\bibinfo {year}
  {2018})}\BibitemShut {NoStop}%
\bibitem [{\citenamefont {Chen}\ \emph {et~al.}(2010)\citenamefont {Chen},
  \citenamefont {Gu},\ and\ \citenamefont {Wen}}]{PhysRevB.82.155138}%
  \BibitemOpen
  \bibfield  {author} {\bibinfo {author} {\bibfnamefont {X.}~\bibnamefont
  {Chen}}, \bibinfo {author} {\bibfnamefont {Z.-C.}\ \bibnamefont {Gu}}, \ and\
  \bibinfo {author} {\bibfnamefont {X.-G.}\ \bibnamefont {Wen}},\ }\href
  {\doibase 10.1103/PhysRevB.82.155138} {\bibfield  {journal} {\bibinfo
  {journal} {Phys. Rev. B}\ }\textbf {\bibinfo {volume} {82}},\ \bibinfo
  {pages} {155138} (\bibinfo {year} {2010})}\BibitemShut {NoStop}%
\bibitem [{\citenamefont {Chen}\ \emph {et~al.}(2011)\citenamefont {Chen},
  \citenamefont {Gu},\ and\ \citenamefont {Wen}}]{PhysRevB.83.035107}%
  \BibitemOpen
  \bibfield  {author} {\bibinfo {author} {\bibfnamefont {X.}~\bibnamefont
  {Chen}}, \bibinfo {author} {\bibfnamefont {Z.-C.}\ \bibnamefont {Gu}}, \ and\
  \bibinfo {author} {\bibfnamefont {X.-G.}\ \bibnamefont {Wen}},\ }\href
  {\doibase 10.1103/PhysRevB.83.035107} {\bibfield  {journal} {\bibinfo
  {journal} {Phys. Rev. B}\ }\textbf {\bibinfo {volume} {83}},\ \bibinfo
  {pages} {035107} (\bibinfo {year} {2011})}\BibitemShut {NoStop}%
\bibitem [{\citenamefont {McArdle}\ \emph {et~al.}(2019)\citenamefont
  {McArdle}, \citenamefont {Jones}, \citenamefont {Endo}, \citenamefont {Li},
  \citenamefont {Benjamin},\ and\ \citenamefont
  {Yuan}}]{mcardle2019variational}%
  \BibitemOpen
  \bibfield  {author} {\bibinfo {author} {\bibfnamefont {S.}~\bibnamefont
  {McArdle}}, \bibinfo {author} {\bibfnamefont {T.}~\bibnamefont {Jones}},
  \bibinfo {author} {\bibfnamefont {S.}~\bibnamefont {Endo}}, \bibinfo {author}
  {\bibfnamefont {Y.}~\bibnamefont {Li}}, \bibinfo {author} {\bibfnamefont
  {S.~C.}\ \bibnamefont {Benjamin}}, \ and\ \bibinfo {author} {\bibfnamefont
  {X.}~\bibnamefont {Yuan}},\ }\href@noop {} {\bibfield  {journal} {\bibinfo
  {journal} {npj Quantum Information}\ }\textbf {\bibinfo {volume} {5}},\
  \bibinfo {pages} {1} (\bibinfo {year} {2019})}\BibitemShut {NoStop}%
\bibitem [{\citenamefont {Szalay}\ \emph {et~al.}(2015)\citenamefont {Szalay},
  \citenamefont {Pfeffer}, \citenamefont {Murg}, \citenamefont {Barcza},
  \citenamefont {Verstraete}, \citenamefont {Schneider},\ and\ \citenamefont
  {Legeza}}]{szalay2015tensor}%
  \BibitemOpen
  \bibfield  {author} {\bibinfo {author} {\bibfnamefont {S.}~\bibnamefont
  {Szalay}}, \bibinfo {author} {\bibfnamefont {M.}~\bibnamefont {Pfeffer}},
  \bibinfo {author} {\bibfnamefont {V.}~\bibnamefont {Murg}}, \bibinfo {author}
  {\bibfnamefont {G.}~\bibnamefont {Barcza}}, \bibinfo {author} {\bibfnamefont
  {F.}~\bibnamefont {Verstraete}}, \bibinfo {author} {\bibfnamefont
  {R.}~\bibnamefont {Schneider}}, \ and\ \bibinfo {author} {\bibfnamefont
  {{\"O}.}~\bibnamefont {Legeza}},\ }\href@noop {} {\bibfield  {journal}
  {\bibinfo  {journal} {International Journal of Quantum Chemistry}\ }\textbf
  {\bibinfo {volume} {115}},\ \bibinfo {pages} {1342} (\bibinfo {year}
  {2015})}\BibitemShut {NoStop}%
\bibitem [{\citenamefont {Worth}\ and\ \citenamefont
  {Cederbaum}(2004)}]{worth2004beyond}%
  \BibitemOpen
  \bibfield  {author} {\bibinfo {author} {\bibfnamefont {G.~A.}\ \bibnamefont
  {Worth}}\ and\ \bibinfo {author} {\bibfnamefont {L.~S.}\ \bibnamefont
  {Cederbaum}},\ }\href@noop {} {\bibfield  {journal} {\bibinfo  {journal}
  {Annu. Rev. Phys. Chem.}\ }\textbf {\bibinfo {volume} {55}},\ \bibinfo
  {pages} {127} (\bibinfo {year} {2004})}\BibitemShut {NoStop}%
\bibitem [{\citenamefont {Reiher}\ and\ \citenamefont
  {Wolf}(2014)}]{reiher2014relativistic}%
  \BibitemOpen
  \bibfield  {author} {\bibinfo {author} {\bibfnamefont {M.}~\bibnamefont
  {Reiher}}\ and\ \bibinfo {author} {\bibfnamefont {A.}~\bibnamefont {Wolf}},\
  }\href@noop {} {\emph {\bibinfo {title} {Relativistic quantum chemistry: the
  fundamental theory of molecular science}}}\ (\bibinfo  {publisher} {John
  Wiley \& Sons},\ \bibinfo {year} {2014})\BibitemShut {NoStop}%
\bibitem [{\citenamefont {Domcke}\ \emph {et~al.}(2004)\citenamefont {Domcke},
  \citenamefont {Yarkony},\ and\ \citenamefont {K\"oppel}}]{domcke2004conical}%
  \BibitemOpen
  \bibfield  {author} {\bibinfo {author} {\bibfnamefont {W.}~\bibnamefont
  {Domcke}}, \bibinfo {author} {\bibfnamefont {D.~R.}\ \bibnamefont {Yarkony}},
  \ and\ \bibinfo {author} {\bibfnamefont {H.}~\bibnamefont {K\"oppel}},\
  }\href {\doibase 10.1142/5406} {\emph {\bibinfo {title} {Conical
  Intersections}}}\ (\bibinfo  {publisher} {WORLD SCIENTIFIC},\ \bibinfo {year}
  {2004})\ \Eprint
  {http://arxiv.org/abs/https://www.worldscientific.com/doi/pdf/10.1142/5406}
  {https://www.worldscientific.com/doi/pdf/10.1142/5406} \BibitemShut {NoStop}%
\bibitem [{\citenamefont {Domcke}\ and\ \citenamefont
  {Yarkony}(2012)}]{domcke2012spectroscopy}%
  \BibitemOpen
  \bibfield  {author} {\bibinfo {author} {\bibfnamefont {W.}~\bibnamefont
  {Domcke}}\ and\ \bibinfo {author} {\bibfnamefont {D.~R.}\ \bibnamefont
  {Yarkony}},\ }\href {\doibase 10.1146/annurev-physchem-032210-103522}
  {\bibfield  {journal} {\bibinfo  {journal} {Annu. Rev. Phys. Chem.}\ }\textbf
  {\bibinfo {volume} {63}},\ \bibinfo {pages} {325} (\bibinfo {year} {2012})},\
  \bibinfo {note} {pMID: 22475338},\ \Eprint
  {http://arxiv.org/abs/https://doi.org/10.1146/annurev-physchem-032210-103522}
  {https://doi.org/10.1146/annurev-physchem-032210-103522} \BibitemShut
  {NoStop}%
\bibitem [{\citenamefont {Ryabinkin}\ \emph {et~al.}(2017)\citenamefont
  {Ryabinkin}, \citenamefont {Joubert-Doriol},\ and\ \citenamefont
  {Izmaylov}}]{ryabinkin2017conical}%
  \BibitemOpen
  \bibfield  {author} {\bibinfo {author} {\bibfnamefont {I.~G.}\ \bibnamefont
  {Ryabinkin}}, \bibinfo {author} {\bibfnamefont {L.}~\bibnamefont
  {Joubert-Doriol}}, \ and\ \bibinfo {author} {\bibfnamefont {A.~F.}\
  \bibnamefont {Izmaylov}},\ }\href {\doibase 10.1021/acs.accounts.7b00220}
  {\bibfield  {journal} {\bibinfo  {journal} {Acc. Chem. Res.}\ }\textbf
  {\bibinfo {volume} {50}},\ \bibinfo {pages} {1785} (\bibinfo {year}
  {2017})},\ \bibinfo {note} {pMID: 28665584},\ \Eprint
  {http://arxiv.org/abs/https://doi.org/10.1021/acs.accounts.7b00220}
  {https://doi.org/10.1021/acs.accounts.7b00220} \BibitemShut {NoStop}%
\bibitem [{\citenamefont {Garlatti}\ \emph {et~al.}(2017)\citenamefont
  {Garlatti}, \citenamefont {Guidi}, \citenamefont {Ansbro}, \citenamefont
  {Santini}, \citenamefont {Amoretti}, \citenamefont {Ollivier}, \citenamefont
  {Mutka}, \citenamefont {Timco}, \citenamefont {Vitorica-Yrezabal},
  \citenamefont {Whitehead} \emph {et~al.}}]{garlatti2017portraying}%
  \BibitemOpen
  \bibfield  {author} {\bibinfo {author} {\bibfnamefont {E.}~\bibnamefont
  {Garlatti}}, \bibinfo {author} {\bibfnamefont {T.}~\bibnamefont {Guidi}},
  \bibinfo {author} {\bibfnamefont {S.}~\bibnamefont {Ansbro}}, \bibinfo
  {author} {\bibfnamefont {P.}~\bibnamefont {Santini}}, \bibinfo {author}
  {\bibfnamefont {G.}~\bibnamefont {Amoretti}}, \bibinfo {author}
  {\bibfnamefont {J.}~\bibnamefont {Ollivier}}, \bibinfo {author}
  {\bibfnamefont {H.}~\bibnamefont {Mutka}}, \bibinfo {author} {\bibfnamefont
  {G.}~\bibnamefont {Timco}}, \bibinfo {author} {\bibfnamefont
  {I.}~\bibnamefont {Vitorica-Yrezabal}}, \bibinfo {author} {\bibfnamefont
  {G.}~\bibnamefont {Whitehead}},  \emph {et~al.},\ }\href@noop {} {\bibfield
  {journal} {\bibinfo  {journal} {Nature communications}\ }\textbf {\bibinfo
  {volume} {8}},\ \bibinfo {pages} {1} (\bibinfo {year} {2017})}\BibitemShut
  {NoStop}%
\bibitem [{\citenamefont {Timco}\ \emph {et~al.}(2009)\citenamefont {Timco},
  \citenamefont {Carretta}, \citenamefont {Troiani}, \citenamefont {Tuna},
  \citenamefont {Pritchard}, \citenamefont {Muryn}, \citenamefont {McInnes},
  \citenamefont {Ghirri}, \citenamefont {Candini}, \citenamefont {Santini}
  \emph {et~al.}}]{timco2009engineering}%
  \BibitemOpen
  \bibfield  {author} {\bibinfo {author} {\bibfnamefont {G.~A.}\ \bibnamefont
  {Timco}}, \bibinfo {author} {\bibfnamefont {S.}~\bibnamefont {Carretta}},
  \bibinfo {author} {\bibfnamefont {F.}~\bibnamefont {Troiani}}, \bibinfo
  {author} {\bibfnamefont {F.}~\bibnamefont {Tuna}}, \bibinfo {author}
  {\bibfnamefont {R.~J.}\ \bibnamefont {Pritchard}}, \bibinfo {author}
  {\bibfnamefont {C.~A.}\ \bibnamefont {Muryn}}, \bibinfo {author}
  {\bibfnamefont {E.~J.}\ \bibnamefont {McInnes}}, \bibinfo {author}
  {\bibfnamefont {A.}~\bibnamefont {Ghirri}}, \bibinfo {author} {\bibfnamefont
  {A.}~\bibnamefont {Candini}}, \bibinfo {author} {\bibfnamefont
  {P.}~\bibnamefont {Santini}},  \emph {et~al.},\ }\href@noop {} {\bibfield
  {journal} {\bibinfo  {journal} {Nature Nanotechnology}\ }\textbf {\bibinfo
  {volume} {4}},\ \bibinfo {pages} {173} (\bibinfo {year} {2009})}\BibitemShut
  {NoStop}%
\bibitem [{\citenamefont {Gu}\ \emph {et~al.}(2017)\citenamefont {Gu},
  \citenamefont {Qi},\ and\ \citenamefont {Stanford}}]{Gu:2016oyy}%
  \BibitemOpen
  \bibfield  {author} {\bibinfo {author} {\bibfnamefont {Y.}~\bibnamefont
  {Gu}}, \bibinfo {author} {\bibfnamefont {X.-L.}\ \bibnamefont {Qi}}, \ and\
  \bibinfo {author} {\bibfnamefont {D.}~\bibnamefont {Stanford}},\ }\href
  {\doibase 10.1007/JHEP05(2017)125} {\bibfield  {journal} {\bibinfo  {journal}
  {JHEP}\ }\textbf {\bibinfo {volume} {05}},\ \bibinfo {pages} {125} (\bibinfo
  {year} {2017})},\ \Eprint {http://arxiv.org/abs/1609.07832} {arXiv:1609.07832
  [hep-th]} \BibitemShut {NoStop}%
\bibitem [{\citenamefont {Hayden}\ and\ \citenamefont
  {Preskill}(2007)}]{Hayden:2007cs}%
  \BibitemOpen
  \bibfield  {author} {\bibinfo {author} {\bibfnamefont {P.}~\bibnamefont
  {Hayden}}\ and\ \bibinfo {author} {\bibfnamefont {J.}~\bibnamefont
  {Preskill}},\ }\href {\doibase 10.1088/1126-6708/2007/09/120} {\bibfield
  {journal} {\bibinfo  {journal} {JHEP}\ }\textbf {\bibinfo {volume} {09}},\
  \bibinfo {pages} {120} (\bibinfo {year} {2007})},\ \Eprint
  {http://arxiv.org/abs/0708.4025} {arXiv:0708.4025 [hep-th]} \BibitemShut
  {NoStop}%
\bibitem [{\citenamefont {Maldacena}\ \emph {et~al.}(2017)\citenamefont
  {Maldacena}, \citenamefont {Stanford},\ and\ \citenamefont
  {Yang}}]{Maldacena:2017axo}%
  \BibitemOpen
  \bibfield  {author} {\bibinfo {author} {\bibfnamefont {J.}~\bibnamefont
  {Maldacena}}, \bibinfo {author} {\bibfnamefont {D.}~\bibnamefont {Stanford}},
  \ and\ \bibinfo {author} {\bibfnamefont {Z.}~\bibnamefont {Yang}},\ }\href
  {\doibase 10.1002/prop.201700034} {\bibfield  {journal} {\bibinfo  {journal}
  {Fortsch. Phys.}\ }\textbf {\bibinfo {volume} {65}},\ \bibinfo {pages}
  {1700034} (\bibinfo {year} {2017})},\ \Eprint
  {http://arxiv.org/abs/1704.05333} {arXiv:1704.05333 [hep-th]} \BibitemShut
  {NoStop}%
\bibitem [{\citenamefont {Maldacena}\ and\ \citenamefont
  {Qi}(2018)}]{Maldacena:2018lmt}%
  \BibitemOpen
  \bibfield  {author} {\bibinfo {author} {\bibfnamefont {J.}~\bibnamefont
  {Maldacena}}\ and\ \bibinfo {author} {\bibfnamefont {X.-L.}\ \bibnamefont
  {Qi}},\ }\href@noop {} {\  (\bibinfo {year} {2018})},\ \Eprint
  {http://arxiv.org/abs/1804.00491} {arXiv:1804.00491 [hep-th]} \BibitemShut
  {NoStop}%
\bibitem [{\citenamefont {Yoshida}\ and\ \citenamefont
  {Kitaev}(2017)}]{Yoshida:2017non}%
  \BibitemOpen
  \bibfield  {author} {\bibinfo {author} {\bibfnamefont {B.}~\bibnamefont
  {Yoshida}}\ and\ \bibinfo {author} {\bibfnamefont {A.}~\bibnamefont
  {Kitaev}},\ }\href@noop {} {\  (\bibinfo {year} {2017})},\ \Eprint
  {http://arxiv.org/abs/1710.03363} {arXiv:1710.03363 [hep-th]} \BibitemShut
  {NoStop}%
\bibitem [{\citenamefont {Maldacena}\ and\ \citenamefont
  {Stanford}(2016)}]{Maldacena:2016hyu}%
  \BibitemOpen
  \bibfield  {author} {\bibinfo {author} {\bibfnamefont {J.}~\bibnamefont
  {Maldacena}}\ and\ \bibinfo {author} {\bibfnamefont {D.}~\bibnamefont
  {Stanford}},\ }\href {\doibase 10.1103/PhysRevD.94.106002} {\bibfield
  {journal} {\bibinfo  {journal} {Phys. Rev. D}\ }\textbf {\bibinfo {volume}
  {94}},\ \bibinfo {pages} {106002} (\bibinfo {year} {2016})},\ \Eprint
  {http://arxiv.org/abs/1604.07818} {arXiv:1604.07818 [hep-th]} \BibitemShut
  {NoStop}%
\bibitem [{\citenamefont {Maldacena}(1999)}]{Maldacena:1997re}%
  \BibitemOpen
  \bibfield  {author} {\bibinfo {author} {\bibfnamefont {J.~M.}\ \bibnamefont
  {Maldacena}},\ }\href {\doibase 10.1023/A:1026654312961} {\bibfield
  {journal} {\bibinfo  {journal} {Int. J. Theor. Phys.}\ }\textbf {\bibinfo
  {volume} {38}},\ \bibinfo {pages} {1113} (\bibinfo {year} {1999})},\ \Eprint
  {http://arxiv.org/abs/hep-th/9711200} {arXiv:hep-th/9711200} \BibitemShut
  {NoStop}%
\bibitem [{\citenamefont {Ma}\ \emph {et~al.}(2020)\citenamefont {Ma},
  \citenamefont {Govoni},\ and\ \citenamefont {Galli}}]{ma2020quantum}%
  \BibitemOpen
  \bibfield  {author} {\bibinfo {author} {\bibfnamefont {H.}~\bibnamefont
  {Ma}}, \bibinfo {author} {\bibfnamefont {M.}~\bibnamefont {Govoni}}, \ and\
  \bibinfo {author} {\bibfnamefont {G.}~\bibnamefont {Galli}},\ }\href@noop {}
  {\bibfield  {journal} {\bibinfo  {journal} {npj Computational Materials}\
  }\textbf {\bibinfo {volume} {6}},\ \bibinfo {pages} {1} (\bibinfo {year}
  {2020})}\BibitemShut {NoStop}%
\bibitem [{\citenamefont {Zhou}\ \emph {et~al.}(2017)\citenamefont {Zhou},
  \citenamefont {Kanoda},\ and\ \citenamefont {Ng}}]{zhou2017quantum}%
  \BibitemOpen
  \bibfield  {author} {\bibinfo {author} {\bibfnamefont {Y.}~\bibnamefont
  {Zhou}}, \bibinfo {author} {\bibfnamefont {K.}~\bibnamefont {Kanoda}}, \ and\
  \bibinfo {author} {\bibfnamefont {T.-K.}\ \bibnamefont {Ng}},\ }\href@noop {}
  {\bibfield  {journal} {\bibinfo  {journal} {Reviews of Modern Physics}\
  }\textbf {\bibinfo {volume} {89}},\ \bibinfo {pages} {025003} (\bibinfo
  {year} {2017})}\BibitemShut {NoStop}%
\bibitem [{\citenamefont {Liu}\ \emph {et~al.}(2019{\natexlab{a}})\citenamefont
  {Liu}, \citenamefont {Lundgren}, \citenamefont {Titum}, \citenamefont
  {Pagano}, \citenamefont {Zhang}, \citenamefont {Monroe},\ and\ \citenamefont
  {Gorshkov}}]{liu2019confined}%
  \BibitemOpen
  \bibfield  {author} {\bibinfo {author} {\bibfnamefont {F.}~\bibnamefont
  {Liu}}, \bibinfo {author} {\bibfnamefont {R.}~\bibnamefont {Lundgren}},
  \bibinfo {author} {\bibfnamefont {P.}~\bibnamefont {Titum}}, \bibinfo
  {author} {\bibfnamefont {G.}~\bibnamefont {Pagano}}, \bibinfo {author}
  {\bibfnamefont {J.}~\bibnamefont {Zhang}}, \bibinfo {author} {\bibfnamefont
  {C.}~\bibnamefont {Monroe}}, \ and\ \bibinfo {author} {\bibfnamefont {A.~V.}\
  \bibnamefont {Gorshkov}},\ }\href@noop {} {\bibfield  {journal} {\bibinfo
  {journal} {Physical review letters}\ }\textbf {\bibinfo {volume} {122}},\
  \bibinfo {pages} {150601} (\bibinfo {year} {2019}{\natexlab{a}})}\BibitemShut
  {NoStop}%
\bibitem [{\citenamefont {Sato}\ and\ \citenamefont
  {Ando}(2017)}]{sato2017topological}%
  \BibitemOpen
  \bibfield  {author} {\bibinfo {author} {\bibfnamefont {M.}~\bibnamefont
  {Sato}}\ and\ \bibinfo {author} {\bibfnamefont {Y.}~\bibnamefont {Ando}},\
  }\href@noop {} {\bibfield  {journal} {\bibinfo  {journal} {Reports on
  Progress in Physics}\ }\textbf {\bibinfo {volume} {80}},\ \bibinfo {pages}
  {076501} (\bibinfo {year} {2017})}\BibitemShut {NoStop}%
\bibitem [{\citenamefont {Chubukov}\ \emph {et~al.}(2008)\citenamefont
  {Chubukov}, \citenamefont {Efremov},\ and\ \citenamefont
  {Eremin}}]{chubukov2008magnetism}%
  \BibitemOpen
  \bibfield  {author} {\bibinfo {author} {\bibfnamefont {A.~V.}\ \bibnamefont
  {Chubukov}}, \bibinfo {author} {\bibfnamefont {D.}~\bibnamefont {Efremov}}, \
  and\ \bibinfo {author} {\bibfnamefont {I.}~\bibnamefont {Eremin}},\
  }\href@noop {} {\bibfield  {journal} {\bibinfo  {journal} {Physical Review
  B}\ }\textbf {\bibinfo {volume} {78}},\ \bibinfo {pages} {134512} (\bibinfo
  {year} {2008})}\BibitemShut {NoStop}%
\bibitem [{\citenamefont {Lutchyn}\ \emph {et~al.}(2018)\citenamefont
  {Lutchyn}, \citenamefont {Bakkers}, \citenamefont {Kouwenhoven},
  \citenamefont {Krogstrup}, \citenamefont {Marcus},\ and\ \citenamefont
  {Oreg}}]{lutchyn2018majorana}%
  \BibitemOpen
  \bibfield  {author} {\bibinfo {author} {\bibfnamefont {R.~t.}\ \bibnamefont
  {Lutchyn}}, \bibinfo {author} {\bibfnamefont {E.}~\bibnamefont {Bakkers}},
  \bibinfo {author} {\bibfnamefont {L.~P.}\ \bibnamefont {Kouwenhoven}},
  \bibinfo {author} {\bibfnamefont {P.}~\bibnamefont {Krogstrup}}, \bibinfo
  {author} {\bibfnamefont {C.}~\bibnamefont {Marcus}}, \ and\ \bibinfo {author}
  {\bibfnamefont {Y.}~\bibnamefont {Oreg}},\ }\href@noop {} {\bibfield
  {journal} {\bibinfo  {journal} {Nature Reviews Materials}\ }\textbf {\bibinfo
  {volume} {3}},\ \bibinfo {pages} {52} (\bibinfo {year} {2018})}\BibitemShut
  {NoStop}%
\bibitem [{\citenamefont {You}\ \emph {et~al.}(2018)\citenamefont {You},
  \citenamefont {Devakul}, \citenamefont {Burnell},\ and\ \citenamefont
  {Sondhi}}]{you2018subsystem}%
  \BibitemOpen
  \bibfield  {author} {\bibinfo {author} {\bibfnamefont {Y.}~\bibnamefont
  {You}}, \bibinfo {author} {\bibfnamefont {T.}~\bibnamefont {Devakul}},
  \bibinfo {author} {\bibfnamefont {F.~J.}\ \bibnamefont {Burnell}}, \ and\
  \bibinfo {author} {\bibfnamefont {S.~L.}\ \bibnamefont {Sondhi}},\
  }\href@noop {} {\bibfield  {journal} {\bibinfo  {journal} {Physical Review
  B}\ }\textbf {\bibinfo {volume} {98}},\ \bibinfo {pages} {035112} (\bibinfo
  {year} {2018})}\BibitemShut {NoStop}%
\bibitem [{\citenamefont {Alicea}(2012)}]{alicea2012new}%
  \BibitemOpen
  \bibfield  {author} {\bibinfo {author} {\bibfnamefont {J.}~\bibnamefont
  {Alicea}},\ }\href@noop {} {\bibfield  {journal} {\bibinfo  {journal}
  {Reports on progress in physics}\ }\textbf {\bibinfo {volume} {75}},\
  \bibinfo {pages} {076501} (\bibinfo {year} {2012})}\BibitemShut {NoStop}%
\bibitem [{\citenamefont {Sau}\ \emph {et~al.}(2010)\citenamefont {Sau},
  \citenamefont {Tewari}, \citenamefont {Lutchyn}, \citenamefont {Stanescu},\
  and\ \citenamefont {Sarma}}]{sau2010non}%
  \BibitemOpen
  \bibfield  {author} {\bibinfo {author} {\bibfnamefont {J.~D.}\ \bibnamefont
  {Sau}}, \bibinfo {author} {\bibfnamefont {S.}~\bibnamefont {Tewari}},
  \bibinfo {author} {\bibfnamefont {R.~M.}\ \bibnamefont {Lutchyn}}, \bibinfo
  {author} {\bibfnamefont {T.~D.}\ \bibnamefont {Stanescu}}, \ and\ \bibinfo
  {author} {\bibfnamefont {S.~D.}\ \bibnamefont {Sarma}},\ }\href@noop {}
  {\bibfield  {journal} {\bibinfo  {journal} {Physical Review B}\ }\textbf
  {\bibinfo {volume} {82}},\ \bibinfo {pages} {214509} (\bibinfo {year}
  {2010})}\BibitemShut {NoStop}%
\bibitem [{\citenamefont {Beenakker}(2013)}]{beenakker2013search}%
  \BibitemOpen
  \bibfield  {author} {\bibinfo {author} {\bibfnamefont {C.}~\bibnamefont
  {Beenakker}},\ }\href@noop {} {\bibfield  {journal} {\bibinfo  {journal}
  {Annu. Rev. Condens. Matter Phys.}\ }\textbf {\bibinfo {volume} {4}},\
  \bibinfo {pages} {113} (\bibinfo {year} {2013})}\BibitemShut {NoStop}%
\bibitem [{\citenamefont {Kim}\ and\ \citenamefont
  {Swingle}(2017)}]{kim2017robust}%
  \BibitemOpen
  \bibfield  {author} {\bibinfo {author} {\bibfnamefont {I.~H.}\ \bibnamefont
  {Kim}}\ and\ \bibinfo {author} {\bibfnamefont {B.}~\bibnamefont {Swingle}},\
  }\href@noop {} {\bibfield  {journal} {\bibinfo  {journal} {arXiv preprint
  arXiv:1711.07500}\ } (\bibinfo {year} {2017})}\BibitemShut {NoStop}%
\bibitem [{\citenamefont {Kim}(2017{\natexlab{a}})}]{kim2017noise}%
  \BibitemOpen
  \bibfield  {author} {\bibinfo {author} {\bibfnamefont {I.~H.}\ \bibnamefont
  {Kim}},\ }\href@noop {} {\bibfield  {journal} {\bibinfo  {journal} {arXiv
  preprint arXiv:1703.00032}\ } (\bibinfo {year}
  {2017}{\natexlab{a}})}\BibitemShut {NoStop}%
\bibitem [{\citenamefont {Kim}(2017{\natexlab{b}})}]{kim2017holographic}%
  \BibitemOpen
  \bibfield  {author} {\bibinfo {author} {\bibfnamefont {I.~H.}\ \bibnamefont
  {Kim}},\ }\href@noop {} {\bibfield  {journal} {\bibinfo  {journal} {arXiv
  preprint arXiv:1702.02093}\ } (\bibinfo {year}
  {2017}{\natexlab{b}})}\BibitemShut {NoStop}%
\bibitem [{\citenamefont {Liu}\ \emph {et~al.}(2019{\natexlab{b}})\citenamefont
  {Liu}, \citenamefont {Zhang}, \citenamefont {Wan},\ and\ \citenamefont
  {Wang}}]{PhysRevResearch.1.023025}%
  \BibitemOpen
  \bibfield  {author} {\bibinfo {author} {\bibfnamefont {J.-G.}\ \bibnamefont
  {Liu}}, \bibinfo {author} {\bibfnamefont {Y.-H.}\ \bibnamefont {Zhang}},
  \bibinfo {author} {\bibfnamefont {Y.}~\bibnamefont {Wan}}, \ and\ \bibinfo
  {author} {\bibfnamefont {L.}~\bibnamefont {Wang}},\ }\href {\doibase
  10.1103/PhysRevResearch.1.023025} {\bibfield  {journal} {\bibinfo  {journal}
  {Phys. Rev. Research}\ }\textbf {\bibinfo {volume} {1}},\ \bibinfo {pages}
  {023025} (\bibinfo {year} {2019}{\natexlab{b}})}\BibitemShut {NoStop}%
\bibitem [{\citenamefont {Foulkes}\ \emph {et~al.}(2001)\citenamefont
  {Foulkes}, \citenamefont {Mitas}, \citenamefont {Needs},\ and\ \citenamefont
  {Rajagopal}}]{RevModPhys.73.33}%
  \BibitemOpen
  \bibfield  {author} {\bibinfo {author} {\bibfnamefont {W.~M.~C.}\
  \bibnamefont {Foulkes}}, \bibinfo {author} {\bibfnamefont {L.}~\bibnamefont
  {Mitas}}, \bibinfo {author} {\bibfnamefont {R.~J.}\ \bibnamefont {Needs}}, \
  and\ \bibinfo {author} {\bibfnamefont {G.}~\bibnamefont {Rajagopal}},\ }\href
  {\doibase 10.1103/RevModPhys.73.33} {\bibfield  {journal} {\bibinfo
  {journal} {Rev. Mod. Phys.}\ }\textbf {\bibinfo {volume} {73}},\ \bibinfo
  {pages} {33} (\bibinfo {year} {2001})}\BibitemShut {NoStop}%
\bibitem [{\citenamefont {Carlson}\ \emph {et~al.}(2015)\citenamefont
  {Carlson}, \citenamefont {Gandolfi}, \citenamefont {Pederiva}, \citenamefont
  {Pieper}, \citenamefont {Schiavilla}, \citenamefont {Schmidt},\ and\
  \citenamefont {Wiringa}}]{RevModPhys.87.1067}%
  \BibitemOpen
  \bibfield  {author} {\bibinfo {author} {\bibfnamefont {J.}~\bibnamefont
  {Carlson}}, \bibinfo {author} {\bibfnamefont {S.}~\bibnamefont {Gandolfi}},
  \bibinfo {author} {\bibfnamefont {F.}~\bibnamefont {Pederiva}}, \bibinfo
  {author} {\bibfnamefont {S.~C.}\ \bibnamefont {Pieper}}, \bibinfo {author}
  {\bibfnamefont {R.}~\bibnamefont {Schiavilla}}, \bibinfo {author}
  {\bibfnamefont {K.~E.}\ \bibnamefont {Schmidt}}, \ and\ \bibinfo {author}
  {\bibfnamefont {R.~B.}\ \bibnamefont {Wiringa}},\ }\href {\doibase
  10.1103/RevModPhys.87.1067} {\bibfield  {journal} {\bibinfo  {journal} {Rev.
  Mod. Phys.}\ }\textbf {\bibinfo {volume} {87}},\ \bibinfo {pages} {1067}
  (\bibinfo {year} {2015})}\BibitemShut {NoStop}%
\bibitem [{\citenamefont {Carleo}\ and\ \citenamefont
  {Troyer}(2017)}]{carleo2017solving}%
  \BibitemOpen
  \bibfield  {author} {\bibinfo {author} {\bibfnamefont {G.}~\bibnamefont
  {Carleo}}\ and\ \bibinfo {author} {\bibfnamefont {M.}~\bibnamefont
  {Troyer}},\ }\href@noop {} {\bibfield  {journal} {\bibinfo  {journal}
  {Science}\ }\textbf {\bibinfo {volume} {355}},\ \bibinfo {pages} {602}
  (\bibinfo {year} {2017})}\BibitemShut {NoStop}%
\bibitem [{\citenamefont {Li}(2017)}]{li2017deep}%
  \BibitemOpen
  \bibfield  {author} {\bibinfo {author} {\bibfnamefont {Y.}~\bibnamefont
  {Li}},\ }\href@noop {} {\bibfield  {journal} {\bibinfo  {journal} {arXiv
  preprint arXiv:1701.07274}\ } (\bibinfo {year} {2017})}\BibitemShut {NoStop}%
\bibitem [{\citenamefont {Peng}\ \emph {et~al.}(2019)\citenamefont {Peng},
  \citenamefont {Harrow}, \citenamefont {Ozols},\ and\ \citenamefont
  {Wu}}]{peng2019simulating}%
  \BibitemOpen
  \bibfield  {author} {\bibinfo {author} {\bibfnamefont {T.}~\bibnamefont
  {Peng}}, \bibinfo {author} {\bibfnamefont {A.}~\bibnamefont {Harrow}},
  \bibinfo {author} {\bibfnamefont {M.}~\bibnamefont {Ozols}}, \ and\ \bibinfo
  {author} {\bibfnamefont {X.}~\bibnamefont {Wu}},\ }\href@noop {} {\bibfield
  {journal} {\bibinfo  {journal} {arXiv preprint arXiv:1904.00102}\ } (\bibinfo
  {year} {2019})}\BibitemShut {NoStop}%
\bibitem [{\citenamefont {Bravyi}\ \emph {et~al.}(2016)\citenamefont {Bravyi},
  \citenamefont {Smith},\ and\ \citenamefont {Smolin}}]{bravyi2016trading}%
  \BibitemOpen
  \bibfield  {author} {\bibinfo {author} {\bibfnamefont {S.}~\bibnamefont
  {Bravyi}}, \bibinfo {author} {\bibfnamefont {G.}~\bibnamefont {Smith}}, \
  and\ \bibinfo {author} {\bibfnamefont {J.~A.}\ \bibnamefont {Smolin}},\
  }\href@noop {} {\bibfield  {journal} {\bibinfo  {journal} {Physical Review
  X}\ }\textbf {\bibinfo {volume} {6}},\ \bibinfo {pages} {021043} (\bibinfo
  {year} {2016})}\BibitemShut {NoStop}%
\bibitem [{\citenamefont {Sun}\ \emph {et~al.}(2021)\citenamefont {Sun},
  \citenamefont {Endo}, \citenamefont {Lin}, \citenamefont {Hayden},
  \citenamefont {Vedral},\ and\ \citenamefont {Yuan}}]{sun2021perturbative}%
  \BibitemOpen
  \bibfield  {author} {\bibinfo {author} {\bibfnamefont {J.}~\bibnamefont
  {Sun}}, \bibinfo {author} {\bibfnamefont {S.}~\bibnamefont {Endo}}, \bibinfo
  {author} {\bibfnamefont {H.}~\bibnamefont {Lin}}, \bibinfo {author}
  {\bibfnamefont {P.}~\bibnamefont {Hayden}}, \bibinfo {author} {\bibfnamefont
  {V.}~\bibnamefont {Vedral}}, \ and\ \bibinfo {author} {\bibfnamefont
  {X.}~\bibnamefont {Yuan}},\ }\href@noop {} {\enquote {\bibinfo {title}
  {Perturbative quantum simulation},}\ } (\bibinfo {year} {2021}),\ \Eprint
  {http://arxiv.org/abs/2106.05938} {arXiv:2106.05938 [quant-ph]} \BibitemShut
  {NoStop}%
\bibitem [{\citenamefont {Mitarai}\ and\ \citenamefont
  {Fujii}(2019)}]{mitarai2019constructing}%
  \BibitemOpen
  \bibfield  {author} {\bibinfo {author} {\bibfnamefont {K.}~\bibnamefont
  {Mitarai}}\ and\ \bibinfo {author} {\bibfnamefont {K.}~\bibnamefont
  {Fujii}},\ }\href@noop {} {\bibfield  {journal} {\bibinfo  {journal} {arXiv
  preprint arXiv:1909.07534}\ } (\bibinfo {year} {2019})}\BibitemShut {NoStop}%
\bibitem [{\citenamefont {{Mitarai}}\ and\ \citenamefont
  {{Fujii}}(2020)}]{2020arXiv200611174M}%
  \BibitemOpen
  \bibfield  {author} {\bibinfo {author} {\bibfnamefont {K.}~\bibnamefont
  {{Mitarai}}}\ and\ \bibinfo {author} {\bibfnamefont {K.}~\bibnamefont
  {{Fujii}}},\ }\href@noop {} {\bibfield  {journal} {\bibinfo  {journal} {arXiv
  e-prints}\ ,\ \bibinfo {eid} {arXiv:2006.11174}} (\bibinfo {year} {2020})},\
  \Eprint {http://arxiv.org/abs/2006.11174} {arXiv:2006.11174 [quant-ph]}
  \BibitemShut {NoStop}%
\bibitem [{\citenamefont {Arute}\ \emph {et~al.}(2019)\citenamefont {Arute},
  \citenamefont {Arya}, \citenamefont {Babbush}, \citenamefont {Bacon},
  \citenamefont {Bardin}, \citenamefont {Barends}, \citenamefont {Biswas},
  \citenamefont {Boixo}, \citenamefont {Brandao}, \citenamefont {Buell} \emph
  {et~al.}}]{arute2019quantum}%
  \BibitemOpen
  \bibfield  {author} {\bibinfo {author} {\bibfnamefont {F.}~\bibnamefont
  {Arute}}, \bibinfo {author} {\bibfnamefont {K.}~\bibnamefont {Arya}},
  \bibinfo {author} {\bibfnamefont {R.}~\bibnamefont {Babbush}}, \bibinfo
  {author} {\bibfnamefont {D.}~\bibnamefont {Bacon}}, \bibinfo {author}
  {\bibfnamefont {J.~C.}\ \bibnamefont {Bardin}}, \bibinfo {author}
  {\bibfnamefont {R.}~\bibnamefont {Barends}}, \bibinfo {author} {\bibfnamefont
  {R.}~\bibnamefont {Biswas}}, \bibinfo {author} {\bibfnamefont
  {S.}~\bibnamefont {Boixo}}, \bibinfo {author} {\bibfnamefont {F.~G.}\
  \bibnamefont {Brandao}}, \bibinfo {author} {\bibfnamefont {D.~A.}\
  \bibnamefont {Buell}},  \emph {et~al.},\ }\href@noop {} {\bibfield  {journal}
  {\bibinfo  {journal} {Nature}\ }\textbf {\bibinfo {volume} {574}},\ \bibinfo
  {pages} {505} (\bibinfo {year} {2019})}\BibitemShut {NoStop}%
\bibitem [{\citenamefont {Huang}\ \emph {et~al.}(2020)\citenamefont {Huang},
  \citenamefont {Zhang}, \citenamefont {Newman}, \citenamefont {Cai},
  \citenamefont {Gao}, \citenamefont {Tian}, \citenamefont {Wu}, \citenamefont
  {Xu}, \citenamefont {Yu}, \citenamefont {Yuan}, \citenamefont {Szegedy},
  \citenamefont {Shi},\ and\ \citenamefont {Chen}}]{huang2020classical}%
  \BibitemOpen
  \bibfield  {author} {\bibinfo {author} {\bibfnamefont {C.}~\bibnamefont
  {Huang}}, \bibinfo {author} {\bibfnamefont {F.}~\bibnamefont {Zhang}},
  \bibinfo {author} {\bibfnamefont {M.}~\bibnamefont {Newman}}, \bibinfo
  {author} {\bibfnamefont {J.}~\bibnamefont {Cai}}, \bibinfo {author}
  {\bibfnamefont {X.}~\bibnamefont {Gao}}, \bibinfo {author} {\bibfnamefont
  {Z.}~\bibnamefont {Tian}}, \bibinfo {author} {\bibfnamefont {J.}~\bibnamefont
  {Wu}}, \bibinfo {author} {\bibfnamefont {H.}~\bibnamefont {Xu}}, \bibinfo
  {author} {\bibfnamefont {H.}~\bibnamefont {Yu}}, \bibinfo {author}
  {\bibfnamefont {B.}~\bibnamefont {Yuan}}, \bibinfo {author} {\bibfnamefont
  {M.}~\bibnamefont {Szegedy}}, \bibinfo {author} {\bibfnamefont
  {Y.}~\bibnamefont {Shi}}, \ and\ \bibinfo {author} {\bibfnamefont
  {J.}~\bibnamefont {Chen}},\ }\href@noop {} {\  (\bibinfo {year} {2020})},\
  \Eprint {http://arxiv.org/abs/2005.06787} {arXiv:2005.06787 [quant-ph]}
  \BibitemShut {NoStop}%
\bibitem [{\citenamefont {Fujii}\ \emph {et~al.}(2020)\citenamefont {Fujii},
  \citenamefont {Mitarai}, \citenamefont {Mizukami},\ and\ \citenamefont
  {Nakagawa}}]{fujii2020deep}%
  \BibitemOpen
  \bibfield  {author} {\bibinfo {author} {\bibfnamefont {K.}~\bibnamefont
  {Fujii}}, \bibinfo {author} {\bibfnamefont {K.}~\bibnamefont {Mitarai}},
  \bibinfo {author} {\bibfnamefont {W.}~\bibnamefont {Mizukami}}, \ and\
  \bibinfo {author} {\bibfnamefont {Y.~O.}\ \bibnamefont {Nakagawa}},\
  }\href@noop {} {\  (\bibinfo {year} {2020})},\ \Eprint
  {http://arxiv.org/abs/2007.10917} {arXiv:2007.10917 [quant-ph]} \BibitemShut
  {NoStop}%
\bibitem [{\citenamefont {Or{\'u}s}(2014)}]{orus2014practical}%
  \BibitemOpen
  \bibfield  {author} {\bibinfo {author} {\bibfnamefont {R.}~\bibnamefont
  {Or{\'u}s}},\ }\href@noop {} {\bibfield  {journal} {\bibinfo  {journal}
  {Annals of Physics}\ }\textbf {\bibinfo {volume} {349}},\ \bibinfo {pages}
  {117} (\bibinfo {year} {2014})}\BibitemShut {NoStop}%
\bibitem [{\citenamefont {Endo}\ \emph {et~al.}(2020)\citenamefont {Endo},
  \citenamefont {Sun}, \citenamefont {Li}, \citenamefont {Benjamin},\ and\
  \citenamefont {Yuan}}]{endo2020variational}%
  \BibitemOpen
  \bibfield  {author} {\bibinfo {author} {\bibfnamefont {S.}~\bibnamefont
  {Endo}}, \bibinfo {author} {\bibfnamefont {J.}~\bibnamefont {Sun}}, \bibinfo
  {author} {\bibfnamefont {Y.}~\bibnamefont {Li}}, \bibinfo {author}
  {\bibfnamefont {S.~C.}\ \bibnamefont {Benjamin}}, \ and\ \bibinfo {author}
  {\bibfnamefont {X.}~\bibnamefont {Yuan}},\ }\href@noop {} {\bibfield
  {journal} {\bibinfo  {journal} {Physical Review Letters}\ }\textbf {\bibinfo
  {volume} {125}},\ \bibinfo {pages} {010501} (\bibinfo {year}
  {2020})}\BibitemShut {NoStop}%
\bibitem [{\citenamefont {Kitaev}(2001)}]{kitaev2001unpaired}%
  \BibitemOpen
  \bibfield  {author} {\bibinfo {author} {\bibfnamefont {A.~Y.}\ \bibnamefont
  {Kitaev}},\ }\href@noop {} {\bibfield  {journal} {\bibinfo  {journal}
  {Physics-Uspekhi}\ }\textbf {\bibinfo {volume} {44}},\ \bibinfo {pages} {131}
  (\bibinfo {year} {2001})}\BibitemShut {NoStop}%
\bibitem [{\citenamefont {Lutchyn}\ \emph {et~al.}(2010)\citenamefont
  {Lutchyn}, \citenamefont {Sau},\ and\ \citenamefont
  {Sarma}}]{lutchyn2010majorana}%
  \BibitemOpen
  \bibfield  {author} {\bibinfo {author} {\bibfnamefont {R.~M.}\ \bibnamefont
  {Lutchyn}}, \bibinfo {author} {\bibfnamefont {J.~D.}\ \bibnamefont {Sau}}, \
  and\ \bibinfo {author} {\bibfnamefont {S.~D.}\ \bibnamefont {Sarma}},\
  }\href@noop {} {\bibfield  {journal} {\bibinfo  {journal} {Physical review
  letters}\ }\textbf {\bibinfo {volume} {105}},\ \bibinfo {pages} {077001}
  (\bibinfo {year} {2010})}\BibitemShut {NoStop}%
\bibitem [{\citenamefont {Jordan}\ \emph {et~al.}(2012)\citenamefont {Jordan},
  \citenamefont {Lee},\ and\ \citenamefont {Preskill}}]{Jordan:2011ne}%
  \BibitemOpen
  \bibfield  {author} {\bibinfo {author} {\bibfnamefont {S.~P.}\ \bibnamefont
  {Jordan}}, \bibinfo {author} {\bibfnamefont {K.~S.}\ \bibnamefont {Lee}}, \
  and\ \bibinfo {author} {\bibfnamefont {J.}~\bibnamefont {Preskill}},\ }\href
  {\doibase 10.1126/science.1217069} {\bibfield  {journal} {\bibinfo  {journal}
  {Science}\ }\textbf {\bibinfo {volume} {336}},\ \bibinfo {pages} {1130}
  (\bibinfo {year} {2012})},\ \Eprint {http://arxiv.org/abs/1111.3633}
  {arXiv:1111.3633 [quant-ph]} \BibitemShut {NoStop}%
\bibitem [{\citenamefont {Jordan}\ \emph {et~al.}(2014)\citenamefont {Jordan},
  \citenamefont {Lee},\ and\ \citenamefont {Preskill}}]{Jordan:2011ci}%
  \BibitemOpen
  \bibfield  {author} {\bibinfo {author} {\bibfnamefont {S.~P.}\ \bibnamefont
  {Jordan}}, \bibinfo {author} {\bibfnamefont {K.~S.}\ \bibnamefont {Lee}}, \
  and\ \bibinfo {author} {\bibfnamefont {J.}~\bibnamefont {Preskill}},\
  }\href@noop {} {\bibfield  {journal} {\bibinfo  {journal} {Quant. Inf.
  Comput.}\ }\textbf {\bibinfo {volume} {14}},\ \bibinfo {pages} {1014}
  (\bibinfo {year} {2014})},\ \Eprint {http://arxiv.org/abs/1112.4833}
  {arXiv:1112.4833 [hep-th]} \BibitemShut {NoStop}%
\bibitem [{\citenamefont {Gao}\ \emph {et~al.}(2017)\citenamefont {Gao},
  \citenamefont {Jafferis},\ and\ \citenamefont {Wall}}]{Gao:2016bin}%
  \BibitemOpen
  \bibfield  {author} {\bibinfo {author} {\bibfnamefont {P.}~\bibnamefont
  {Gao}}, \bibinfo {author} {\bibfnamefont {D.~L.}\ \bibnamefont {Jafferis}}, \
  and\ \bibinfo {author} {\bibfnamefont {A.~C.}\ \bibnamefont {Wall}},\ }\href
  {\doibase 10.1007/JHEP12(2017)151} {\bibfield  {journal} {\bibinfo  {journal}
  {JHEP}\ }\textbf {\bibinfo {volume} {12}},\ \bibinfo {pages} {151} (\bibinfo
  {year} {2017})},\ \Eprint {http://arxiv.org/abs/1608.05687} {arXiv:1608.05687
  [hep-th]} \BibitemShut {NoStop}%
\bibitem [{\citenamefont {Balasubramanian}\ \emph {et~al.}(2014)\citenamefont
  {Balasubramanian}, \citenamefont {Hayden}, \citenamefont {Maloney},
  \citenamefont {Marolf},\ and\ \citenamefont
  {Ross}}]{balasubramanian2014multiboundary}%
  \BibitemOpen
  \bibfield  {author} {\bibinfo {author} {\bibfnamefont {V.}~\bibnamefont
  {Balasubramanian}}, \bibinfo {author} {\bibfnamefont {P.}~\bibnamefont
  {Hayden}}, \bibinfo {author} {\bibfnamefont {A.}~\bibnamefont {Maloney}},
  \bibinfo {author} {\bibfnamefont {D.}~\bibnamefont {Marolf}}, \ and\ \bibinfo
  {author} {\bibfnamefont {S.~F.}\ \bibnamefont {Ross}},\ }\href@noop {}
  {\bibfield  {journal} {\bibinfo  {journal} {Classical and Quantum Gravity}\
  }\textbf {\bibinfo {volume} {31}},\ \bibinfo {pages} {185015} (\bibinfo
  {year} {2014})}\BibitemShut {NoStop}%
\bibitem [{\citenamefont {Maldacena}\ and\ \citenamefont
  {Susskind}(2013)}]{Maldacena:2013xja}%
  \BibitemOpen
  \bibfield  {author} {\bibinfo {author} {\bibfnamefont {J.}~\bibnamefont
  {Maldacena}}\ and\ \bibinfo {author} {\bibfnamefont {L.}~\bibnamefont
  {Susskind}},\ }\href {\doibase 10.1002/prop.201300020} {\bibfield  {journal}
  {\bibinfo  {journal} {Fortsch. Phys.}\ }\textbf {\bibinfo {volume} {61}},\
  \bibinfo {pages} {781} (\bibinfo {year} {2013})},\ \Eprint
  {http://arxiv.org/abs/1306.0533} {arXiv:1306.0533 [hep-th]} \BibitemShut
  {NoStop}%
\bibitem [{\citenamefont {Shenker}\ and\ \citenamefont
  {Stanford}(2014)}]{Shenker:2013pqa}%
  \BibitemOpen
  \bibfield  {author} {\bibinfo {author} {\bibfnamefont {S.~H.}\ \bibnamefont
  {Shenker}}\ and\ \bibinfo {author} {\bibfnamefont {D.}~\bibnamefont
  {Stanford}},\ }\href {\doibase 10.1007/JHEP03(2014)067} {\bibfield  {journal}
  {\bibinfo  {journal} {JHEP}\ }\textbf {\bibinfo {volume} {03}},\ \bibinfo
  {pages} {067} (\bibinfo {year} {2014})},\ \Eprint
  {http://arxiv.org/abs/1306.0622} {arXiv:1306.0622 [hep-th]} \BibitemShut
  {NoStop}%
\bibitem [{\citenamefont {Yoshida}\ and\ \citenamefont
  {Yao}(2019)}]{Yoshida:2018vly}%
  \BibitemOpen
  \bibfield  {author} {\bibinfo {author} {\bibfnamefont {B.}~\bibnamefont
  {Yoshida}}\ and\ \bibinfo {author} {\bibfnamefont {N.~Y.}\ \bibnamefont
  {Yao}},\ }\href {\doibase 10.1103/PhysRevX.9.011006} {\bibfield  {journal}
  {\bibinfo  {journal} {Phys. Rev. X}\ }\textbf {\bibinfo {volume} {9}},\
  \bibinfo {pages} {011006} (\bibinfo {year} {2019})},\ \Eprint
  {http://arxiv.org/abs/1803.10772} {arXiv:1803.10772 [quant-ph]} \BibitemShut
  {NoStop}%
\bibitem [{\citenamefont {Brown}\ \emph {et~al.}(2019)\citenamefont {Brown},
  \citenamefont {Gharibyan}, \citenamefont {Leichenauer}, \citenamefont {Lin},
  \citenamefont {Nezami}, \citenamefont {Salton}, \citenamefont {Susskind},
  \citenamefont {Swingle},\ and\ \citenamefont {Walter}}]{Brown:2019hmk}%
  \BibitemOpen
  \bibfield  {author} {\bibinfo {author} {\bibfnamefont {A.~R.}\ \bibnamefont
  {Brown}}, \bibinfo {author} {\bibfnamefont {H.}~\bibnamefont {Gharibyan}},
  \bibinfo {author} {\bibfnamefont {S.}~\bibnamefont {Leichenauer}}, \bibinfo
  {author} {\bibfnamefont {H.~W.}\ \bibnamefont {Lin}}, \bibinfo {author}
  {\bibfnamefont {S.}~\bibnamefont {Nezami}}, \bibinfo {author} {\bibfnamefont
  {G.}~\bibnamefont {Salton}}, \bibinfo {author} {\bibfnamefont
  {L.}~\bibnamefont {Susskind}}, \bibinfo {author} {\bibfnamefont
  {B.}~\bibnamefont {Swingle}}, \ and\ \bibinfo {author} {\bibfnamefont
  {M.}~\bibnamefont {Walter}},\ }\href@noop {} {\  (\bibinfo {year} {2019})},\
  \Eprint {http://arxiv.org/abs/1911.06314} {arXiv:1911.06314 [quant-ph]}
  \BibitemShut {NoStop}%
\end{thebibliography}%

\appendix
\newpage
\widetext

\tableofcontents


\begin{figure*}[b]\centering
  \includegraphics[width=.6\linewidth]{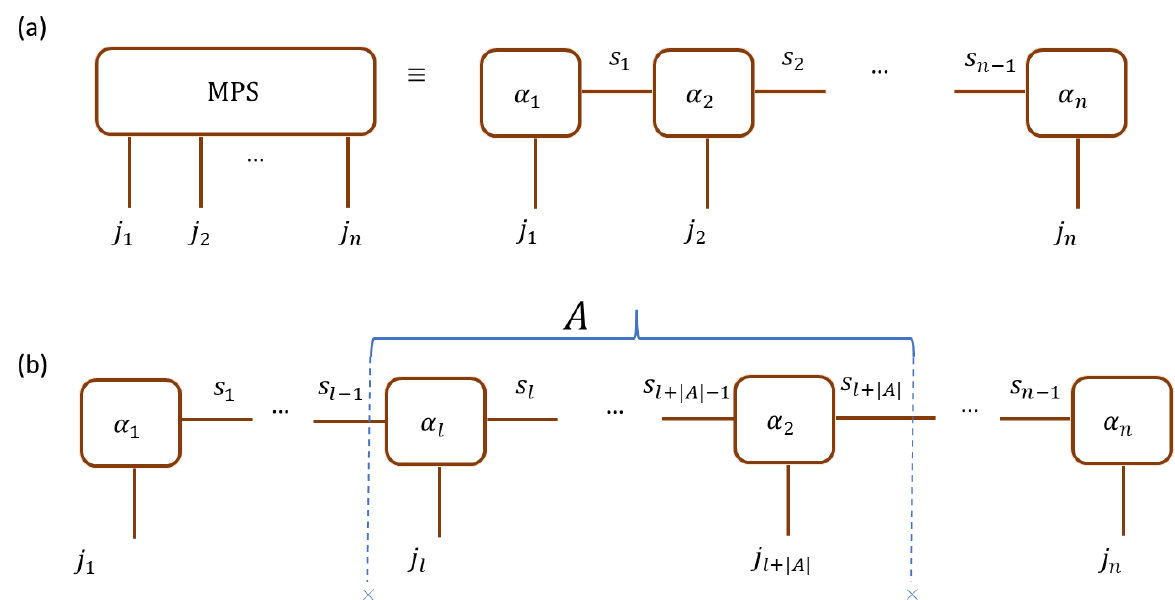}
  \caption{Illustration of a typical classical tensor network --- the matrix product state.}
   \label{fig:FigMPS}
\end{figure*}

\section{Hybrid tensor networks}\label{SM:frame}
As shown in the main text, the hybrid tensor network consists of classical and quantum tensors, whose mathematical definition is consistent with that of the conventional tensor network. That is, tensor contractions are mathematically defined the same for classical and quantum tensors. Nevertheless, we distinguish them because operationally, classical tensors are contracted classically via tensor multiplication, while quantum tensors are contracted via measuring a quantum state on a quantum computer. In the following, we elaborate on the detailed definition of classical and quantum tensors, the definition of tensor contraction and its meaning, the way to measure local observables, and the application in quantum simulation.

\subsection{Classical and quantum tensors}
A general rank-$n$ tensor is a multi-dimensional array  with $n$ indices denoted as $T_{i_1,i_2,\dots,i_n}$. In quantum mechanics, it represents the wave function of an $n$-partite quantum state in the computational basis,
\begin{equation}	\label{eqn:quantum_state}
	\ket{\psi}=\sum_{j_1,j_2,\dots,j_n}\psi_{j_1,j_2,\dots,j_n}\ket{j_1}\ket{j_2}\dots\ket{j_n}.
\end{equation}
We can see that directly storing a general quantum state in a classical memory is highly inefficient, which in general costs exponential space resources with respect to the number of parties. This thus motivates us to find more efficient ways to represent quantum states.

The deep observation of physicists is that quantum states in nature may only lie in a small subset of the whole Hilbert space, where the area law scaling may exist, for example, with the ground state of certain gaped local Hamiltonians. It thus enables the possibility of efficient classical representation of these quantum states. The overall idea is to decompose the rank-$n$ tensor into a network of low-rank tensors. 
Take the matrix product state (MPS) ansatz as an example, as shown in Fig.~\ref{fig:FigMPS}(a), the rank-$n$ tensor is now decomposed into $n$ low-rank tensors as
\begin{equation}\label{SM:MPS}
	\ket{\psi} = \sum_{j_1,j_2,\dots,j_n} \tr[\alpha_1^{j_1}\dots \alpha_n^{j_n}]\ket{j_1}\ket{j_2}\dots\ket{j_n}.
\end{equation}
Here each $\alpha_k$ is a rank-3 tensor (except for $\alpha_1$ and $\alpha_n$ whose rank is 2), and the index $j_k$ is the physical index, with dimension $2$ for the qubit case. The other two indices (or one index for $\alpha_1, \alpha_n$) are called the bond indices with dimension $\kappa$, normally being a constant number. Here the trace operation in Eq.~\eqref{SM:MPS} is for the bond indices.

It is clear that the entanglement of any local subsystem is upper bounded by $2\kappa$, where $2$ accounts for the two boundaries and $\kappa$ for the contribution from each boundary, as shown in Fig.~\ref{fig:FigMPS}(b). 
Note that the MPS representation thus compresses the space of the $n$-partite state from $O(2^n)$ to $O(n\kappa^2)$, which is from exponential to linear with the particle number $n$. This tremendous reduction is based on the pre-knowledge of the weakly entangled state under the geometrically local interactions. However, typically the quantum state in the large Hilbert space could be highly entangled, such as excited eigenstates and states after quenched dynamics of the chaotic Hamiltonian. Many different classical tensor networks have been proposed for different problems. Nevertheless, it would be likely that certain quantum systems, such as \rm the electronic structure in chemistry and the Hubbard model, may not be efficiently described via any classical method.  

It thus motivates the idea of quantum simulation, i.e., using a controllable quantum system to simulate a target quantum problem. A quantum state generated from applying a unitary circuit to a certain initial state forms an intrinsic large-rank quantum tensor and can be naturally stored and manipulated with a quantum computer. As an alternative approach, several quantum algorithms have been proposed to solve either static or dynamic problems of a general many-body problem. 
In literature, classical tensor network theory and quantum simulation are generally used as separate techniques in classical and quantum computing. Here, we introduce quantum tensors to be general $n$-partite quantum states prepared by a quantum computer and classical tensors to low-rank tensors stored in a classical computer and show the combination of quantum and classical tensors as a hybrid tenor network. 

Suppose we generate an $n$-partite quantum state by applying a unitary $U_{\psi}$ to an initial state $\ket{\bar 0}$ as $\ket{\psi}=U_{\psi}\ket{\bar 0}$. As shown in Eq.~\eqref{eqn:quantum_state}, the quantum state can be regarded as a rank-$n$ tensor in the computational basis. 
We can also introduce classical index to the quantum state by applying different unitary gates as
\begin{equation}
 	\ket{\psi^i}=U_{\psi^i}\ket{\bar 0}=\sum_{j_1,j_2,\dots,j_n}\psi^i_{j_1,j_2,\dots,j_n}\ket{j_1}\ket{j_2}\dots\ket{j_n}.
\end{equation}
Alternatively, we can also apply the same unitary but to different initial states as
\begin{equation}
	\ket{\psi^i}=U\ket{\bar 0^i}=\sum_{j_1,j_2,\dots,j_n}\psi^i_{j_1,j_2,\dots,j_n}\ket{j_1}\ket{j_2}\dots\ket{j_n},
\end{equation}
where the classical index $i$ indicates the different unitaries or different initial states. As a result, it as a whole forms a rank-$(n+1)$ tensor $\Psi_{j_1,j_2,\dots,j_n}^i$. 
For simplicity, we only introduce one classical index here, and it is clear that there is no restriction to introduce more classical indices.  
We regard all these cases as quantum tensors, and the network connected with quantum tensors and classical tensors as a hybrid tensor network. 
Hereafter, we put indices corresponding to classical labels and quantum basis to the superscript and subscript of the tensor, respectively.

\subsection{Hybrid tensor networks}\label{SM:subContract}
Here, we show how to connect quantum and classical tensors to form a hybrid tensor. When connecting two tensors, being either classical or quantum, we follow the conventional rule for tensor contraction. 
While the mathematical definition of tensor contraction of a hybrid tensor network is consistent with the conventional definition, its practical meaning can be different.  
Depending on whether the tensor and the index are quantum or classical, there are five different cases under contraction, as shown in Fig. \ref{fig:QansatzContraction}. To ease the explanation, without loss of generality, we consider the contraction of rank-2 classical tensors and rank-($n+1$) quantum tensors. We also use $\dots$ to abbreviate the quantum indices when they are not contracted. 

\begin{itemize}
    \item Case 1: quantum tensor (contracted index: classical) \& classical tensor (contracted index: classical).

First, we combine a quantum tensor $\psi_{\dots}^{i_1}$ with a classical tensor $\alpha^{i_1,i_2}$ to form a new rank-$(n+1)$ tensor,
\begin{equation}\label{Eq:Q-C-C}
	\tilde{\psi}_{\dots}^{i_2}  = \sum_{i_1} \psi_{\dots}^{i_1} \cdot \alpha^{i_1,i_2},
\end{equation}
where the contracted index from the quantum and classical tensors is a classical label. 
To understand the meaning of Eq.~\eqref{Eq:Q-C-C}, we regard the quantum tensor $\psi_{\dots}^{i_1}$ as a set of independent quantum states $\{\ket{\psi^{i_1}}\}$ and the tensor $\tilde{\psi}_{\dots}^{i_2}$ represents a new set of states
\begin{equation}\label{Eq:Q-C-C2}
	\ket{\tilde{\psi}^{i_2}}  = \sum_{i_1}  \alpha^{i_1,i_2}\ket{\psi^{i_1}},
\end{equation}
where each one is now a superposition of the original states $\{\ket{\psi^{i_1}}\}$. As a special case, when the classical tensor is rank-1, $\alpha^{i_1}$, the output tensor is
\begin{equation}\label{Eq:Q-C-C222}
	\ket{\tilde{\psi}}  = \sum_{i_1}  \alpha^{i_1}\ket{\psi^{i_1}}.
\end{equation}
Therefore, we can connect a classical tensor to the classical index of a quantum tensor to effectively represent a superposition of quantum states.

\begin{figure*}[t]\centering
  \includegraphics[width=.97\linewidth]{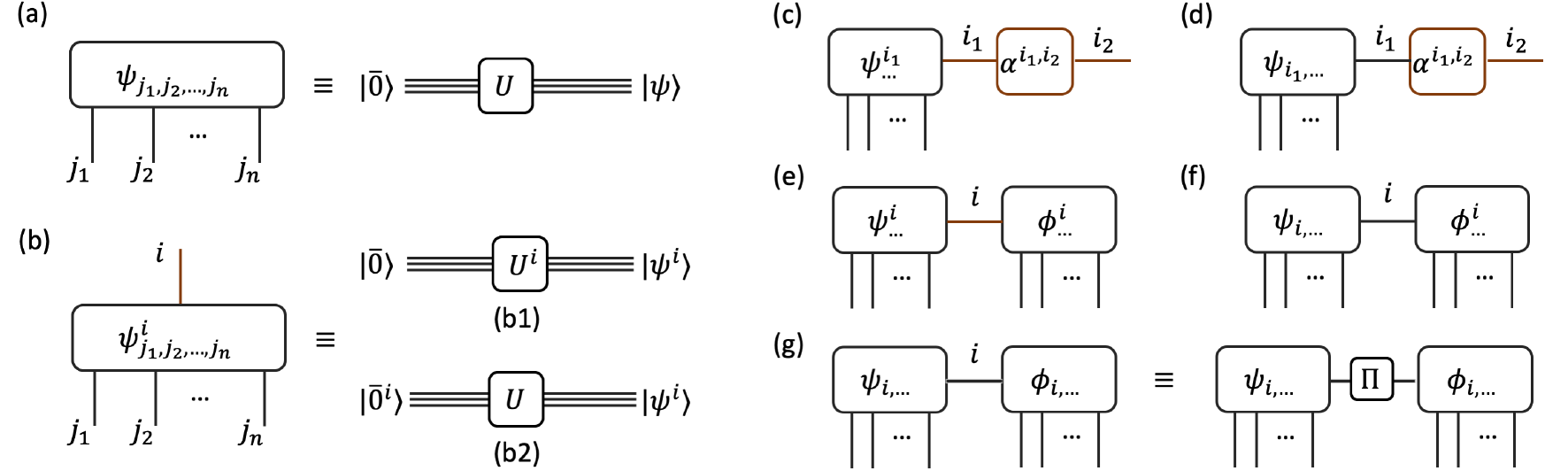}
  \caption{Tensor network representation of quantum states and tensor contractions. 
   (a) A general $n$-partite quantum state can be regarded as a rank-$n$ tensor. 
   (b) We add a classical index to an $n$-partite quantum state to generate a rank-$(n+1)$ tensor with $n$ indices representing $n$ quantum systems and $1$ classical index. With a quantum circuit, it is equivalent to preparing different states $\ket{\psi^i}=U^i\ket{i}$ with  (b1) different unitary operations as $\ket{\psi^i}=U^i\ket{\bar 0}$ or (b2) simply the same unitary but different initial states as $\ket{\psi^i}=U\ket{\bar 0^i}$. 
   (c, d)  Tensor contractions between a quantum tensor and a classical tensor. 
   (e, f, g) Tensor contractions between two quantum tensors. 
   (c, e) The contracted index of both tensors corresponds to a classical index. 
   (d, f) The contracted index corresponds to a classical index for one tensor and a quantum index for another tensor. 
   (g) The contracted index of both tensors corresponds to a quantum index. The tensor $\Pi$ is equivalent to a projective measurement $\sum_{i=i'} \ket{i}\bra{i}\otimes\ket{i'}\bra{i'}$.}
   \label{fig:QansatzContraction}
\end{figure*}

    \item Case 2: quantum tensor (contracted index: quantum) \& classical tensor (contracted index: classical).
    
When the contracted index $i_1$ of the quantum tensor $\psi_{i_1, \dots}$ corresponds to a quantum system, the tensor contraction is similarly defined as
\begin{equation}\label{SM:Q-C2}
	\tilde{\psi}_{\dots}^{i_2}  = \sum_{i_1} \psi_{i_1,\dots} \cdot \alpha^{i_1,i_2}.
 \end{equation}
When considering quantum states, the contraction transforms an input state $\ket{\psi}$ to a set of output states $\{\ket{\tilde{\psi}^{i_2} }\}$ as
 \begin{equation}
 	\ket{\tilde{\psi}^{i_2} }  = \sum_{i_1} \alpha^{i_1,i_2} \braket{i_1|\psi},
\end{equation}
which is equivalent to projecting the contracted system onto $\ket{i_1}$ to form a set of un-normalized states $\ket{\psi^{i_1}}=\braket{i_1|\psi}$ and re-combining them with coefficients $\alpha^{i_1,i_2}$. 
Actually, if we regard $\alpha$ as a unitary gate with $i_2$ representing a quantum system, it corresponds to a local unitary transformation of the state. 

    \item Case 3: quantum tensor (contracted index: classical) \& quantum tensor (contracted index: classical).
    
Next, we consider the contraction of two quantum tensors with the contracted index being classical for both tensors. Suppose the two quantum tensors are $\psi_{\dots}^{i}$ and $\phi_{\dots}^{i}$, the contraction of index $i$ gives
\begin{equation}
	\tilde{\psi}_{\dots} = \sum_{i} \psi_{\dots}^{i}\cdot \phi_{\dots}^{i}.
 \end{equation}
Considering quantum states, the contraction transforms two sets of states $\{\ket{\psi^{i}}\}$ and $\{\ket{\phi^{i}}\}$ to an un-normalized state
 \begin{equation}
	\ket{\tilde{\psi}} = \sum_i \ket{\psi^i}\otimes \ket{\phi^i}.
\end{equation}
By contracting two quantum tensors, we can thus effectively entangle two quantum systems. We can also add a classical tensor in between so that the amplitude for each $\ket{\psi^i}\otimes \ket{\phi^i}$ is different. 

    \item Case 4: quantum tensor (contracted index: quantum) \& quantum tensor (contracted index: classical).

When one of the contracted indices corresponds to a quantum system, the contraction is similarly defined as
 \begin{equation}
 	\tilde{\psi}_{\dots} = \sum_{i} \psi_{i,\dots}\cdot \phi_{\dots}^{i}.
\end{equation}
Considering quantum states, the contraction converts $\ket{\psi}$ and $\{\ket{\phi^i}\}$ to
\begin{equation}
 	\ket{\tilde{\psi}} = \sum_i  \braket{i|\psi}\otimes \ket{\phi^i}.
\end{equation}
Again, this is equivalent to applying a projection to $\ket{\psi}$ to get a set of states $\{\ket{\psi^i}=\braket{i|\psi}\}$ and then connecting the classical indices of the two quantum tensors.

    \item Case 5: quantum tensor (contracted index: quantum) \& quantum tensor (contracted index: quantum).

When both contracted indices represent for quantum systems, we contract two quantum tensors $\psi_{i,\dots}$ and $\phi_{i,\dots}$ as
\begin{equation}
	\tilde{\psi}_{\dots} = \sum_{i} \psi_{i,\dots}\cdot \phi_{i,\dots}.
 \end{equation}
In the quantum state language, it is equivalent to
\begin{equation}
 	\ket{\tilde{\phi}}= \sum_{i} \braket{i|\psi}\otimes\braket{i|\phi}
 	=\sum_{i} \bra{i,i}\ket{\psi}\otimes\ket{\phi}.
 \end{equation}
which accounts for a Bell state projection on the contracted systems. Note that since it is a measurement on both states, the success probability could be less than 1. Then if there are multiple contractions of quantum indices, the overall probability will be exponentially small. Therefore, we only allow a constant number of contractions of two quantum indices in the hybrid tensor network. 

\end{itemize}

For a general hybrid tensor network consisting of classical and quantum tensors, the tensor contraction rule and its meaning with respect to the quantum state follow similarly by sequentially applying the above cases.

\subsection{Calculation of expectation values of local observables}\label{SM:Ob}
Given a hybrid tensor network representation of a quantum state, we now show how to measure the expectation values of tensor products of local observables. 
Here we only show the basic rules and whether the calculation is efficient highly depends on the structure of the hybrid tensor network, the same as the scenario of the conventional tensor network. The basic rule follows the same mathematics of tensor contraction.
Therefore, for classical tensors, the expectation value is calculated in the same way as conventional tensor networks. While for quantum tensors, we can no longer calculate the expectation value via tensor contraction since it involves matrix multiplication of a rank-$n$ tensor. Instead, we calculate the expectation value with a quantum computer by preparing the state and measuring it.

\begin{figure*}[t]\centering
  \includegraphics[width=1\linewidth]{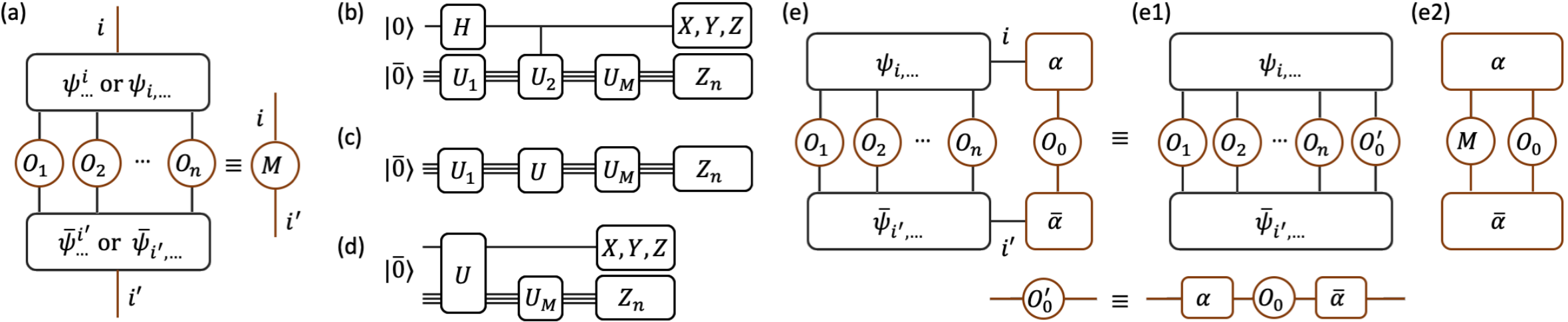}
  \caption{Measuring expectation values of a quantum tensor. 
  (a) Consider a rank-$(n+1)$ quantum tensor, which could be either an $n$-partite quantum state with a classical index $i$ or an $(n+1)$-partite quantum state with a quantum basis index $i$. The expectation value of the $n$ quantum systems gives a hermitian observable $M^{i',i}=\bra{\psi^{i'}}O_1\otimes O_2\otimes \cdots\otimes O_n\ket{\psi^i}$ on the open indices. Each element $M^{i',i}$ can be measured with a quantum circuit of (b), (c), or (d).
  (b) Suppose the index $i$ is classical and $\ket{\psi^i}=U^i\ket{\bar 0}$, we choose $U_1=U^i$, $U_2=U^{i'}(U^i)^\dag$, and $U_M$ to be the unitary that rotates the eigenstates of the observables to the computational basis. We get each $M^{i',i}$ by measuring the ancillary qubit in the $X$, $Y$, $Z$ bases and the other $n$ qubits in the the computational $Z$ basis. 
  (c) Suppose the index $i$ is classical and $\ket{\psi^i}=U\ket{\bar 0^i}$, we use $U_1$ to prepare four input states $\ket{\bar0^i}, \ket{\bar0^{i'}}, (\ket{\bar 0^i}+ \ket{\bar 0^{i'}})/\sqrt{2}, (\ket{\bar 0^i}+i \ket{\bar 0^{i'}})/\sqrt{2}$ and each $M^{i',i}$ corresponds to a linear combination of the measurement results. 
  (d) Suppose the index $i$ is quantum, after applying the unitary $U$ for preparing the state $\ket{\psi}=U\ket{\bar 0}$, we measure $n$ qubits in the computational basis and the qubit with index $i$ in the Pauli $X$, $Y$, and $Z$ bases. 
  (e) Tensor contraction can have different orders. With a rank-$(n+1)$ quantum tensor connected to a classical tensor, we can either (e1) first calculate the expectation value of the classical tensor and then measure the $(n+1)$-partite quantum state or (e2) first measure the $n$ systems via (d) and then do classical tensor contraction.  }\label{fig:FigExpectationsup}
\end{figure*}

As shown in Fig.~\ref{fig:FigExpectationsup}(a), we consider the expectation value on the $n$ quantum systems of a rank-$(n+1)$ quantum tensor. This tensor can be either an $n$-partite quantum state with a classical index $i$ or an $(n+1)$-partite quantum state with a quantum basis index $i$. By measuring the $n$ systems, it gives a new rank 2 tensor $M^{i',i}$ with two open indices $i$ and $i'$,
\begin{equation}\label{eq:smmii}
 M^{i',i}=\bra{\psi^{i'}}O_1\otimes O_2\otimes \cdots\otimes O_n\ket{\psi^i}.
\end{equation}
Here the definition is the same if we measure an $(n+1)$-partite quantum state. We always put the indices of $M^{i',i}$ to the superscript, because the measurement observable is always a classical low-rank tensor. 
Note that the matrix $M^{i',i}$ is always hermitian so that it can be measured when the indices $i$ and $i'$ are contracted to another quantum tensor. 
Now we show how to get $M^{i',i}$ for  different cases.
\begin{itemize}
    \item The rank-$(n+1)$ quantum tensor is an $n$-partite quantum state with a classical index $i$. 
    
   \begin{itemize}
       \item  Suppose $\ket{\psi^i}=U^i\ket{\bar 0}$, $U_1=U^i$. 
       We measure $M^{i',i}$ with the quantum circuit in Fig.~\ref{fig:FigExpectationsup}(b). Consider $U_2=U^{i'}(U^i)^\dag$ and $U_M$ to be the unitary that rotates the eigenstates of the observable to the computational basis. The output state before the $U_M$ gate is
       \begin{equation}
     \ket{\tilde\psi}=\frac{1}{\sqrt{2}}\big(\ket{0}\ket{\psi^i}+\ket{1}\ket{\psi^{i'}}\big).
    \end{equation}

When the ancillary qubit measures the Pauli $X$, $Y$, $Z$ operators, and the $n$-partite system measures $M=O_1\otimes O_2\otimes \cdots\otimes O_n$, the expectation values are
\begin{equation}
\begin{aligned}
    \braket{\tilde\psi|X\otimes M|\tilde\psi} &=\frac{1}{2}\big( M^{2,1} + M^{1,2}\big),\\
    \braket{\tilde\psi|Y\otimes M|\tilde\psi} &=\frac{1}{2}\big( iM^{2,1} -i M^{1,2}\big),\\
    \braket{\tilde\psi|Z\otimes M|\tilde\psi} &=\frac{1}{2}\big( M^{1,1} - M^{2,2}\big).\\        
\end{aligned}
\end{equation}
Note that $M^{1,2}$ is the complex conjugate of $M^{2,1}$, and we have
\begin{equation}
\begin{aligned}
    \braket{\tilde\psi|I\otimes M|\tilde\psi} &=\frac{1}{2}\big( M^{1,1} + M^{2,2}\big),\\        
\end{aligned}
\end{equation}
which can be obtained from the measurement of any Pauli basis. Therefore we can exactly solve each term $M^{i,j}$ $(i,j=1,2)$ and construct the 
measurement 
$$\tilde M=
 \left\{ \begin{matrix}
  M^{1,1} & M^{1,2}  \\
   M^{2,1}  &M^{2,2}  
  \end{matrix} \right\}$$.
  
  \item Suppose $\ket{\psi^i}=U\ket{\bar0^i}$. We measure $M^{i',i}$ with the quantum circuit in 
 Fig.~\ref{fig:FigExpectationsup}(c). Now we need to input $(\ket{\bar0^i}\pm \ket{\bar0^{i'}})/\sqrt{2}$ and $(\ket{\bar0^i}\pm i \ket{\bar0^{i'}})/\sqrt{2}$ and the matrix elements can be similarly obtained.
   \end{itemize}
    
    \item The rank-$(n+1)$ quantum tensor is an $n+1$-partite quantum state. 
    
    We need to measure 
    \begin{equation}\label{eq:smmii2}
 M^{i',i}=\bra{\psi}\ket{i'}\otimes O_1\otimes O_2\otimes \cdots\otimes O_n\otimes \bra{i}\ket{\psi}=\bra{\psi}(\ket{i'}\bra{i})\otimes O_1\otimes O_2\otimes \cdots\otimes O_n\ket{\psi}.
\end{equation}
Note that the matrix $\ket{i'}\bra{i}$ can always be represented as a linear combination of the Pauli operators, we can thus instead measure the uncontracted qubit in the three $X,Y,Z$ Pauli bases to equivalently get any $M^{i',i}$ as shown in Fig.~\ref{fig:FigExpectationsup}(d). Suppose $\ket{\psi}=U\ket{\bar0}$, we denote
\begin{equation}\label{eq:}
 E(\sigma)=\bra{\psi}\sigma\otimes O_1\otimes O_2\otimes \cdots\otimes O_n\ket{\psi},
\end{equation}
and we can reconstruct the measurement $\tilde M$ as
\begin{equation}
\tilde M= \frac{1}{2}\big(E(I)I+ E(X)X - E(Y)Y+ E(Z)Z\big), 
\end{equation}
where $E(X)$, $E(I)$, $E(Y)$, $E(Z)$ are the obtained expectation values with Pauli measurements $I, X,Y, Z$. 

\end{itemize}

Calculating the expectation value of a general hybrid tensor network follows the above basic rules for classical and quantum tensors. Nevertheless, similar to the conventional tensor network, different orders of tensor contraction could have different procedures and complexities. For example, say that we are considering the hybrid tensor shown in Fig.~\ref{fig:FigExpectationsup}(e), which consists of a rank-$(n+1)$ quantum tensor and a classical tensor. We could first contract the right classical observable $O_0$ with classical tensor $\alpha$, and obtain a new observable $O_0^{'}$. Then we measure the $n+1$-partite quantum state to get the final expectation value. Here we need classical contraction and a single local measurement with repetition samples $\mathcal{M}$. This procedure is shown in Fig.~\ref{fig:FigExpectationsup}(e1). Alternatively, we can also use the circuit in Fig.~\ref{fig:FigExpectationsup}(d) to reconstruct observable $M$ by measuring the $n+1$-partite quantum state and then contract the classical tensors. This procedure requires three local measurement settings ($X$, $Y$, and $Z$ on the first qubit) with total repetition samples of $3\mathcal{M}$.

\subsection{Application in variational quantum simulation}
The hybrid tensor network provides a way to more efficiently represent quantum states with fewer quantum resources. When using the hybrid tensor network, it can be applied in variational quantum simulation for solving static energy spectra and simulating the dynamic time evolution of a quantum system.  

We consider a many-body Hamiltonian 
\begin{equation}
    H=\sum_i \lambda_i h_i
\end{equation}
with coefficients $\lambda_i$ and tensor products of Pauli matrices $h_i$. To find the ground state of the Hamiltonian, we consider a parameterized hybrid quantum tensor network, which corresponds to a possibly un-normalized state $\ket{\psi(\vec x)}$. Here $\vec x$ denotes the parameters that can be changed in the hybrid tensor network, which includes the parameterized angles in the quantum circuit and parameters in the classical tensors. 
Then we can measure the average energy as 
\begin{equation}
    E(\vec x) = \frac{\bra{\psi(\vec x)}H\ket{\psi(\vec x)}}{\braket{\psi(\vec x)|\psi(\vec x)}} = \frac{\sum_i \lambda_i \bra{\psi(\vec x)}h_i\ket{\psi(\vec x)}}{\braket{\psi(\vec x)|\psi(\vec x)}},
\end{equation}
where each $\bra{\psi(\vec x)}h_i\ket{\psi(\vec x)}$ or the normalization $\braket{\psi(\vec x)|\psi(\vec x)}$ can be obtained by calculating the expectation value of the hybrid tensor network with the method we discussed in the previous section. Having measured $E(\vec x)$ for any $\vec x$, we can then optimize the parameter space via the classical algorithm to minimize $E(\vec x)$ to search for the ground state. We note that the whole optimization procedure is identical to the conventional approach called variational quantum eigensolver. The difference lies in the usage of the hybrid tensor network, which may enable quantum simulation of large systems with small quantum processors. We can also use the hybrid tensor network for simulating Hamiltonian dynamics. The circuit for the implementation of variational Hamiltonian simulation with hybrid tensor networks is slightly more complicated, and we leave the discussion to follow-up works. In the following, we mainly focus on using the hybrid tensor network for finding the ground state of Hamiltonians.

\section{Hybrid tree tensor networks}\label{SM:ansatzs}
Calculating a general hybrid tensor network can be costly. Here we expand the discussion of the main text and focus on hybrid tensor networks with a tree structure. 
We first consider examples of hybrid tree tensor networks (TNN) and discuss its application in representing correlations of the multipartite quantum state. We then study the cost of calculating the expectation values of a general hybrid TNN. We also study the entanglement property and correlation of the quantum state represented by the hybrid TNN.

\begin{figure*}[t]\centering
  \includegraphics[width=.9\linewidth]{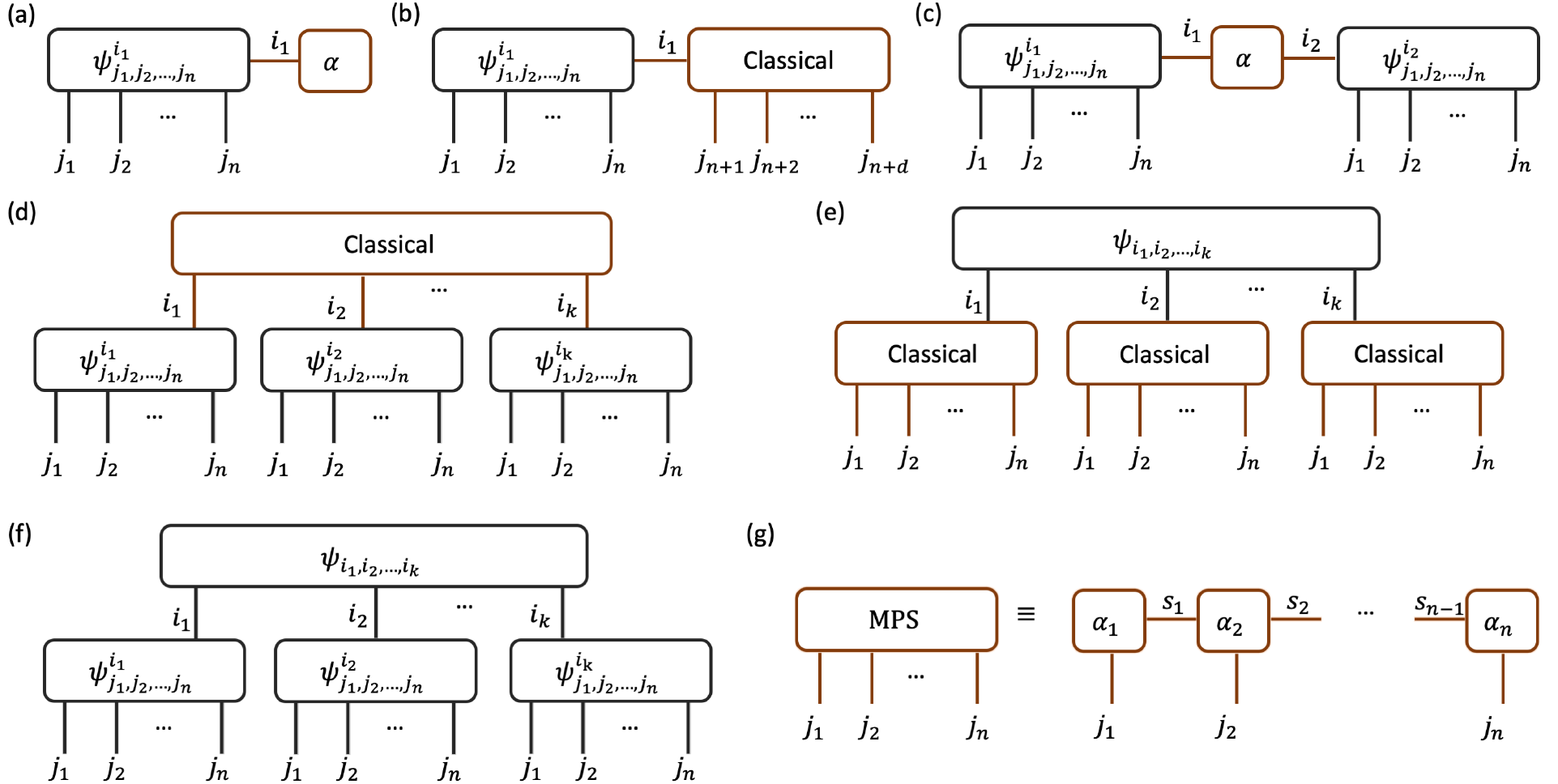}
  \caption{Hybrid quantum-classical tensor network. 
  (a) We can extend the power of a quantum state by adding a classical tensor as in Eq.~\eqref{Eq:generalansatz}. 
  (b) We can combine a quantum state and a classical tensor to represent a state in a larger Hilbert space as in Eq.~\eqref{SM:QplusC}. 
  (c) We can use a classical tensor to connect two quantum states as in Eq.~\eqref{SM:QCQ}. 
  (d) A quantum-classical hybrid tensor as in Eq.~\eqref{SM:QC2tree}.
  (e) A classical-quantum hybrid tensor as in Eq.~\eqref{SM:CQ2tree}.
  (f) A quantum-quantum hybrid tensor as in Eq.~\eqref{SM:QQ2tree}.
  (g) A commonly-used classical tensor network MPS.
  }\label{fig:ansatzsup}
\end{figure*}

\subsection{Example of hybrid tree tensor networks}
We show several examples of hybrid tree tensor networks in Fig.~\ref{fig:ansatzsup}. In the following, we discuss the application of each tensor network and its connection with existing results. For each $n$-partite state, we assume it is an $n$-qubit parameterized state $\ket{\psi(\vec\theta)}$, obtained by applying a sequence of local gates as $\ket{\psi(\vec\theta)} = \prod_jU_j(\theta_j)\ket{\bar 0}$ with an initial state $\ket{\bar 0}$ and parameters $\vec\theta=\{\theta_j\}$.

\subsubsection{Extending the power of the quantum state}
Suppose we use the quantum state $\ket{\psi(\vec \theta)}$ as a potential solution to the $n$-qubit problem. We can regard the quantum state as a pure rank-$n$ quantum tensor. A simple way to extend the capability of the rank-$n$ quantum tensor is to concatenate a classical tensor $\alpha^i$ to it as
\begin{equation}\label{Eq:generalansatz}
	\ket{\psi(\vec x)} = \sum_i \alpha^i\ket{\psi^i(\vec \theta_i)},
\end{equation}
where each $\ket{\psi^i(\vec \theta_i)}$ can be regarded as different rank-$n$ quantum tensors and $\vec x=(\alpha^1,\dots, \vec \theta_1,\dots)$ are the total parameter setting. Such a concatenation corresponds to the hybrid tensor network in  Fig.~\ref{fig:ansatzsup}~(a). To find the ground state of Hamiltonian $H$, we can obtain the energy as
\begin{equation}\label{Eq:brakeths}
	E(\vec x) = \frac{\bra{\psi(\vec x)}H\ket{\psi(\vec x)}}{\braket{\psi(\vec x)|\psi(\vec x)}} = \frac{\sum_{i,j} \bar\alpha^i\alpha^j\bra{\psi^i(\vec \theta_i)}H\ket{\psi^j(\vec \theta_j)}}{\sum_{i,j} \bar\alpha^i\alpha^j\braket{\psi^i(\vec \theta_i)|\psi^j(\vec \theta_j)}},
\end{equation} 
and a minimization over the parameter space could lead to the solution. 

We can see that such a hybrid tensor network contains the subspace expansion method as a special case. In particular, suppose we fix the parameters of the quantum tensors $\ket{\psi^i(\vec \theta_i)}$ and denote $\ket{\psi^i(\vec \theta_i)} = \ket{\psi^i(\vec \theta)}$, then we can analytically solve the minimization of $E(\vec x)$ as follows.
Denote 
\begin{equation}\label{SM:crossTerm}
	H^{i,j} = \bra{\psi^i(\vec \theta)}H\ket{\psi^j(\vec \theta)}, \, S^{i,j} = \braket{\psi^i(\vec \theta)|\psi^j(\vec \theta)}.
\end{equation}
Suppose we consider the subspace with $\braket{\psi(\vec x)|\psi(\vec x)}=1$, then it is equivalent to optimize $E(\vec x) = \bra{\psi(\vec x)}H\ket{\psi(\vec x)}$, or the function $E'(\vec x) = E(\vec x)-\lambda\braket{\psi(\vec x)|\psi(\vec x)}$ with a Lagrangian multiplier $\lambda$. Variation of of the new function $E'(\vec x)$ gives 
\begin{equation}
\begin{aligned}
	\delta E'(\vec x)  = \sum_{i,j} (\alpha^j\delta \bar\alpha^i +\bar\alpha^i\delta \alpha^j) H^{i,j} - \lambda\sum_{i,j} (\alpha^j\delta \bar\alpha^i +\bar\alpha^i\delta \alpha^j)S^{i,j},
\end{aligned}
\end{equation}
and a local minimum solution requires $	\delta E'(\vec x)=0$, which is equivalent to
\begin{equation}
	H^{i,j}\alpha^j = \lambda S^{i,j}\alpha^j. 
\end{equation}
Writing the equation in the matrix form, it is equivalent to
\begin{equation}
	H\vec{\alpha} = \lambda S\vec{\alpha},
\end{equation}
which coincides with the subspace expansion method. 

In practice, we can optimize all the parameters in both quantum and classical tensors. We can simultaneously optimize them by treating $E(\vec x)$ as a black box cost function. Alternatively, we can first optimize the parameters of the quantum tensor and then fix them and directly solve the optimal parameters of the classical tensor. Since the parameters are not simultaneously optimized, we may need to repeat the procedure several times until observing energy convergence.

\subsubsection{Virtual qubits via classical tensors}
In addition to extending the power of the quantum circuits, we can also use the classical tensor to represent physical quantum systems, similar to classical tensor network. As shown in Fig.~\ref{fig:ansatzsup}(b), we connect a rank-$(n+1)$ quantum tensor to a rank $d+1$ classical tensor network to represent a system of $n+d$ qubits. Here we assume the classical tensor network consists of low-rank classical tensors and admits efficient contraction, such as the matrix product state (MPS) as defined in Eq.~\eqref{SM:MPS}. In the remainder of the Appendix, we consider MPS as an example of the classical tensor network. Note that the discussion applies to general contractable classical tensor networks, such as tree tensor networks and the multi-scale entanglement renormalization ansatz (MERA). 

Suppose the rank-$(n+1)$ quantum tensor $\psi^{i_1}_{\dots}$ represents a set of $n$-qubit quantum states $\{\ket{\psi^{i_1}}\}$ and the classical tensor is given by
$\alpha^{i_1,j_{n+1},\dots j_{n+d}} = \tr[\alpha_1^{i_1,j_{n+1}}\alpha_2^{j_{n+2}}\dots \alpha_n^{j_{n+d}}]$, then the hybrid tensor of Fig.~\ref{fig:ansatzsup}(b) represents a quantum state
\begin{equation}\label{SM:QplusC}
    \ket{\tilde\psi} = \sum_{i_1,j_1,\dots,j_{n+d}} \alpha^{i_1,j_{n+1},\dots j_{n+d}}\ket{\psi^{i_1}} \ket{j_{n+1}}\dots\ket{j_{n+d}}.
\end{equation}
For any tensor products of local observables $M=O_1\otimes\dots \otimes O_{n+d}$, we have
\begin{equation}
    \braket{\tilde\psi|M|\tilde\psi} = \sum_{i_1,i_1'} \braket{\psi^{i'_1}|O_1\otimes\dots O_{n}|\psi^{i_1}}
    M^{i'_1,i_1},
\end{equation}
with 
\begin{equation}
    M^{i_1',i_1} = \sum_{j'_{n+1},\dots j'_{n+d}, j_{n+1},\dots j_{n+d}} \bar\alpha^{i'_1,j'_{n+1},\dots j'_{n+d}} \alpha^{i_1,j_{n+1},\dots j_{n+d}} \braket{j'_{n+1}|O_{n+1}|j_{n+1}}\dots \braket{j'_{n+d}|O_{n+d}|j_{n+d}}.
\end{equation}
Here each $\braket{\psi^{i'_1}|O_1\otimes\dots O_{n}|\psi^{i_1}}$ is obtained with a quantum computer and each element $M^{i'_1,i_1}$ is obtained by an efficient tensor contraction of the MPS ansatz.
Note that the dimension of $i_1$ can be chosen as a small number similar to how we decide the bond dimension of the MPS ansatz. The definition also holds when the quantum tensor is an $(n+1)$-partite state, where we can assign multiple qubits to the system that the $i_1$ label represents.

\subsubsection{Local quantum correlation and non-local classical correlation}\label{SM:subQC}
We can also use quantum tensors to represent quantum correlations of local subsystems and classical tensors to represent correlations between the subsystems. For example, consider two subsystems $A$ and $B$ with Hamiltonian $H = H_A + H_B + \lambda h_{A}\otimes h_{B}$ and a small coupling constant $\lambda$. We can use the hybrid tensor network in Fig.~\ref{fig:ansatzsup}(c) to represent its ground state,
\begin{equation}\label{SM:QCQ}
	\ket{\tilde\psi}_{AB} = \sum_{i_1,i_2}\alpha^{i_1,i_2}\ket{\psi^{i_1}}_{A}\otimes \ket{\psi^{i_2}}_{B}.
\end{equation}
Here $\ket{\psi^{i_1}}_{A}$ and $\ket{\psi^{i_2}}_{B}$ represent the state of subsystem $A$ and $B$, respectively, and $\alpha^{i,j}$ is the classical tensor representing the correlation between $A$ and $B$. If the quantum correlation is not too strong, we can set the rank of $\alpha_{i,j}$ to be a small number. The average energy of the Hamiltonian is
\begin{equation}
    E = \frac{\braket{\tilde\psi|H|\tilde\psi}_{AB}}{{\braket{\tilde\psi|\tilde\psi}_{AB}}} = \frac{\sum_{i_1,i_2,i_1',i_2'} \bar\alpha^{i_1',i_2'}\alpha^{i_1,i_2} \left(H_A^{i_1',i_1}S_B^{i_2',i_2} + S_A^{i_1',i_1}H_B^{i_2',i_2} + \lambda h_A^{i_1',i_1}h_B^{i_2',i_2}\right)}{\sum_{i_1,i_2,i_1',i_2'} \bar\alpha^{i_1',i_2'}\alpha^{i_1,i_2} S_A^{i_1',i_1}S_B^{i_2',i_2}},
\end{equation}
where the matrices $H_A^{i_1',i_1}$, $S_B^{i_2',i_2}$, $S_A^{i_1',i_1}$, $H_B^{i_2',i_2}$, $h_A^{i_1',i_1}$, $h_B^{i_2',i_2}$ are defined in a general way as in Eq.~\eqref{SM:crossTerm}, that is, $M_{A(B)}^{i,j} = \braket{\psi^i|M|\psi^j}_{A(B)}$. Then we can get the Hamiltonian by measuring the matrices with a quantum computer and contract the classical tensors classically. Suppose each system $A$ and $B$ consists of $n$ qubits so that the total system size is $2n$ qubits. We note that the energy terms can be obtained by only manipulating states of $n$ qubits instead of $2n$ qubits.

In a similar way, we can extend the hybrid tensor network for two subsystems to $k$ subsystems, as shown in Fig.~\ref{fig:ansatzsup}(d). We use the matrix product state $\alpha^{i_{1},i_2\dots i_{k}} = \tr[\alpha_1^{i_1}\alpha_2^{i_{2}}\dots \alpha_k^{i_{k}}]$ as the description of the correlation between subsystems. Suppose each subsystem is represented by quantum states $\{\ket{\psi^{i_s}}_{s}\}$, the hybrid tensor network of Fig.~\ref{fig:ansatzsup}(d) represents a quantum state 
\begin{equation}\label{SM:QC2tree}
	\ket{\tilde\psi} = \sum_{i_1,i_2,\dots i_k}\alpha^{i_1,i_2,\dots i_k}\ket{\psi^{i_1}}_{1}\otimes \ket{\psi^{i_2}}_{2}\dots \ket{\psi^{i_k}}_{k}.
\end{equation}
To measure the expectation value of $M=O_1\otimes\dots \otimes O_{k}$ with each $O_s$ representing local observable on the $s$-th subsystem, we have
\begin{equation}
    \braket{\tilde\psi|M|\tilde\psi} = \sum_{i_1\dots i_k,i_1',\dots i_k'} \bar\alpha^{i_1',\dots i_k'}\alpha^{i_1,\dots i_k}
    M_1^{i_1',i_1}\dots M_k^{i_k',i_k},
\end{equation}
with 
\begin{equation}
    M_s^{i_s',i_s} = \braket{\psi^{i'_s}|O_s|\psi^{i_s}}_s.
\end{equation}
As a result, we can just use an $n$-qubit system to represent a $kn$-qubit system, and the bipartite version corresponds to $k=2$. The dimension of each index $i_1,\dots i_k$ should be a small number similar to the bond dimension of MPS. This is a general form of the hybrid quantum-MPS tensor network, and one can also consider other classical tensor networks, such as MERA. Note that here indices involved in the contraction between the quantum and classical tensors are both classical ones. Alternatively, one can also make a hybrid contraction, where the index of the quantum tensor is a quantum one, as shown in Eq.~\eqref{SM:Q-C2}.

Here the quantum tensors are used to represent the local $n$-qubit correlation, and the classical rank-$k$ tensor is for the global correlation among these $k$ clusters of qubits. Consequently, this kind of quantum-classicaln tensor (ansatz) is suitable for systems where local correlation dominates over global correlation, such as the weak coupling of $k$ qubit chains.

\subsubsection{Non-local quantum correlation and local classical correlation}\label{SM:subCQ}
Instead of using the quantum processor to represent local correlations, one can also consider the classical-quantum two-depth tree structure in Fig.~\ref{fig:ansatzsup}(e), where the classical tensors are used to represent local correlations of each subsystem, and the quantum tensor is used for representing the non-local correlation between the subsystems.

The idea is that, after we apply the quantum circuit to prepare a $k$-qubit state $\ket{\psi}$, we further connect a classical tensor network to each qubit to transform it to $n$ qubits. Suppose we use the MPS for representing each subsystem as $\alpha^{i_{s},j^s_1\dots j^s_{n}} = \tr[\alpha_1^{i_s,j^s_1}\dots \alpha_n^{j^s_{n}}]$, the state corresponding to Fig.~\ref{fig:ansatzsup}(e) is
\begin{equation}\label{SM:CQ2tree}
    \ket{\tilde \psi} = \sum_{\vec i,\vec j^1,\dots\vec j^k} \alpha^{i_{1},\vec j^1}\dots \alpha^{i_{k},\vec j^k} \psi_{i_1,\dots,i_k}\ket{\vec j^1}\otimes\dots\otimes\ket{\vec j^k},
\end{equation}
where we denote $\vec i=(i_1,\dots i_k)$, $\vec j^s=(j^s_1,\dots,j^s_n)$, and $\psi_{i_1,\dots,i_k}=\braket{\vec i|\psi}$. 
To measure $M=O_1\otimes\dots \otimes O_{k}$ with each $O_s$ representing tensor products of local observables, we have
\begin{equation}
    \braket{\tilde \psi|M|\tilde \psi} = \braket{\psi|\tilde O_1\otimes \dots \otimes \tilde O_k|\psi},
\end{equation}
with each observable $\tilde O_s$ obtained by classical tensor contraction as
\begin{equation}
    \tilde O_s^{i_s',i_s} = \sum_{\vec j^{s'},\vec j^s}\tilde\alpha^{i_{s}',\vec j^{s'}}\alpha^{i_{s},\vec j^s} O_s^{\vec j^{s'},\vec j^s}.
\end{equation}
Again we only use $k$ qubits to represent a system of $nk$ qubits.
Note that each subsystem may have a different number of qubits, and we can use multiple qubits to represent each index $i_s$ to increase the bond dimension. 
When $n>>1$, this kind of hybrid tensor network can represent long-range correlation due to the effect of quantum tensor, and it may be applied to an exotic topological state. When $n$ is a small number, it also represents a normalization of local correlations with the classical tensor.

\subsubsection{Local and non-local quantum correlations}\label{SM:subQQ}
In the previous two cases, we use the classical tensor network to represent either local or non-local correlation and the quantum tensor to represent the other part. Here we show how to use the quantum tensor to represent both the local and non-local correlations. Considering a two-depth tree structure of Fig.~\ref{fig:ansatzsup}(f), it represents a state
\begin{equation}\label{SM:QQ2tree}
    \ket{\tilde \psi} = \sum_{i_1,\dots,i_k} \psi_{i_1,\dots,i_k} \ket{\psi_1^{i_1}}\dots\ket{\psi_k^{i_k}},
\end{equation}
where $\psi_{i_1,\dots,i_k}=\bra{i_1}\dots\bra{i_k}\psi\rangle$ denotes the quantum tensor of the correlation between the subsystems and $\{\ket{\psi_s^{i_s}}\}$ denotes the quantum states for each subsystem $s$. Similar to the previous cases, we can measure the expectation values of local observables. To measure $M=O_1\otimes\dots \otimes O_{k}$ with each $O_s$ representing tensor products of local observables on subsystem $s$, we have
\begin{equation}
    \braket{\tilde \psi|M|\tilde \psi} = \braket{\psi|\tilde O_1\otimes \dots \otimes \tilde O_k|\psi},
  \label{eqn:appendixCon}
\end{equation}
with each observable $\tilde O_s$ being
\begin{equation}
    \tilde O_s^{i_s',i_s} = \braket{\psi_s^{i'_s}|O_s|\psi_k^{i_s}}
\end{equation}
obtained via the method discussed in Sec.~\ref{SM:Ob}. Here we represent a system of $nk$ qubits by controlling a quantum device with up to $\max\{n,k\}$ qubits. We can also use multiple qubits for each index $i_s$ to increase the bond dimension. Suppose the quantum states are generated as $\ket{\psi}=U\ket{ 0}^{\otimes k}_0$ and $\ket{\psi^{i_s}_s}=U_s\ket{i_s}\ket{ 0}^{\otimes (n-1)}_s$, the hybrid tensor network of Fig.~\ref{fig:ansatzsup}(f) can be obtained via a quantum circuit 
\begin{equation}
    \ket{\tilde \psi} =U_k \left(  \dots U_2 \left( U_1 \left(U\ket{0}^{\otimes k}_0 \otimes \ket{ 0}^{\otimes (n-1)}_1 \right)\otimes \ket{ 0}^{\otimes (n-1)}_2 \right)\dots \otimes \ket{ 0}^{\otimes (n-1)}_k \right),
\end{equation}
where each $U_s$ applies to the $s$th qubit of the first $k$ qubits with subscript $0$ and the new $n-1$ qubits with subscript $s$. While such a quantum circuit requires to jointly control $nk$ qubits, our hybrid tensor network allows us to represent the same state by controlling up to $\max\{n,k\}$ qubits. 

\begin{figure}[t]\centering
  \includegraphics[width=.95\linewidth]{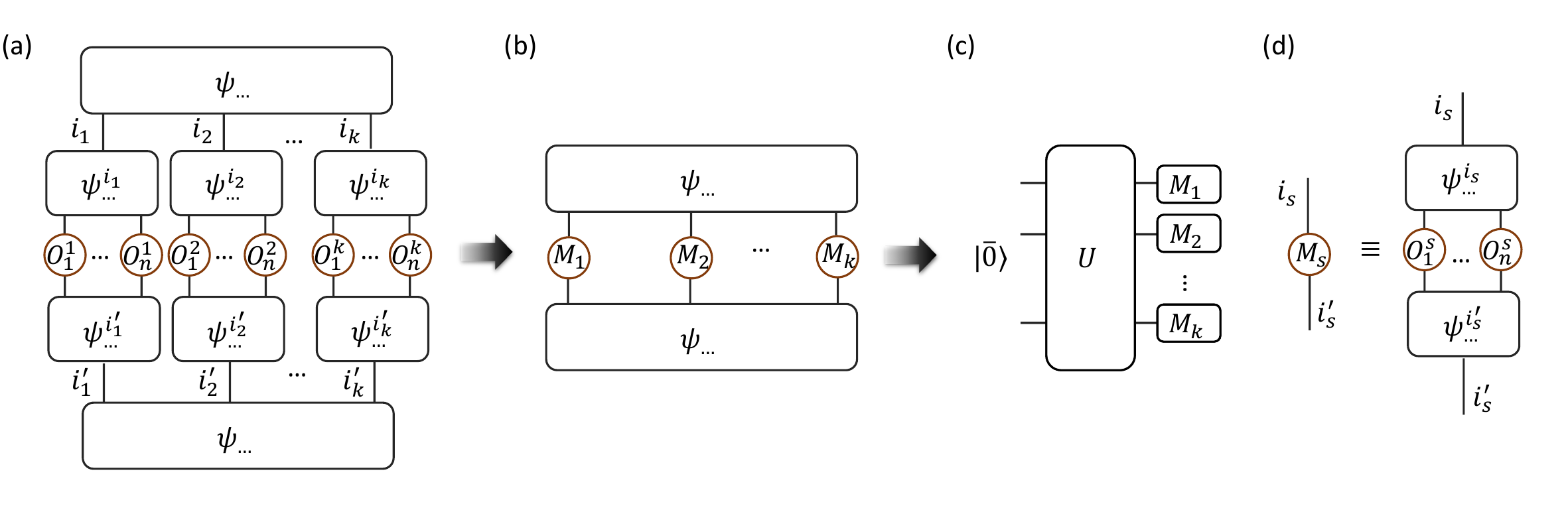}
  \caption{An example for calculating expectation values of a hybrid TTN. Considering a hybrid TTN of Fig.~\ref{fig:ansatzsup}(f), the expectation value of local observables $\otimes_{i=1}^k\otimes_{j=1}^n O_j^i$ corresponds to tensor contraction of (a). We first calculate the observable $M^{i_s',i_s}$ for each tensor on the second layer as (d) with quantum circuits shown in Fig.~\ref{fig:treeexpectation}(b,d). Then tensor contraction of (a) reduces to the contraction of (b), which corresponds to {a quantum circuit representation in} (c) that prepares the state $\ket{\psi}$ with $\ket{\psi}=U\ket{\bar 0}$ and measures the observables $M_1\otimes\dots\otimes M_k$.
  }\label{fig:treeexpectation}
\end{figure}

Note that we use the $2$-layer MPS structure as the example, and other classical tensor networks such as MERA can be similarly used to represent the full quantum state.
In the quantum-quantum network, the contracted indices of the second layer quantum tensors are classical, and we show a quantum circuit representation of this network in Fig.~\ref{fig:treeexpectation}. The expectation values of arbitrary local observables can be efficiently obtained from measurements using quantum circuits.

\subsection{Construction of D-depth tree tensor networks }

{
For quantum systems preserving local renormalization properties, such as $d$-dimensional lattice problem, we can construct a $D$-depth tree tensor network  to represent the many-body quantum state. 
More explicitly, we summarize the procedures as follows. }

\begin{enumerate}
\item Divide the quantum system into  local units  (subsystems) preserving renormalization  properties.
\item Perform a real-space renormalization group transformation to produce a coarse-grained system, where we attach a tensor to connect the original local units. 
\item Repeat procedure $2$ until  the full quantum system is represented.
\item Contract the  $D$-depth tree tensor network using the contraction rules.
\end{enumerate}

The  tensor of upper layer, which connects the subsystems in procedure 2, can be constructed according to the type of interactions between the subsystems. Recall that we can introduce classical index to the quantum state of subsystem by applying  unitary operations as $|\psi^{i}\rangle=U_{\psi^{i}} |\bar{0}\rangle$. 
This indeed forms a representation of the target quantum state in the basis given by $\{U_{\psi^{i}}|\bar{0}\rangle\}$. 
In the above procedure, each node of the tree can be regarded a  coarse-grained quantum state and the degree of connections among all the tree nodes can be bounded by a constant number $t$. We denote $C_j=\max\{n_j,t\}$ with $n_j$ representing maximum number of qubits of the subsystems in the $j$th layer. It is worth noting that even if we construct the tensors of the upper layer (($j$-1)th layer) by considering all operators in the interacting Hamiltonian of the subsystems, the maximum bond dimension of tensors scales as $\mathcal O(\textrm{poly}(n_j))$ and at most $\mathcal O(\log n_j)$ qubits are required to encode the interactions for the subsystem. 
The number of controllable qubits to represent the quantum state with system size $\mathcal{O}(t^{D-1})$ is less than  $\max_{j}(C_j\log n_j)$ qubits, which in practice can be greatly reduced by considering boundary conditions or renormalization properties.

In the main text, we provide an example of $2$-layer tree structure for 1D and 2D systems based on the properties of the full quantum system by using different initial state $\ket{\bar 0}$. We can also specify different unitaries depending on the type of interactions of the target quantum systems. In addition, we may also consider the contracted indices of  quantum tensors are quantum, and we leave the discussion on the transit of quantum information to the upper layer for future work.


{
A recent paper discussed how to use deep variational quantum eigensolver to solve large-scale problems with a small-scale quantum computer in Ref.~\cite{fujii2020deep}. We note that this method could be regarded as a specific scheme within the framework of hybrid tree quantum tensors. Our results could be compared and discussed.}





\subsection{Cost for a general hybrid tree tensor network}
Since conventional tensor network theory is a special case of our hybrid TN framework, contracting an arbitrary hybrid TN is a $\#$P-hard problem. Therefore, we need to consider special networks and here we give a resource estimation for the cost of calculating expectation values of tensor products of local observables on a hybrid tree tensor network. 

\subsubsection{Cost for a tree network}
Starting from a chosen node referred to as root in a tree, we separate other vertices into different layers according to the distance to the root. 
For each node, it can be either a classical tensor network or a quantum tensor. 
In order to efficiently contract the whole tensor network, we only consider classical tensor networks that can be efficiently contracted, for instance, matrix product states (MPS). 
Suppose the tree has a maximal of $D$ layers, and each node has at most $g$ connected, it corresponds to a tree with depth $D$ and maximal degree $g$, and we call a $(D,g)$-tree. The total number of nodes in the tree is upper bounded by $\mathcal O(g^{D})$. 
The $D$-th layer has at most $g^{D-1}$ tensors, and each one is at most a $g$-rank classical tensor network or quantum tensors, with the open index representing at most $g$ qubits. 
Thus a $(D,g)$-tree represents about $\mathcal O(g^D)$ physical qubits.

Now suppose we aim to measure the expectation value of tensor products of local physical observables. 
Without loss of generality, we consider a node with degree $g$ and denote the cost to be $C_q$ or $C_c$ for a quantum or a classical tensor, respectively. In the $i$th layer, denote $n_i^q$ and $n_i^c$ to be the numbers of quantum and classical tensors, respectively, which satisfy $n_i^q+n_i^c\le g^{i-1}$. The cost of contracting the $i$th layer is thus  $C_i= C_q n_i^q +C_c n_i^c\le g^{i-1}(C_q+C_c)$ and the total cost of contracting the whole tree is at most $\mathcal O(g^{D}(C_q+C_c))$ by summing all the layers. 
Note that the number of qubits $N$ represented by the $(D,g)$-hybrid tree tensor network is $N = \mathcal O(g^{D})$, so the cost is also linear to the number of qubits. Here we take the sum of the classical and quantum cost. However, in practice, the cost for quantum tensors is measuring more quantum circuits, while the cost for classical tensors is pure classical computation. We can thus separately use $\mathcal O(NC_q)$ and $\mathcal O(NC_c)$ to be the quantum and classical cost.We summarize the result as follows.

\begin{proposition}
The cost for evaluating the expectation values of local observables  of a $(D,g)$-hybrid tree tensor network is at most  $\mathcal O(g^{D}(C_q+C_c))$ or $\mathcal O(N(C_q+C_c))$. In particular, we need $\mathcal O(NC_q)$ quantum circuits and classical cost $\mathcal O(NC_c)$. Here $N = \mathcal  O(g^{D})$ is the number of qubits represented by the $(D,g)$-hybrid tree tensor network.
\end{proposition}
\noindent  In the following, we further discuss the magnitude of $C_q$ and $C_c$. 

The value of $C_q$ depends on the bond dimension of the index connecting the node to its parent, which quantifies how many measurement settings one needs to contract the quantum tensor to get the effective observable, as illustrated in Fig.~\ref{fig:treeexpectation}~(a). For each observable element, $C_q$ also depends on the number of samples required for suppressing shot noise to a desired accuracy $\varepsilon$. Meanwhile, the value of $C_c$ depends on the choice of the classical tensor network and the bond dimension of the connecting index as well. 
Suppose the bond dimension of each contracted index is upper bounded by $\kappa$, then we have $C_q=\mathcal O(\kappa^2/\varepsilon^2)$ and $C_c=\mathcal O(g\kappa^4)$ for MPS (the cost could be improved with more dedicated contraction methods). Therefore, one has the following detailed contraction cost.
\begin{proposition}
Consider a $(D,g)$-hybrid tree tensor network with quantum tensors and classical MPS tensor networks as the building block with bond dimension at most $\kappa$. The cost for evaluating the expectation values of local observables is 
$\mathcal O(N(\kappa^2/\varepsilon^2 + g\kappa^4))$. In particular, we need to evaluate $\mathcal O(N\kappa^2/\varepsilon^2)$ quantum circuits with $\mathcal O(Ng\kappa^4)$ classical computation cost. 
\end{proposition}

\begin{table}[b]
\begin{tabular}{|c|c|c|c|}
\hline
\hline
Quantum tensor $(C_q)$    & \multicolumn{3}{c|}{Classical tensor $(C_c)$} \\ \hline
\multirow{2}{*}{$\mathcal O(\kappa^2/\varepsilon^2)$} & MPS      & $(D',g')$-Tree    & PEPS      \\ \cline{2-4} 
                  &       $\mathcal O(g\kappa^4)$    &     $\mathcal O(g\kappa^{(g'+2)})$  & 
                  $\mathcal O(g\kappa^8 \tilde \kappa^2)$ (approx.)        \\ \hline
\end{tabular}\caption{Contraction cost for $g$-qubit tensors }
\label{table:cost}
\end{table}

Some remarks are as follows. It looks like that the quantum tensor does not reduce the cost too much compared with the classical tensor, that is, $\kappa^2$ compared with $\kappa^4$. Actually, this is not true. The quantum tensor in our hybrid TTN can express more complicated quantum correlations, as illustrated in Sec.~\ref{SM:EntCo}. If one substitutes this quantum tensor with a classical tensor say MPS, the new bound dimension of the classical tensor denoted by $\kappa'$ could be much larger than $\kappa$, hence leading to a significantly larger cost. This is the major quantum advantage of our hybrid TN framework. Note that here we regard the cost of classical and quantum tensor contractions to be the same and add them together to be the total cost. In practice, classical and quantum computation are independently run on a classical and a quantum processor, so they are totally parallel. If we only focus on the resource cost for the quantum processor, the cost scales as $\mathcal O(N(\kappa^2/\varepsilon^2))$, which is linear to the number of qubits and polynomial to the bound dimension and inverse polynomial to the simulation accuracy. 

Furthermore, one can replace the classical MPS tensors in the hybird TTN with various classical tensors, for instance, the $g$-qubit classical tree tensor, Multi-scale Entanglement Renormalization Ansatz (MERA), and Projected Entangled Pair State (PEPS). For example, one substitutes the MPS block for a $(g',D')$ tree classical tensor \cite{PhysRevA.74.022320}. By contracting the local observable from the deepest layer, one layer by one layer, then one can get the classical cost being $\frac{g'^{D'}-1}{g'-1}(g'+1)\kappa^{(g'+2)}$. Note this tree tensor is on behalf a vertex in the whole hybird TNN, and thus $g=g'^{D'}$. Consequently, the cost for contraction is $\mathcal O(g\kappa^{(g'+2)})$.
For the generalization of MPS to the 2-D, i.e.,  PEPS~\cite{orus2014practical}, there is no efficient way to contract them exactly. There is approximate algorithm running in $\mathcal O(g\kappa^8 \tilde\kappa^2)$ based on the contraction of 1D MPS inside the 2D PEPS, where $\tilde \kappa$ is the bond dimension used to approximately truncate the original tensor.
We summarize the contraction cost for different types of $g$-qubit tensors in the Table~\ref{table:cost}.

\begin{figure*}[t]\centering
  \includegraphics[width=0.8\linewidth]{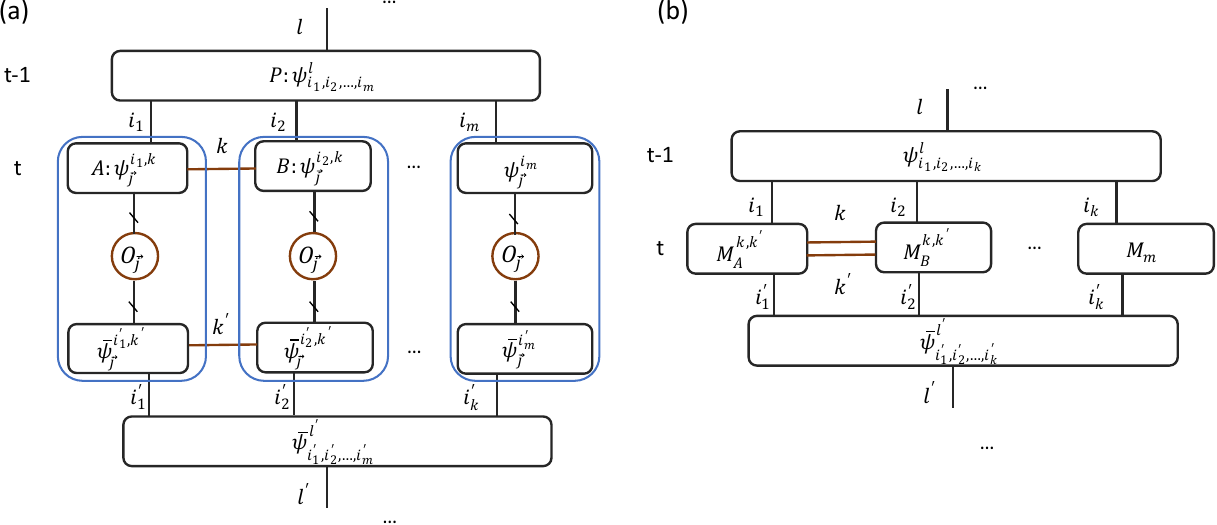}
  \caption{Contraction cost of the loop in the tree hybrid tensor networks. (a) The quantum tensor $A$ and $B$ are in the t-th layer and connected to the quantum tensor $P$ in the (t-1)-th layer by classical index $i_1$ and $i_2$ , and they are also connected to each other by a classical index denoted by $k$ to form a loop. Here $O_{\vec{j}}$ denotes the local observable obtained from the deeper layer, and they are not necessarily same for $A$ and $B$. Contraction of the t-th layer, i.e., the blue circles. Since there are interconnecting indices $k$ and $k'$, one needs $\kappa^4$ measurements to get the operators $M_A^{k,k'}$ and $M_B^{k,k'}$, compared with the previous $\kappa^2$ cost. (b) The effective observables for $(t-1)$-th layer is not a tensor product but in the form $\sum_{k,k'}M_A^{k,k'}\otimes M_B^{k,k'} $, and one needs at most $\kappa^2$ decompositions to measure it.}\label{fig:loopQuantum}
\end{figure*}

\begin{figure*}[t]\centering
  \includegraphics[width=0.7\linewidth]{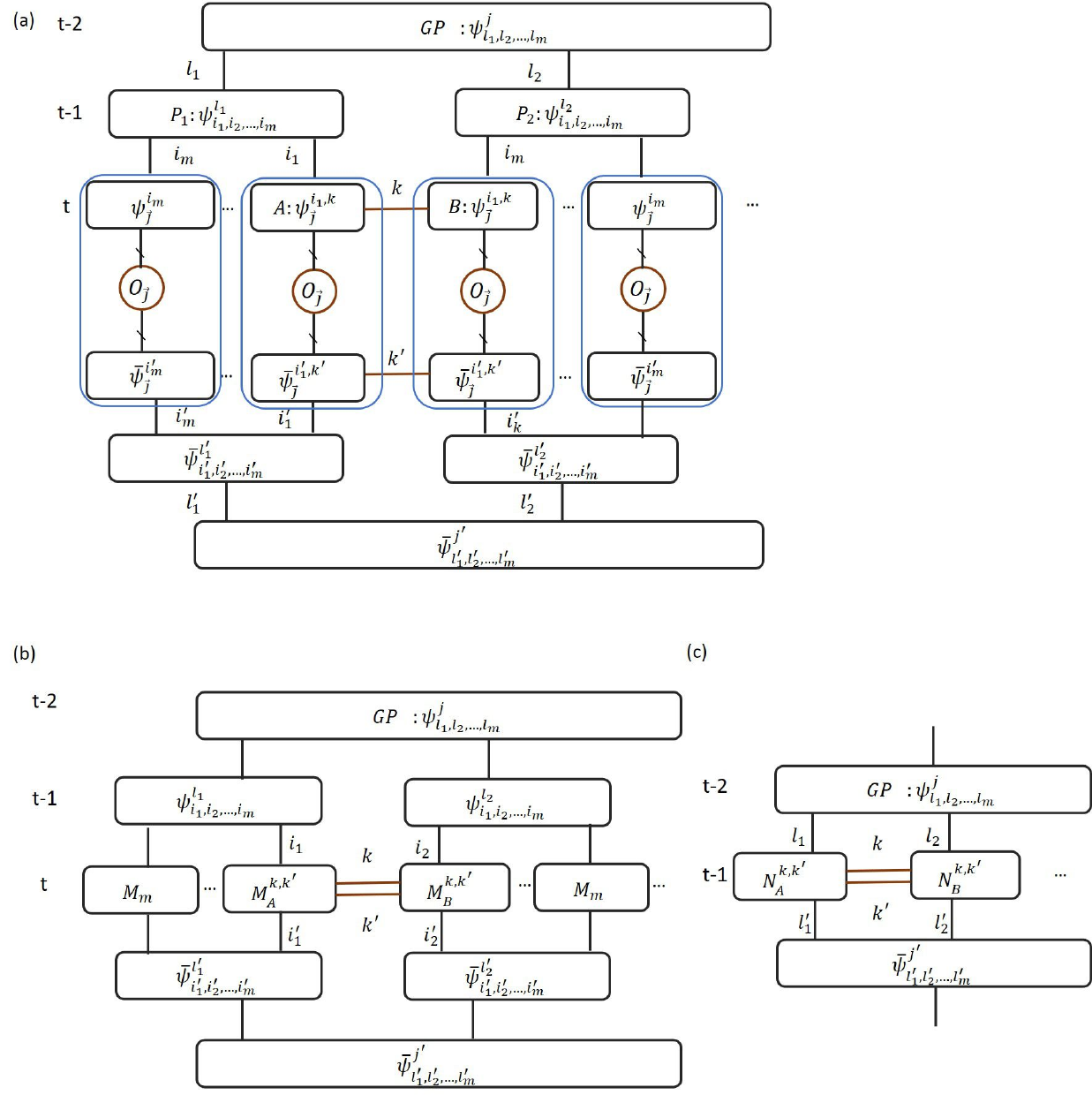}
  \caption{Contraction cost of the loop in the tree hybrid tree network. The tensors $A$ and $B$ are in the t-th layer and have the same grandparent  tensor $GP$ in the (t-2)-th layer. Similar to the case in Fig.\ref{fig:loopQuantum}, we contract the indices in the deeper layer first. Due to the indices $k$ and $k'$, in each step we require  $\kappa^4$ measurements . We contract the effective observables for $t$, $t-1$, and $t-2$ layers in (a), (b), and (c), respectively.} \label{fig:loopQuantum2}
\end{figure*}

\subsubsection{Contraction cost for a tree network with loops}
Next we consider hybrid tensor networks in a generalized tree structure with a small number of loops. 
By adding the edges between the tensors in the same layer, one can introduce a few loops. 
Here we assume these additional edges are classical indices with bond dimension $\kappa$. We calculate the additional cost introduced by adding one loop. We assume that two tensors $A$ and $B$ in the layer $t$ are connected by an additional index denoted by $k$. Depending on the topology, that is, $A$ and $B$ share the same parent tensor $P$ or grandparent tensor $GP$, and also the classical/quantum nature of the these tensors, we discuss the cost as follows. See Fig.~\ref{fig:loopQuantum} and Fig.~\ref{fig:loopQuantum2} for the parent and grandparent cases, respectively.



\begin{itemize}
 \item Same parent: 
Suppose  two tensors $A$ and $B$ in the $t$-th layer are connected to the same tensor $P$ in the $t-1$-th layer and they are also connected with each other by the classical index $k$. We first contract the tensors in the $t$-th layer. If $A$ and $B$ tensors are quantum, due to the interconnecting indices $k,k'$, one needs $C_q=\mathcal O(\kappa^4/\varepsilon^2)$ measurements to obtain the observable for $P$, compared with the previous cost $C_q=\mathcal O(\kappa^2/\varepsilon^2)$. If they are classical, the extra classical cost is $C_c=\mathcal O(\kappa^6)$.  
Then we contract the indices for the parent tensor $P$. 
The link between $A$ and $B$ changes the previous independent measurement on single parties to be joint measurement on two parties. 
One can decompose the measurement as a linear combination of local observables, and the cost comes from this decomposition. With the additional two indices connecting $M_A$ and $M_B$, we have to perform $\kappa^4$ number of local measurements.
Thus the cost here is $C_q=\mathcal O(\kappa^4/\varepsilon^2)$. If the parent tensor $P$ at $t-1$ layer is classical,  we can contract $M_A$ and $M_B$  with $C_c=\mathcal O(\kappa^6)$ extra resource. See Fig.~\ref{fig:loopQuantum} for this procedure.
Thus in this process, one requires $\{C_q=\mathcal O(\kappa^4/\varepsilon^2), C_c=\mathcal O(\kappa^6)\}$ extra cost for the loop. See Fig.~\ref{fig:loopQuantum} for each step.
  \item Same grandparent:
  Suppose that $A$ and $B$ have different parents in the $t-1$-th layer but the same grandparent denoted by $GP$ in the $t-2$-th layer. 
  We first contract the indices in the $t$-th layer and leave the edge connecting $A$ and $B$. Then the additional contraction costs in $t-2$, $t-1$, $t$ layers are   $C_q=\mathcal O(\kappa^4/\varepsilon^2)$ and   $C_c=\mathcal O(\kappa^6)$ for quantum and classical tensors, respectively. See Fig.~\ref{fig:loopQuantum2} for each step. 
 \end{itemize}
From the above analysis, we can see that if there are $L$ number of additional edges that independently acted on different pairs of tensors. The extra resource is 
$\{C_q=\mathcal O(L\kappa^4/\varepsilon^2), C_c=\mathcal O(L\kappa^6)\}$.
However, if the additional edges act on the same tensor, the extra cost could exponentially increase similar to the case of PEPS. In this case, approximate contraction method might be useful to decrease the contraction efficiency and we leave it to a future work.

\begin{figure*}[t]\centering
  \includegraphics[width=0.6\linewidth]{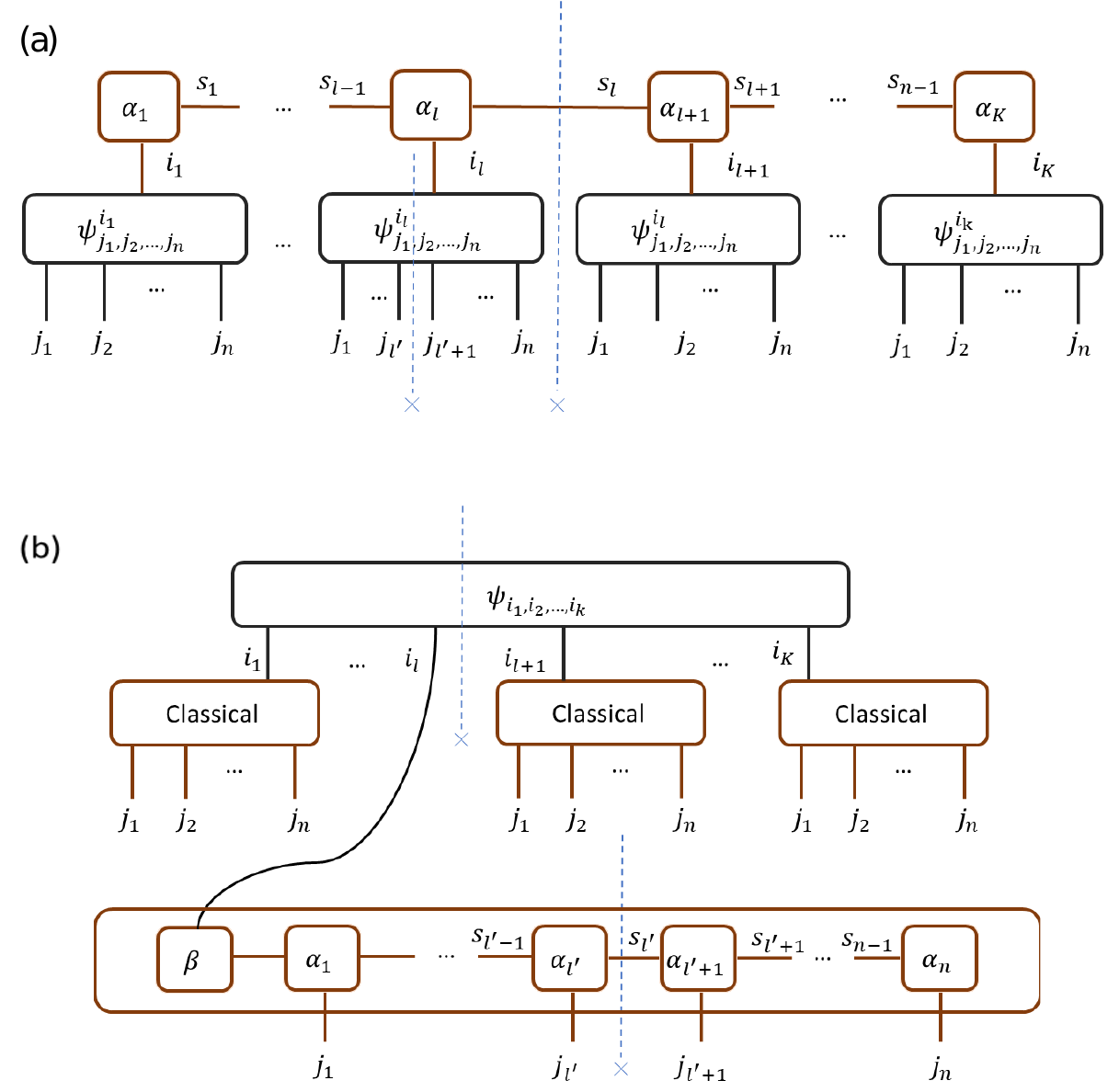}
  \caption{Illustration of the hybrid tensor network with bipartitions: (a) quantum-classical tensor stricture (ansatz) with the classical tensor being a MPS introduced in Sec.~\ref{SM:subQC}; (b) classical-quantum tensor stricture introduced in Sec.~\ref{SM:subCQ}. Here we separate the whole $Kn$-qubit system into two parts, that is, the left and the right subsystems, and the boundary is denoted by the dotted blue line. Without loss of generality, we choose two kinds of boundaries: one is in the bulk of the (classical or quantum) tensor at the first level, where the boundary is between indices $i_l$ and $i_{l+1}$; the other is in the bulk of the tensor at the second level, where the boundary is between indices $j_{l'}$ and $j_{l'+1}$.}\label{fig:FigEntQCCQ}
\end{figure*}

\begin{figure*}[t]\centering
  \includegraphics[width=0.5\linewidth]{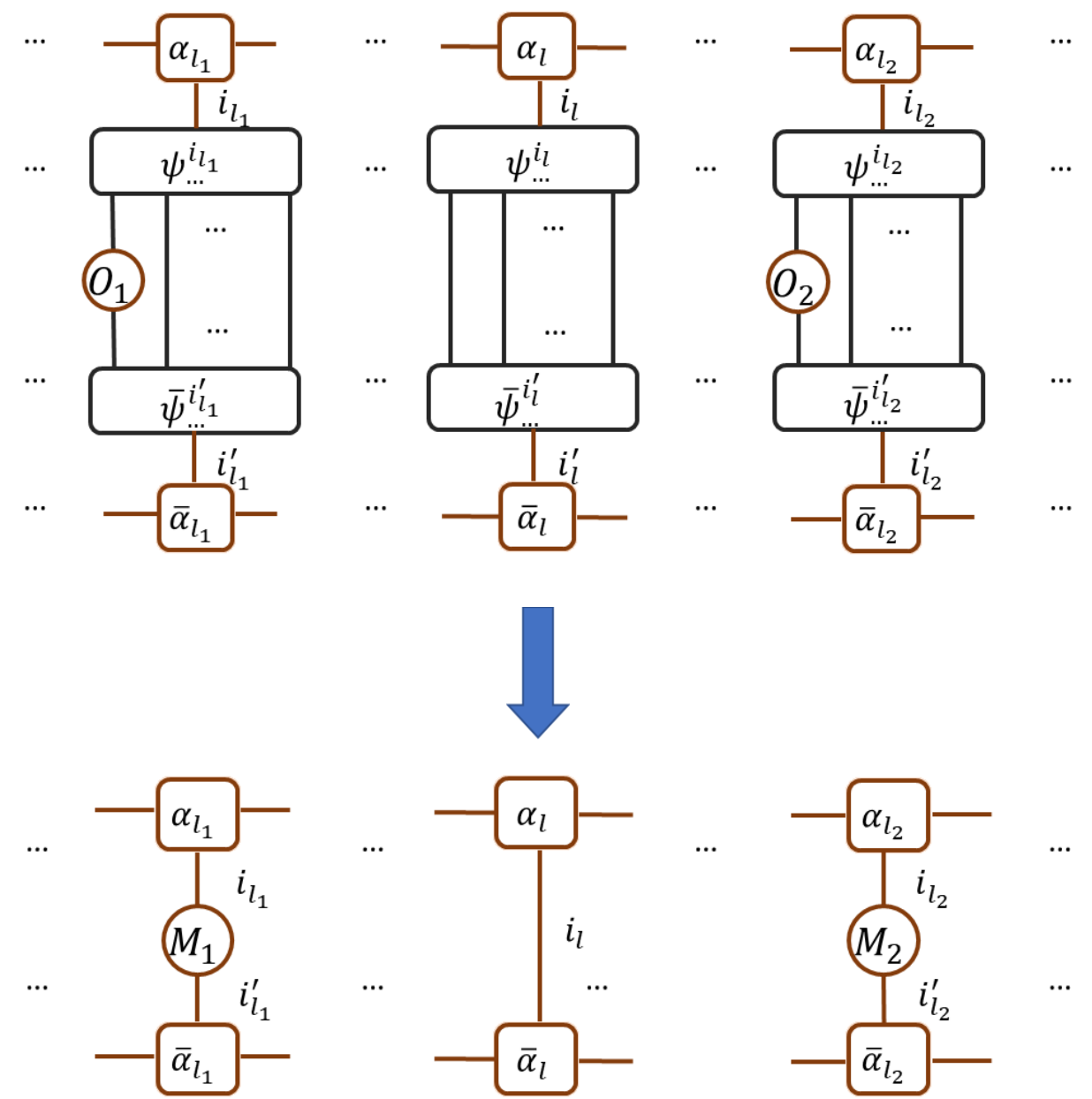}
  \caption{The contraction process to extract the correlation function of the quantum-classical hybrid tensor network introduced in ~\ref{SM:subQC}. Here the classical tensor is an MPS. The two observables $O_1$ and $O_2$ are in two local quantum tensors $\Psi_{\dots}^{i_{l_1}}$ and $\Psi_{\dots}^{i_{l_2}}$, respectively, with $i_{l_1}$ and $i_{l_2}$ are two classical indices. Similar to the extraction of expectation value in Fig.~\ref{fig:FigExpectationsup} (a), We first contract $O_1$ and $O_2$ with the local quantum tensors to get two new observables $M_1$ and $M_2$ for the classical MPS. Here we assume that $\bra{\Psi_{\dots}^{i_l'}}\Psi_{\dots}^{i_l}\rangle=\delta_{\{i_li_l'\}}$ for simplicity, that is, the quantum states indexed by $i_l$ are orthogonal to each other. In this way, the contraction result is the identity for other quantum tensors which do not contain $O_1$ or $O_2$. As a result, the correlation function shows $\langle O_1O_2 \rangle=\langle M_1M_2 \rangle_{\{MPS\}}$, where the second expectation value is on the MPS and shows an exponential decay.}
  \label{figQCcorr}
\end{figure*}

\subsection{Entanglement and correlation of hybrid tensor network}\label{SM:EntCo}

In this section, we discuss the entanglement and correlation properties of the hybrid tree tensor network. We consider a tree structure with two layers and three cases with local or non-local correlations being either classical or quantum. The discussion can be generalized to trees with multiple layers. 

\subsubsection{Local quantum correlation and  non-local classical correlation}
First, we focus on the hybrid tensor network with local quantum correlation and non-local classical correlation introduced in Sec.~\ref{SM:subQC}. Here, we take the classical tensor as MPS for an illustration. As shown in Fig.~\ref{fig:FigEntQCCQ}(a), we separate the whole $kn$ qubits into two subsystems with a blue boundary line and denote the left and the right parts to be $A$ and $\bar{A}$, respectively. 
Without loss of generality, we choose two kinds of boundaries --- one is in the bulk of the classical tensor at the first layer, where the boundary is between indices $i_l$ and $i_{l+1}$ (in the same time between $j_n$ of $\Psi^{i_l}_{\vec{j}}$ and $j_1$ from $\Psi^{i_{l+1}}_{\vec{j}}$); the other is in the bulk of the quantum tensor at the second layer, where the boundary is between indices $j_{l'}$ and $j_{l'+1}$.

For the first kind of bipartition between $i_l$ and $i_{l+1}$, the entanglement of the subsystem is upper bounded by the bond dimension of the index $s_l$ of the MPS, which is normally a constant number that is independent of the subsystem size. As a result, the correlation between these $n$-qubit clusters is weak. The second kind of bipartition between $j_{l'}$ and $j_{l'+1}$ is inside the quantum tensor $\Psi_{\vec{j}}^{i_1}$, which is represented by a general quantum state. Thus the entanglement entropy can be in principle proportional to $\min\{l',n-l'\}$, which is proportional to the subsystem size of the quantum tensor. 

The previous analyses on entanglement can also be revealed by the correlation functions. If we select two local observables $O_1$ and $O_2$ inside the bulk of any local quantum tensors, the correlation function $\langle O_1O_2 \rangle$ could be even a constant. However, if they are located in different local quantum tensors, the correlation suffers a exponential decay $\langle O_1O_2 \rangle\sim \exp(-a|l_2-l_1|)$, where $a$ is some constant depending on the chosen MPS and $|l_2-l_1|$ labels the distance between the two quantum tensors where $O_1$ and $O_2$ are inside. As shown in Fig.~\ref{figQCcorr}, this scaling can be obtained by first contracting the local observables $O_1$ and $O_2$ with the quantum tensors, and then the total result becomes two new observables say $M_1$ and $M_2$ in the MPS tensor network.

\subsubsection{Local classical correlation and non-local quantum correlation}

Next, we study the hybrid tensor network introduced in Sec.~\ref{SM:subCQ}, where one has local classical correlation and non-local quantum correlation. As shown in Fig.~\ref{fig:FigEntQCCQ} (b), the whole $kn$-qubit is separated into two parts by the blue boundary. We also choose two kinds of boundaries --- one is in the bulk of the quantum tensor between indices $i_l$ and $i_{l+1}$; the other is in the bulk of the classical tensor between indices $j_{l'}$ and $j_{l'+1}$.

For the bipartition between $i_l$ and $i_{l+1}$ inside the global quantum tensor $\psi_{i_1,i_2,\cdots i_k}$, the entanglement can be proportional to $\min\{l,k-l\}$, with the the subsystem size being $|A|=nl$ and $|\bar{A}|=n(k-l)$. As a result, when $k \gg 1$ and $n=\mathcal O(1)$, the hybrid tensor network can have a volume law entanglement scaling due to the contribution from the quantum tensor. For the regime $n\sim k$, the entanglement is weaker but still stronger than the area law. From this point of view, the hybrid tree tensor network could represent more complicated entanglement than pure classical tensor networks. For the second kind of bipartition between $j_{l'}$ and $j_{l'+1}$, the entanglement of the subsystem is upper bounded by the bond dimension of the index $s_{l'}$ of the local MPS, which is normally a constant number independent of the subsystem dimension. As a result, the correlations inside these $n$-qubit clusters are weak. 

Similarly, the previous analyses on entanglement can be revealed by the correlation function. If two local observables $O_1$ and $O_2$ are inside the bulk of any local classical tensor, the correlation function $\langle O_1O_2 \rangle$ decays exponentially. However, if they are located in different local classical tensors, the correlation could be some constant or decay polynomially. One can first contract the local observables $O_1$ and $O_2$ with the classical tensors, and then the total result becomes two new observables for the general quantum tensor.

\subsubsection{Local quantum correlation and non-local quantum correlation}

At last, we consider the hybrid tensor network introduced in Sec.~\ref{SM:subQQ}, where both local and non-local correlations are represented by quantum tensors. As a result, it can possess a strong correlation both for local and global correlation. 
The entanglement entropy and correlation function can be analyzed in the same way above.

For the general hybrid tree structure, the analyses are also similar. For the entanglement, one just needs to check the boundary is in the bulk of quantum or classical tensor. For the correlation function, one can obtain its behavior by iterative contractions and check whether the final new observables are in a quantum or classical tensor.

\section{Numerical simulation}

In this section we discuss the implementation and numerical simulation of hybrid TTN in details.  
We discuss  how to implement it with quantum-quantum tensor networks in details. We first review the variational quantum simulation algorithm of imaginary time evolution. Then, we show the simulation results for finding the ground state of 1D and 2D spin systems.

\subsection{Variational quantum algorithms} 
 
We consider Hamiltonian $H=\sum_i \lambda_i h_i$ with coefficients $\lambda_i$ and tensor products of Pauli matrices $h_i$. 
The normalized state at imaginary time $\tau$ is $
|\psi(\tau)\rangle=\frac{e^{-H \tau}|\psi(0)\rangle}{\sqrt{\left\langle\psi(0)\left|e^{-2 H \tau}\right| \psi(0)\right\rangle}}
$ and  
the Wick-rotated Schr\"{o}dinger equation is 
\begin{equation}
\frac{d|\psi(\tau)\rangle}{d \tau}=-\left(H-E_{\tau}\right)|\psi(\tau)\rangle
\end{equation}
where $
E_{\tau_{ }}=\langle\psi(\tau)|H| \psi(\tau)\rangle
$ is the expected energy at imaginary time $\tau$. 
The ground state can be determined from the long time limit of the Wick-rotated Schr\"{o}dinger equation $\ket{\psi}_{\rm GS}=\lim_{\tau\rightarrow \infty }\ket{\psi(\tau)}$. Consider a normalized trial state $\ket{\psi(\vec{\theta}(\tau))}$ with real parameters $\vec \theta$ representing all the parameters,  the imaginary time evolution of the Schr\"{o}dinger  equation on the trial state space is 
\begin{equation}
\sum_{i} \frac{\partial|\psi(\vec{\theta}(\tau))\rangle}{\partial \theta_{i}} \dot{\theta_{i}}=-\left(H-E_{\tau}\right)|\psi(\vec{\theta}(\tau))\rangle.
\end{equation}
Applying the McLachlan’s variational principle, which minimizes the distance between the evolution of trial state $\frac{\partial|\psi(\vec{\theta}(\tau))\rangle}{\partial \tau}$ and $-(H-E_\tau)\ket{\psi(\vec{\theta}(\tau))\rangle}$, we have
\begin{equation}
\delta \|\left(d / d \tau+H-E_{\tau}\right)|\psi(\vec{\theta}(\tau))\rangle \|=0,
\end{equation}
which determines the evolution of the parameters 
\begin{equation}
\sum_{j} A_{i, j} \dot{\theta}_{j}=-C_{i},
\end{equation}
with the matrix elements of $A$ and $C$  given by
\begin{equation}
\begin{aligned} 
A_{i, j} =\Re \left( \frac{\partial\langle\psi(\vec{\theta}(\tau))|}{\partial \theta_{i}} \frac{\partial|\psi(\vec{\theta}(\tau))\rangle}{\partial \theta_{j}} \right),~ C_{i} =\Re \left( \frac{\partial\langle\psi(\vec{\theta}(\tau))|}{\partial \theta_{i}} H|\psi(\vec{\theta}(\tau))\rangle \right).
\end{aligned}
\end{equation}
Therefore, we can effectively simulate imaginary time evolution by tracking the evolution of the parameters.

To variationally simulate imaginary time evolution, we have to be able to calculate $A$ and $C$ for any given parameters.
When $\ket{\psi(\vec \theta)}$ is directly prepared by a quantum circuit, we can obtain the matrix elements by a modified quantum circuit by introducing an ancillary qubit~\cite{mcardle2019variational,yuan2019theory}. When we consider trial states represented by a hybrid tensor network, we can calculate $A$ and $C$ by making use of a similar method for calculating the expectation values of hybrid tensor networks. The main idea is to generalize the circuit to implement the contraction of two quantum tensors. Related work can be found in Refs.~\cite{mcardle2019variational,yuan2019theory,endo2020variational}. We leave the circuit implementation of the matrix elements and the application of the hybrid tensor network in simulating real-time dynamics to future work.

In this work, we calculate the matrix elements by the finite difference method. For example, to calculate each $A_{i,j}$, we can approximate it as
\begin{equation}
\begin{aligned}
    A_{i, j} &=\Re \left (\frac{(\langle\psi(\vec{\theta}+\delta\theta_i)|-\langle\psi(\vec{\theta})|)}{\delta \theta_{i}} \frac{(|\psi(\vec{\theta}+\delta\theta_j)\rangle-\partial|\psi(\vec{\theta}))\rangle}{\delta \theta_{j}} \right),\\
    &=\frac{1}{\delta\theta_i\delta\theta_j} \Re\left(\langle\psi(\vec{\theta}+\delta\theta_i)|\psi(\vec{\theta}+\delta\theta_j)\rangle - \langle\psi(\vec{\theta})|\psi(\vec{\theta}+\delta\theta_j)\rangle - \langle\psi(\vec{\theta}+\delta\theta_i)|\psi(\vec{\theta})\rangle + \langle\psi(\vec{\theta})|\psi(\vec{\theta})\rangle \right).
\end{aligned}
\end{equation}
The last terms correspond to the normalization of the hybrid tensor, which is 1 for the hybrid TTN considered in the simulation. The second two terms are overlap with two different hybrid tensor networks. Again, calculating such overlaps requires quantum circuits similar to calculating expectation values. In our simulation, for simplicity, we obtain the overlaps by directly contracting the quantum tensors by summing over the state vector array. For each $C_i$ element, it can be obtained via the difference of the energy gradient,
\begin{equation}
    C_i = \frac{\braket{\psi(\vec\theta+\delta\theta_i)|H|\psi(\vec\theta+\delta\theta_i)} -\braket{\psi(\vec\theta)|H|\psi(\vec\theta)}}{2\delta \theta_i}.
\end{equation}
Therefore, the $C$ vector can be obtained from the finite difference of energy changes. 
We use the hybrid TTN to represent the trail quantum state as
\begin{equation}
    \ket{\tilde \psi(\vec \theta)} = \sum\limits_{{i_1} \ldots {i_{k}}} {{\alpha _{{i_1}, \ldots ,{i_{k}}}}}(\vec\theta_0) \ket{{\psi _1^{{i_1}}}(\vec\theta_1) }  \otimes  \cdots  \otimes \ket{ {\psi_{k}^{{i_{k}}}}(\vec\theta_k)},
\end{equation}
with $\vec\theta=(\vec\theta_0,\vec\theta_1,\dots,\vec\theta_k)$ representing all the parameters of the tree.  We can optimize the total energy by either minimizing all the parameters as $\min_{\theta_0,\theta_1,...,\theta_k} \braket { \tilde \psi(\vec\theta)|H|\tilde\psi(\vec\theta)}$ or minimizing local subsystem of each layer as $\min_i\min_{\{\theta_i\} } \braket { \tilde \psi(\vec\theta)|H|\tilde\psi(\vec\theta)}$.

Here, we exemplify our hybrid TTN  using a quantum-quantum tensor networks, which can be implemented on a near-term quantum computer using variational quantum algorithms. We note that other hybrid TTN structures with classical tensors can be implemented in a similar way. 

\subsection{Simulation details}



In the main text, we consider the Hamiltonians for 1D and 2D spin systems that admit a general form 
\begin{equation}
   H = \sum\limits_{j = 1}^{k} {{H_j}}  + \lambda {H_{{\rm{int}}}}, 
\end{equation}
where $H_j$ and $H_{{\rm{int}}}$  represents the local Hamiltonian of the $j$th subsystem and their interactions, with interaction strength of subsystems $\lambda$. The topology of the spin systems can be found in Fig.~3 in the main text. 
The interactions between subsystems $\{f_j\}$ or $\{f_{j,i}\}$ are generated randomly from $[0,1]$. In this work, we generated a sequence of random numbers and fixed the strength for comparison. The lattice model considered in the main text, which has local interactions in the subsystems and random interactions between subsystems have been investigated to describe the phenomena in the high energy physics. For example, the generalized Sachdev-Ye-Kitaev model with random nearest-neighbor couplings in the large $N$ limit preserves local criticality in two-point functions, zero temperature entropy, and some unique properties such as diffusive energy transport and propagation of chaos in space, which agrees with the holographic calculation on incoherent black hole~\cite{Gu:2016oyy}. More discussions can be found in the Sec~\ref{SM:example_high_energy}.

To benchmark the simulation accuracy of our method, we compare with the results obtained from conventional tensor networks.
For the 1D case, we compare $E$ to the reference result $E_0=E_{\rm MPS}$ obtained from a standard DMRG implementation with the bond dimension $\kappa= 32$. For the 2D case, we represent the full quantum state using projected entangled-pair state (PEPS). We start from a random tensor product state, and use a standard imaginary
time evolution scheme to find the ground state of the 2D Hamiltonian. To reduce the computational cost, we use the local update method, i.e. the so-called 'simple update' method, and we set the bond dimension $\kappa = 5$ and the  maximum allowed bouncary bond dimension  that approximates the original tensor during
the contraction $\tilde\kappa = 64$. 

\begin{figure*}[htb]\centering \includegraphics[width=1.0\linewidth] {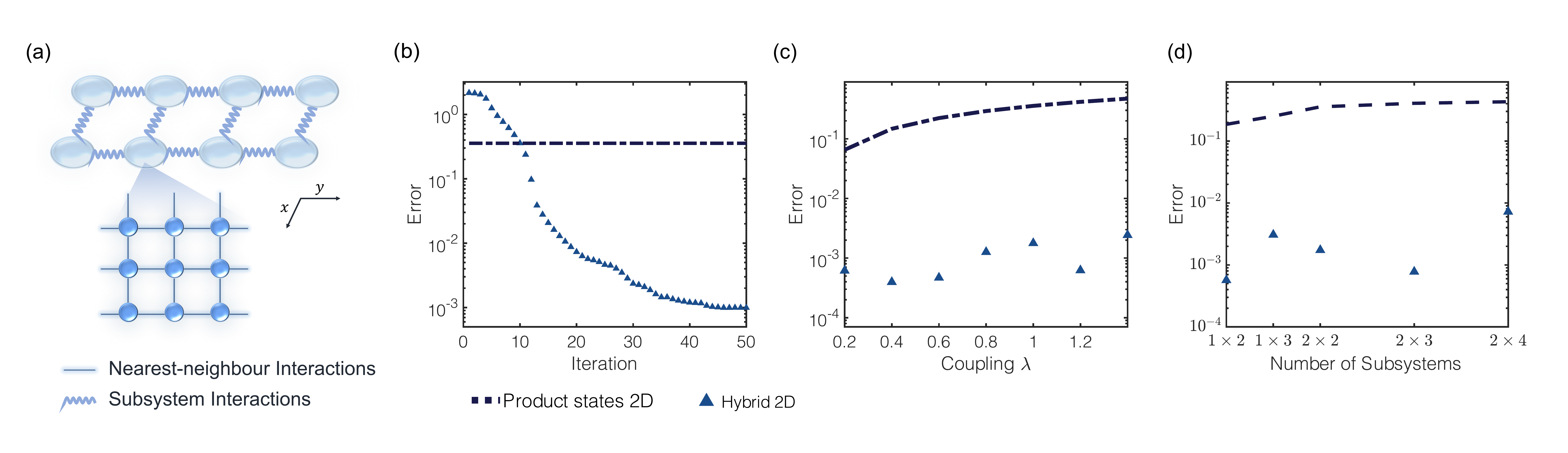}
  \caption{
   Numerical simulation for 2D spin systems on a square lattice using hybrid TTN. 
  (a) Sketch for the 2D spin lattice with nearest-neighbor interactions on the boundary. The interactions of subsystems are represented by thick lines. We group $3 \times 3$  qubits on a square sublattice as subsystems. 
  (b)-(d) Simulation results of the ground state energy comparing to the results from PEPS $E_{\rm PEPS}$. We use the relative error $1-E/E_{\rm PEPS}$ to characterize the calculation accuracy.
The  blue dash-dotted line correspond to the energy when using the tensor products of the ground state of local subsystems. The blue triangle are results obtained with hybrid TNs. 
(b) Convergence towards the ground state for the 2D $9\times 4$ systems with $\lambda=1$.
(c) Error versus different  coupling strength of subsystems $\lambda$ for $9\times 4$ systems. 
(d) Errors with different numbers  of local subsystems  with $9\times k$ qubits and and $\lambda=1$. We consider $k = N_x \times N_y$ for the 2D system.}
  \label{fig:numer_SM}
\end{figure*}

To further test the validity of our method, we numerically test the spin lattice with uniform nearest-neighbor interactions on the boundary.
We consider the uniform coupling regime by setting the interaction strength as a constant $f_{j,i} = \lambda$. The parameters of the local interactions and  external fields are set the same as those in the main text, i.e., $f = 1$, $h =1/\pi= 0.32$ and $g = 0.5$. This model provides a natural partitioning strategy using our method. We show the lattice model and the partitioning strategy in Fig.~\ref{fig:numer_SM}(a). We consider the same quantum circuits as in the main text for identification.
In Fig.~\ref{fig:numer_SM}(b) we study the convergence of ground state energy  both with coupling strength $\lambda=1$ on $9\times 4$ qubits.
In Fig.~\ref{fig:numer_SM}(c, d), we study how the coupling strength or the number of subsystems affect the efficacy of hybrid TTN, respectively.
From the simulation results, we demonstrate that we can decrease the error to a relatively low level, which indicates the effectiveness of our method in a proof-of-principle way. In practice, we can use different optimization method and circuit to further decrease the errors. We may also  simulate other models to explore interesting physics behind these models. Examples can be found in the next section, and we leave it to dedicated readers.  

\section{Applications}
Hybrid tensor networks may have wide applications in quantum computing and quantum simulation for solving different physics problems. The key benefit of a hybrid tensor network is to more efficiently represent a multipartite quantum state so that the required quantum resource is significantly reduced with the help of classical computers. 
The hybrid tensor network could extend the power of near-term quantum computers so that the limitation on the number of controllable qubits and the circuit depth  could be greatly alleviated. As such, the hybrid TTN approach might be useful to resolve the barren plateau issues on NISQ era quantum computing  \cite{wang2020noise,cerezo2020impact}, and the discussions could be an interesting future direction. Meanwhile, hybrid tensor networks may find their applications in fault-tolerant quantum computing as well, where the number of logical qubits could also be limited owing to the huge overhead for error correction. 
In this section, we discuss potential applications of the hybrid tensor network in practical problems in chemistry, condensed matter physics, quantum field theory, and quantum gravity thought experiments.

\subsection{Chemistry}

The most promising application of the hybrid TN is for clustered subsystems with weak subsystem-wise interactions. Consider that the whole quantum system is divided into several subsystems, where particles in the same subsystem have strongly interacted, and the particles from different subsystems have weakly interacted. Since each subsystem has strong interaction, the whole system, in general, could be classically hard to solve. On the contrary, our hybrid TN only uses the classical TN to represent the cluster-wise interactions and uses the quantum computer to represent the strongly interacted subsystems, hence surpassing the classical power TNs meanwhile using a small quantum computer. Note that here each subsystem does not need to be 1D or 2D, but systems with general interaction topology structures.

\begin{figure}[t]\centering \includegraphics[width=0.5\linewidth] {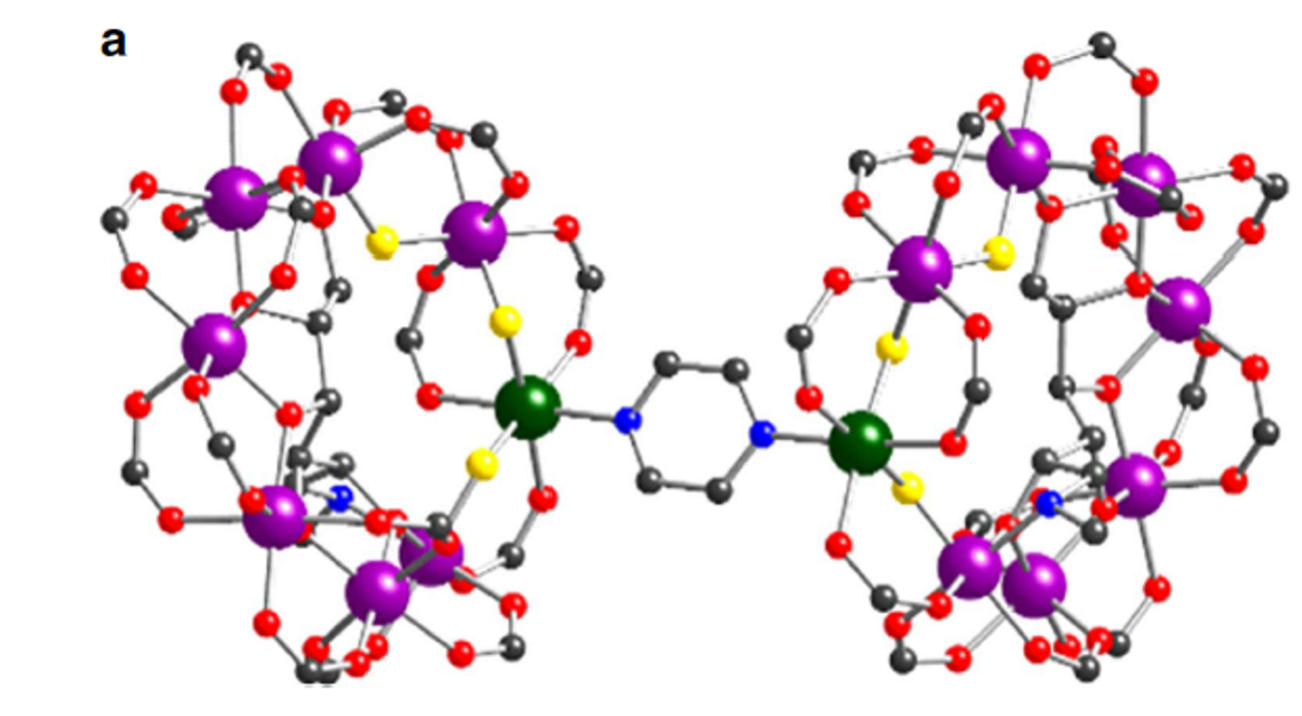}
\caption{
  Molecular structure of the molecular dimer, consisting of two purple-(Cr$_7$Ni) antiferromagnetic rings. The two Cr$_7$Ni rings are linked
through a pyrazine unit, which provides two N-donor atoms
binding to Ni centers in different rings. This leads to a weak
exchange coupling between the Ni ions. The figure is adapted from Ref.~\cite{garlatti2017portraying}. }
  \label{fig:dimer}
\end{figure}

The physical systems in the real world admit the interaction form considered in our framework.
Several chemical molecules can be discribed by the interacting Hamiltonian that has a  similar form as that in our simulation, providing a natural application of our method.
The examples include the molecular rings, such as (Cr$_7$Ni)$_2$ dimers, consisting of two purple-(Cr$_7$Ni) antiferromagnetic rings. The two Cr$_7$Ni rings are linked
through a pyrazine unit as shown in Fig.~\ref{fig:dimer}, which provides two donor atoms
binding to Ni centers in the nearest-neighbor rings. This leads to a weak exchange coupling between the Ni ions~\cite{garlatti2017portraying,timco2009engineering}.
in which the subsystem has strong correlations, and the two subsystems are weakly interacted by the boundary spins.

Aside from the physical systems featuring these cluster properties, we can use the hybrid TNs to represent different degrees of freedom. 
Here, we focus on the application in chemistry for solving the molecular vibronic spectra. 
The vibrational and electronic structure of a \rm molecule generally assumes the Born-Oppenheimer approximation by treating the electrons and nuclei separately. Here, we show how to go beyond the Born-Oppenheimer with the hybrid tensor network method. Consider the \rm molecular Hamiltonian in atomic units as
\begin{equation}
\begin{aligned}
	\Hamilt_{\rm mol} =& -\sum_i\frac{\nabla^2_i}{2}  -\sum_I\frac{\nabla^2_I}{2M_I} - \sum_{i,I}\frac{Z_I}{|\mathbf{r}_i-\mathbf{R}_I|}+\frac{1}{2}\sum_{i\neq j}\frac{1}{|\mathbf{r}_i-\mathbf{r}_j|}+\frac{1}{2}\sum_{I\neq J}\frac{Z_IZ_J}{|\mathbf{R}_I-\mathbf{R}_J|}.
\end{aligned}
\end{equation}
with $M_I$, $\mathbf{R}_I$, and $Z_I$ being the mass, position, and charge of nuclei $I$, respectively, and $\mathbf{r}_i$ being the position of electron $i$. Given the location of the nucleus, the electronic Hamiltonian is
\begin{equation}
\begin{aligned}
	\Hamilt_e(\mathbf{R}_I) &=  -\sum_i\frac{\nabla^2_i}{2} + \sum_{i,I}\frac{Z_I}{|\mathbf{r}_i-\mathbf{R}_I|}+\frac{1}{2}\sum_{i\neq j}\frac{1}{|\mathbf{r}_i-\mathbf{r}_j|},
\end{aligned}
\end{equation}
and the total Hamiltonian can be represented as
\begin{equation}
\Hamilt_{\rm mol} =  -\sum_I\frac{\nabla^2_I}{2M_I} +\frac{1}{2}\sum_{I\neq J}\frac{Z_IZ_J}{|\mathbf{R}_I-\mathbf{R}_J|} + \Hamilt_e(\mathbf{R}_I).
\end{equation}
Under the Born-Oppenheimer approximation, we assume the electrons and nuclei are in a product state, 
\begin{equation}
	\ket{\psi} = \ket{\psi}_n\ket{\psi}_e,
\end{equation}
and the ground state energy under the Born-Oppenheimer approximation is solved by
\begin{equation}
E_0 = \min_{\ket{\psi}_n}\min_{\ket{\psi}_e} \bra{\psi}_n\bra{\psi}_e \Hamilt_{\rm mol} \ket{\psi}_n\ket{\psi}_e.
\end{equation}
Because only the electronic Hamiltonian $\Hamilt_e(\mathbf{R}_I)$ depends on electronic state $\ket{\psi}_e$, the minimisation over the electronic state $\ket{\psi}_e$ is equivalent to finding the ground state of the electronic Hamiltonian $\Hamilt_e(\mathbf{R}_I)$. Suppose we solve the electronic structure for any $\Hamilt_e(\mathbf{R}_I)$  by finding 
\begin{equation}
V_{0}^e(\mathbf{R}_I) = \min_{\ket{\psi}_e} \bra{\psi}_e \Hamilt_e(\mathbf{R}_I) \ket{\psi}_e, 
\end{equation}
then the ground state of $\Hamilt_{\rm mol}$ can be found by solving the ground state of $H_0$,
\begin{equation}
\Hamilt_0 = -\sum_I\frac{\nabla^2_I}{2M_I} +\frac{1}{2}\sum_{I\neq J}\frac{Z_IZ_J}{|\mathbf{R}_I-\mathbf{R}_J|} + V_{0}^e(\mathbf{R}_I).
\end{equation}
The Born-Oppenheimer approximation enables us to solve the \rm molecular Hamiltonian by separately solving the electronic Hamiltonian and the nuclei Hamiltonian. We thus only need to operate a quantum system either for the electronic Hamiltonian or the nuclei Hamiltonian. 

The conventional approach to go beyond the Born-Oppenheimer approximation is to consider the electrons and nuclei together as a whole system and directly solve the Hamiltonian $H_{\rm mol}$. However, this requires to store the joint entangled state of electrons and nuclei, making it harder to simulate with near-term quantum computers. Since the nuclei is much heavier than the electrons, even though the Born-Oppenheimer approximation breaks, the entanglement between electrons and nuclei may still be small. Therefore, we can use the hybrid tree tensor network to represent the whole state. Suppose the tensor for the electrons and nuclei are  $\{\ket{\psi^i_e(\vec\theta_e)}\}$ and $\{\ket{\phi^i_n(\vec\theta_n)}\}$, respectively. Then a hybrid tensor network representation of the joint state is 
\begin{equation}
    \ket{\tilde\psi} = \sum_i \alpha_i \ket{\psi^i_e(\vec\theta_e)}\ket{\phi^i_n(\vec\theta_n)},
\end{equation}
and we use it to represent the ground state of the molecule by only controlling states of either the electrons or the nuclei. We can also apply the hybrid tensor network for representing the electrons or the nuclei to further reduce the size of the quantum system we need to control.


Our method could also be applied for representing virtual qubits.
In ~Ref.~\cite{PhysRevX.10.011004}, the authors considered to prepare the quantum state in the active space as  a reference state $\ket{\psi_{\textrm{Ref}}}$. They choose a set of expansion operators $\{O_i\}$, which act on this reference, to describe the excitations in a virtual space as $O_i\ket{\psi_{\textrm{Ref}}}$. This form a representation of the excitation operator in the basis given by $ \{O_i\ket{\psi_{\textrm{Ref}}}\}$.
The ground state and low-lying eigenstates can be obtained by solving the generalized eigenvalue problem in the well conditioned subspace as
\begin{equation}
    HC = SCE
\end{equation}
with the matrix elements of $H$ and $S$ given by 
\begin{equation}
    H^{i j}=\langle\psi_{\mathrm{ref}}|O_{i}^{\dagger} H O_{j}| \psi_{\mathrm{ref}}\rangle ~~ S^{i j}=\langle\psi_{\text {ref }}|O_{i}^{\dagger} O_{j}| \psi_{\text {ref }}\rangle.
\end{equation}
In our framework, we show analytically in Sec. II A1 that the subspace expansion method is one special case of our method.
More specifically, we could choose the quantum tensor as $\ket{\psi_i} = U_i \ket{\psi_{\textrm{Ref}}}$, where $U_i$ could be prepared by a quantum circuit, and it reduces to the original method when  $U_i$  is selected as the single and double excitation operators.
We can add a classical tensor to further increase the representation capability in the chemistry problems. As the subspace expansion method could potentially improve the accuracy of the ground state and provide approximations to excited states, we expect our method applicable to these problems.

\subsection{Condensed matter physics}

Many interesting quantum phenomena could be captured by the model of weakly- or medium-coupled subsystems, as the models we considered in Fig.~3(a) in the main context. However, the interactions of local subsystems could generate complex multipartite entanglement and lead to effective quasi-particle transportation between the quantum systems, making it hard to simulate classically. 
In this section, we discuss the use of variational quantum simulation and the hybrid tensor network approach to search for Majorana zero-modes and the topological phase transition in correlated materials.

We first consider a spinless one-dimensional tight-binding spin chain representation of a $p$-wave superconductor introduced by Kitaev~\cite{kitaev2001unpaired},
\begin{equation}
    {H}=-\sum_{i=1}^{N-1}\left(t c_{i}^{\dagger} c_{i+1}+\Delta c_{i} c_{i+1}+h . c .\right)-\mu \sum_{i=1}^{N} n_{i},
\end{equation}
where $t$ is the nearest-neighbor hopping amplitude, $\mu$ is the chemical potential, and $\Delta=|\Delta|e^{i\theta}$ is the induced superconducting gap.
This toy model of $p$-wave superconductor is in contrast to standard $s$-wave pairing since it couples electrons with the same spin. When tuning the hopping amplitude and chemical potential to $
|\Delta|=t>0$, $\mu=0$,   
unpaired Majorana fermions appear at the boundary of the chain, which results in a topologically non-trivial phase.
Kitaev's quantum wire bridge proposal provides guidance for realizing the topological $p$-wave superconductors. 
In practice, this can be difficult to be physically realized because asides from the necessary condition for the unpaired Majorana, which requires an energy gap in the excitation spectrum, i.e., superconductivity in the bulk, the ground state of the connected chain has to be degenerate, where Majorana fermions still exist at the ends of the spin chain, and this is the parity condition.

When inducing the $p$-wave superconductivity,  Zeeman coupling can also be proximity induced in the film by an adjacent magnetic insulator.  
Aside from the proximity induced $p$-wave superconductivity in the film of the topological insulator and $s$-wave superconductor, we can replace the magnetic insulator by other materials. \textcite{sau2010non} showed that  tuning the Zeeman coupling of spins in the spin-orbit-coupled systems could also induce a topological phase transition, and
for the Zeeman coupling above the critical value, there are localized Majorana zero-energy modes at the two ends of a semiconducting quantum nanowire. 
This provides a proposal for searching for the Majorana zero-energy modes in materials with strong spin-orbit coupling.
The spinless toy model could not describe the conventional materials in which electrons have spin $1/2$, and more importantly, electrons in correlated materials inherently have multiple degrees of freedom, which can be difficult to simulate.
Nevertheless, we are able to use the variational quantum algorithms and our hybrid approach to determine the energy spectra of the bulk materials. For instance, we can store the degrees of freedom of electrons in the bulk materials with a quantum processor and also use the classical  (quantum)  tensor to represent the coupling effect.
We are able to solve the ground state and low lying excited eigenstates of strongly correlated materials involving spinless fermions that hop along a certain translationally invariant one-dimensional spin chain or a general spin model with spin-orbit interactions.  
By tuning the coupling and bulk properties, such as spin-orbit coupling, chemical potential $\mu$, external magnetic field, etc., we could drive the system to topologically non-trivial phase \cite{lutchyn2018majorana,lutchyn2010majorana,alicea2012new,beenakker2013search}. As the boundary condition of the superconductor heterostructures is usually much simpler than the bulk, the tensor-network-type algorithms would be suitable to resolve this category of problems.
The quantum simulation of the topological phase might be able to provide an avenue for the systematic search of topological superconductivity from heterostructures consisting of strongly correlated materials. We could represent multiple degrees of freedom of  correlated materials in a similar way, such as in the transition metals.

\subsection{Quantum field theories}

Since we are interested in simulating systems with large degrees of freedom using the hybrid tensor network, a perfect physical application might be the quantum simulation of quantum field theories. If we wish to simulate the quantum field theory process, for instance, the scattering process in a collider physics setup, we could consider using the hybrid tensor network to simulate it in the near-term quantum device. For instance, one may consider using the setup of the Jordan-Lee-Preskill algorithm \cite{Jordan:2011ne,Jordan:2011ci}, and try to use the hybrid tensor network to perform state preparation and time evolution. In particular, we could imagine split the whole system into small subsystems. For coupling constants and entanglement, both with an intermediate amount, our hybrid tensor network might be useful. 

Now, we make some more precise suggestions. Say that we are simulating a specific non-integrable, local, latticed-version of quantum field theory in a near-term quantum device, for instance, the $\lambda \phi^4$ theory in 1+1 dimension, with the Lagrangian density
\begin{align}
{\cal L}(\phi ) = \frac{1}{2}\left[ {{\partial ^\mu }\phi {\partial _\mu }\phi  - {m^2}{\phi ^2}} \right] - \frac{\lambda }{{4!}}{\phi ^4}.
\end{align}
Here $\phi$ is the scalar field where the whole local Hilbert space will be truncated, $\mu$ is the spacetime coordinate indices, $m$ is the mass and $\lambda$ is the coupling. 

We could compute the initial states and the time evolution process using variational quantum simulation. Say that we are taking $N$ sites in total. One could consider making a variational ansatz by dividing the whole system by two. A superposition of several product states will cover a large amount of the whole Hilbert space with bounded entanglement. One could deal with the system when the coupling is not weak, but also not super-strong using the above ansatz. The superposition coefficients might be treated quantumly, where the hybrid tensor network might play an important role. This is similar to the situation discussed in the numerical example we present in this paper. For instance, the ansatz could be naturally assigned when we are considering the time evolution of a two-particle scattering event
\begin{align}
\left| {{\rm{ansatz}}} \right\rangle  = \sum\limits_{{\rm{superposition}}} {\left| {{\rm{left}}} \right\rangle  \otimes \left| {{\rm{right}}} \right\rangle },
\end{align}
where the left and right Hilbert spaces could be split naturally. Note that we should be careful about providing enough entanglement towards the superposition since quantum field theory itself provides fruitful vacuum entanglement.  

One could provide another possibility where the hybrid tensor network might play an important role. Say that we wish to take a $d$-dimensional local Hilbert space in each site. One could consider the superposition of low energy states as the variational ansatz, where the high energy sector could be less excited. This corresponds to the hybrid tensor network of Fig.~\ref{fig:ansatzsup}(f). This suggestion might be helpful for a system with strong coupling when the system is approaching a critical point that is scale-free.

\subsection{Quantum gravity thought experiments}
\label{SM:example_high_energy}
Aside from the quantum simulation of quantum field theories, we could also consider using a hybrid tensor network to simulate quantum gravity thought experiments. There are, of course, many possible quantum simulation problems that are not solved by conventional methods. We will take a specific example here, the traversable wormhole. 

How to make a wormhole traversable, which is typically not allowed in general relativity due to the energy condition? Recently people found a beautiful thought experiment as a solution to this problem with the help of quantum information theory and holography (see \cite{Gao:2016bin}, and also \cite{Maldacena:2017axo,Maldacena:2018lmt,Yoshida:2017non,balasubramanian2014multiboundary,Maldacena:2016hyu}). This thought experiment is set in light of the dual descriptions between quantum entanglement and Einstein-Rosen bridge (ER=EPR) \cite{Maldacena:2013xja}, where a thermofield double state in two conformal field theories (CFTs) is dual to a two-sided wormhole in the dual gravity theory. With a coupling between two boundary CFTs, one is allowed to send a gravitational shockwave \cite{Shenker:2013pqa} into the bulk with negative energy, which introduces the time advance instead of Shapiro time delay when the ingoing boundary signal is passing through the spacetime discontinuity of the shockwave, making the wormhole traversable.

This thought experiment could be interpreted as a modification of Hayden-Preskill protocol \cite{Hayden:2007cs} for extracting quantum information dropping inside a black hole from Hawking radiation, but without introducing unknown Planckian physics to solve the no-cloning paradox where information is never duplicated in the bulk. It is also allowed for us to explore more physics about black hole interior with some concrete boundary theories assuming ER=EPR.

To introduce this formalism more concretely, we start from a simple classical, non-relativistic analog of the traversable wormhole, as illustrated by \cite{Maldacena:2017axo}. Imagine that there are two identical systems, $L$, and $R$, with no interaction initially. In both systems there are $N$ particles with position $\bar{x}^{L(R)}_i$ and momentum $\bar{p}^{L(R)}_i$. At the time $t=0$, they are set to have the same positions but opposite momenta,
\begin{align}
& \bar{x}_{i}^{L}(0)=\bar{x}_{i}^{R}(0)~,\nonumber\\
& \bar{p}_{i}^{L}(0)=-\bar{p}_{i}^{R}(0)~,
\end{align}
to simulate a thermofield double state for two boundary CFTs. Now, we are considering adding a small perturbation $\delta x_{\alpha}^{R}(t_R)$ at the original trajectory $\bar{x}_{\alpha}^{R}(t_R)$ for $t_R<0$ and a specific particle $\alpha$. According to the perturbation, generically, the location of other particles will be affected, especially when the system is chaotic. Thus, we obtain a perturbation for another particle $\beta$, at time $t=0$ in this system, $\delta x_{\beta}^{R}(0)$. Then we choose a coupling between $L$ and $R$ system at time $t=0$, for simplicity, with the potential
\begin{align}
V=\frac{1}{2}k_\text{in}{{(x_{\beta }^{R}-x_{\beta }^{L})}^{2}}~,
\end{align}
where $x_i^{L(R)}$ denotes the quantities after the perturbation. After turning on this potential for a very small time interval $g$, it leads to a perturbation over the momentum of $\beta$ in $L$, $\delta p_{\beta }^{L}(0^+)$, where $0^+$ denotes a time slightly larger than $t=0$ which could be set as $g$. Now turn off the interaction and evolve the system $L$ to a time scale $t_L$. Because of the perturbation of the momentum $\delta p_{\beta }^{L}(0^+)$ we get the perturbation of momentum for $\alpha$ in $L$ at time $t_L$, $\delta p_{\alpha }^{L}(t_L)$. This formalism could be regarded as teleportation from the perturbation of location for a specific particle in the system $R$ at time $t_R$, to the perturbation of momentum for the dual particle in the dual system $L$ at time $t_L$. If the dependence between $\delta x_{\alpha}^{R}(t_R)$ and $\delta p_{\alpha }^{L}(t_L)$ is simple, and some features are universal for different initial conditions of the system, we might say that the teleportation is successful.

One can generalize this logic to the quantum information  scenario~\cite{Maldacena:2017axo}. Consider the thermofield double state of two identical quantum systems, $L$ and $R$. Similarly, we set two times $t_L$ and $t_R$, and we add a small interaction $V=O_L(0)O_R(0)$ between two systems. Namely, we add the following factor into the path integral when evaluating correlation functions
\begin{align}
\exp (igV)=\exp (ig{{O}_{L}}(0){{O}_{R}}(0))~,
\end{align}
where $g$ is a small number. Now consider an operator $\phi_R$ and its dual operator $\phi_L$. Add a small unitary perturbation on $\phi_R$ at time $t_R$, $\exp(i\epsilon_R\phi_R(t_R))$, we measure $\phi_L$ at $t_L$ as
\begin{align}
   \left\langle {{e}^{-i{{\epsilon }_{R}}{{\phi }_{R}}({{t}_{R}})}}{{e}^{-igV}}{{\phi }_{L}}({{t}_{L}}){{e}^{igV}}{{e}^{i{{\epsilon }_{R}}{{\phi }_{R}}({{t}_{R}})}} \right\rangle 
  =\left\langle {{e}^{-igV}}{{\phi }_{L}}({{t}_{L}}){{e}^{igV}} \right\rangle -i{{\epsilon }_{R}}{{\left\langle \left[ {{\phi }_{R}}({{t}_{R}}),{{\phi }_{L}}({{t}_{L}}) \right] \right\rangle }_{V}}+\mathcal{O}(\epsilon _{R}^{2})~,
\end{align}
where
\begin{align}
{{\left\langle \left[ {{\phi }_{R}}({{t}_{R}}),{{\phi }_{L}}({{t}_{L}}) \right] \right\rangle }_{V}}\equiv \left\langle \left[ {{\phi }_{R}}({{t}_{R}}),{{e}^{-igV}}{{\phi }_{L}}({{t}_{L}}){{e}^{igV}} \right] \right\rangle~.
\end{align}
Thus, the non-vanishing value of ${{\left\langle \left[ {{\phi }_{R}}({{t}_{R}}),{{\phi }_{L}}({{t}_{L}}) \right] \right\rangle }_{V}}$ shows that there are some messages that have been teleported from $R$ to $L$. In the quantum system dual to the wormhole geometry, this quantity could measure the traversability of the wormhole, and the whole process has a clear geometric picture as introduced before. For explicit holographic models, for instance, the Sachdev-Ye-Kitaev (SYK) model \cite{Maldacena:2016hyu}, it is investigated in the very detail in \cite{Maldacena:2018lmt}.

It is known that an honest holographic model representing features of emergent gravity requires boundary systems to have a large number of degrees of freedom \cite{Maldacena:1997re}. Thus, to construct quantum gravitational dynamics in the bulk, strong computational power is needed to simulate complicated boundary quantum dynamics, lightning possible opportunities for quantum simulation. About quantum simulation of traversable wormholes, until now, people mostly use analog quantum simulation in cold-atomic systems to study them (see, for instance,  \cite{Yoshida:2018vly} or \cite{Brown:2019hmk}). Works based on analog simulations are very important for many quantum-mechanical problems at large scale, especially for those who could obtain dynamical quantities with real physical meanings. However, the variational quantum simulation might also be suitable and important, and it has the following certain advantages. Firstly, variational quantum simulation has more degrees of freedom to operate in the near-term digital quantum computer instead of a cold-atomic device without universal gates. Secondly, it is hard to implement a strongly-coupled model with Majonara fermions and all-to-all interactions in a cold-atomic system, while a digital setup could easily overcome this problem with certain encoding protocols. 

The hybrid tensor network would be specifically useful for traversable wormhole simulations for near-term quantum computing. For specific chaotic, holographic models, the algorithm will involve a state preparation of a thermofield double at a certain temperature, where the entanglement between two sides is around an intermediate scale. Since the coupling between the two systems is weak, we could use the hybrid tree tensor network, where the whole system is divided into two subsystems. We could use time-dependent variational quantum simulation with a hybrid tree tensor network using existing algorithms (for instance, see \cite{yuan2019theory}) to compute correlation functions between two sides. We leave those interesting possibilities and actual simulation in future research.

\end{document}